\documentclass[aps,prx,twocolumn,amsmath,amssymb,superscriptaddress,nolongbibliography,floatfix]{revtex4-2}
\usepackage{graphicx}
\usepackage{dcolumn}
\usepackage{bm}
\usepackage{hyperref}
\usepackage[mathlines]{lineno}
\usepackage{algorithm}
\usepackage[noend]{algpseudocode}
\usepackage{xcolor}
\usepackage{pifont}
\usepackage{tikz}
\usepackage[normalem]{ulem}
\usetikzlibrary{matrix}

\usepackage{float}

\usepackage{soul}
\usepackage{changes}

\begin{document}
\preprint{arXiv}

\title{Origin and fate of the pseudogap in the doped Hubbard model}

\author{Fedor \v{S}imkovic IV}
\affiliation{CPHT, CNRS, Ecole Polytechnique, Institut Polytechnique de Paris, Route de Saclay, 91128 Palaiseau, France}
\affiliation{Coll\`ege de France, 11 place Marcelin Berthelot, 75005 Paris, France}
\author{Riccardo Rossi}
\affiliation{Institute of Physics, \'Ecole Polytechnique F\'ed\'erale de Lausanne (EPFL), CH-1015 Lausanne, Switzerland}
\affiliation{
Laboratoire de Physique Th\'eorique de la Mati\`ere Condens\'ee, UPMC,
CNRS UMR 7600, Sorbonne Universit\'e, 4, place Jussieu, 75252 Paris Cedex 05, France}
\author{Antoine Georges}
\affiliation{Coll\`ege de France, 11 place Marcelin Berthelot, 75005 Paris, France}
\affiliation{Center for Computational Quantum Physics, Flatiron Institute, 162 Fifth Avenue, New York, NY 10010, USA}
\affiliation{CPHT, CNRS, Ecole Polytechnique, Institut Polytechnique de Paris, Route de Saclay, 91128 Palaiseau, France}
\affiliation{DQMP, Universit{\'e} de Gen{\`e}ve, 24 quai Ernest Ansermet, CH-1211 Gen{\`e}ve, Suisse}
\author{Michel Ferrero}
\affiliation{CPHT, CNRS, Ecole Polytechnique, Institut Polytechnique de Paris, Route de Saclay, 91128 Palaiseau, France}
\affiliation{Coll\`ege de France, 11 place Marcelin Berthelot, 75005 Paris, France}
\date{\today}

\begin{abstract}

We investigate the doped two-dimensional Hubbard model at finite temperature using controlled diagrammatic Monte Carlo calculations allowing for the computation of spectral properties in the infinite-size 
limit and, crucially, with arbitrary momentum resolution.
We show that three distinct regimes are found as a function of doping and interaction strength, 
corresponding to a weakly correlated metal with properties close to those of the non-interacting system, 
a correlated metal with strong interaction effects including a reshaping of the Fermi surface,
and a pseudogap regime at low doping in which quasiparticle excitations are selectively destroyed near the 
antinodal regions of momentum space.
We study the physical mechanism leading to the pseudogap and show that it forms both at weak coupling when the magnetic correlation length is large and at strong coupling when it is shorter. 
In both cases, we show that spin-fluctuation theory can be modified in order to account for the behavior of the 
non-local component of the self-energy.
We discuss the fate of the pseudogap as temperature goes to zero and 
show that, remarkably, this regime extrapolates precisely to the ordered stripe phase found 
by ground-state methods.
This handshake between finite temperature and ground-state
results significantly advances the elaboration
of a comprehensive picture of the physics of the doped Hubbard model.
\end{abstract}

\maketitle

The discovery of high-temperature superconductivity in copper oxide (cuprate) 
compounds~\cite{bednorz1986possible} has 
put into full light the relevance and urgency of the program outlined by Dirac in 1929~\cite{dirac1929prsl}, namely the need to develop 
practical methods of calculating and predicting the properties of large quantum systems of interacting particles. 
In this context, the Hubbard model~\cite{qin2021hubbard,arovas2021hubbard} 
quickly established itself~\cite{anderson1987resonating} as a fundamental and 
paradigmatic model. Although not fully realistic on a microscopic level, it captures important phenomena which are central to the `strong correlation' problem in a broad range of materials~ 
\cite{imada_rmp_1998}.

Especially fascinating among those phenomena is the highly unconventional nature of the 
non-superconducting `normal' state of the cuprate materials. At elevated temperature, this metallic state 
displays a partial destruction of the Fermi surface associated with the formation of a `pseudogap' corresponding to 
a depletion of the number of excitations available to the system. 
At low temperature, a rich diversity 
of phases with different kinds of intertwined long-range order are observed, 
most notably charge density waves. 
This raises a fundamental and still widely open question. 
Is the pseudogap state a fundamentally new kind of metallic state that could in principle 
be stabilized down to zero temperature without encountering an ordering instability, 
or is it a finite-temperature intermediate state  
which is always unstable to various kinds of long-range ordering?

Interrogating the Hubbard model about this fundamental question has proven to be a daunting challenge. 
In recent years, significant progress has been made in understanding the physical properties of this model 
through the development and use of controlled and accurate computational methods.  
However, a dichotomy largely exists among those computational studies. Wave-function 
based methods have addressed the nature of the ground-state and demonstrated that it is characterized 
by spin and charge ordering forming stripe patterns at low doping 
levels~\cite{white1998density,jianground2020,wietekstripes2021} 
as proposed early on in the context of mean-field studies~\cite{zaanencharged1989,schulz1990incommensurate,machida1989magnetism}.
Methods aimed at non-zero temperatures, on the other hand, have revealed that the Hubbard model 
hosts a pseudogap regime associated 
with magnetic correlations~\cite{maier2005review,tremblay2006,kotliar2006rmp,macridin_prl_2006,gull2010momentum,gunnarsson2015fluctuation,wu2018pseudogap,krien2021spin}.
Understanding the fate of the pseudogap state as temperature is lowered and how it connects to 
ground-states with long-range order calls for a `handshake' between different families 
of established computational methods and the development of new ones. 

Here, we provide an answer to some of these outstanding questions. 
Using an unbiased computational method, we identify the 
crossovers between the different regimes of the two-dimensional Hubbard model. 
We show that the pseudogap originates from magnetic correlations and by following its 
temperature dependence we provide evidence that it eventually evolves 
into a ground state with long-range spin and charge stripe order.

We study the doped repulsive two-dimensional Hubbard model defined by
$\hat{H} = \sum_{\mathbf{k}, \sigma} \epsilon_{\mathbf{k}} \,
    \hat{c}_{\mathbf{k} \sigma}^\dagger
    \hat{c}_{\mathbf{k} \sigma}^{\phantom{\dagger}}+U\sum_{\mathbf{r}} \hat{n}_{\mathbf{r}\uparrow}\,
    \hat{n}_{\mathbf{r} \downarrow} 
    -\mu \sum_{\mathbf{r},\sigma}\hat{n}_{\mathbf{r}\sigma}$ 
    with $\epsilon_{\mathbf{k}}=-2t\left(\cos k_x +\cos k_y\right)$. 
We investigate the temperature range $0.07t\leq T\leq0.25t$ and
coupling strengths up to $U=8.5t$.
We employ diagrammatic Monte Carlo, which is an unbiased, numerically exact technique 
formulated directly in the thermodynamic limit~\cite{prokof2008fermi,kozik2010diagrammatic,van2012feynman, wu_controlling,cdet,chen2019combined,vsimkovic2020extended}. 
It allows us to obtain physical quantities with arbitrary momentum resolution, 
which is a crucial advance over other available methods for the purposes of this study. 
We also compare our results with dynamical mean-field theory~\cite{antoine_dmft} 
and its 16-site dynamical cluster approximation 
extension~\cite{maier2005review,tremblay2006,kotliar2006rmp,gull2010momentum}, 
see Supplementary Information (SI). 
In the following, the hopping amplitude $t$ is used as the unit of energy and temperature. 

\subsection{Finite temperature phase diagram and crossovers}

\begin{figure*}[t]
\includegraphics[width=0.99\textwidth]{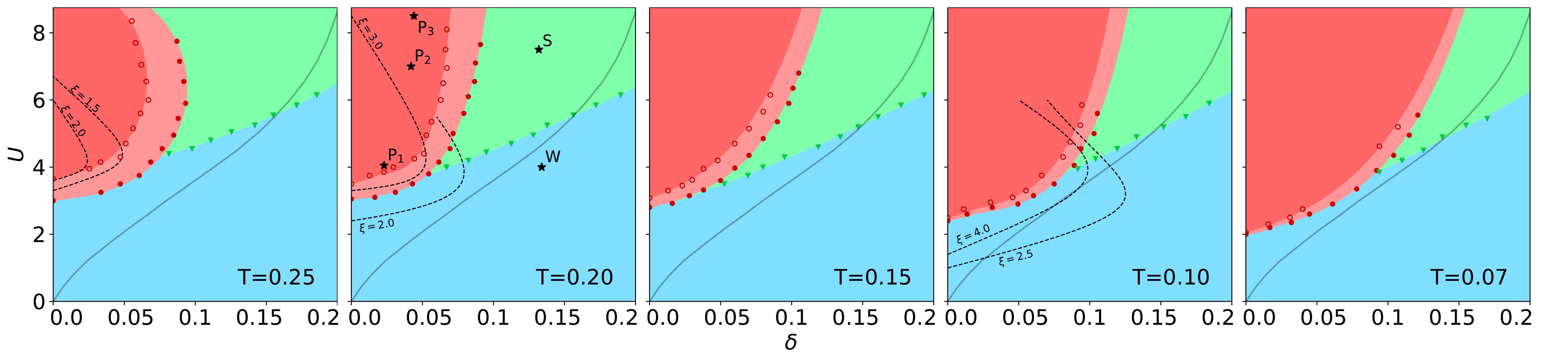}
\caption{
{\it Distinct regimes of the doped two-dimensional Hubbard model and their evolution as a function of temperature.} 
The blue and green regions correspond to a weakly and strongly correlated metal, respectively (see text). 
The red regions correspond to the regime where a pseudogap is present at the antinode. 
Dashed lines indicate contours of constant spin correlation length. 
The plain gray line, reproduced from Ref.~\cite{xu2021stripes},
indicates the region where the ground state displays long-range spin/charge stripe ordering.}
\label{fig_PD}
\end{figure*}

By analyzing our data, we identify in Fig.~\ref{fig_PD} three distinct physical regimes 
separated by well-defined crossovers, as a function of interaction strength $U\leq 8$ 
and hole-doping levels up to $\delta=1-n=20\%$ with $n$ the average electronic density 
per site. The five panels in Fig.~\ref{fig_PD} display how these crossovers evolve 
as a function of temperature $T=\{0.07,0.1,0.15, 0.2,0.25\}$. 
We describe these regimes here in qualitative terms, see SI for details. 
%

    The first regime (blue regions in Fig.~\ref{fig_PD}) corresponds to a weakly correlated metal. We identify the Fermi surface (FS) from the maximum of the spectral function proxy $A(\mathbf{k}) = -\frac{1}{\pi} \mathrm{Im} G(\mathbf{k}, i\omega_0)$, 
    where $\omega_0=\pi T$ is the lowest $p=0$ 
    Matsubara frequency $\omega_p=(2p+1)\pi T$~\footnote{Let us emphasize that, throughout this article, we {\it do not} perform 
    numerical analytic continuations but rather base our study on a direct analysis of imaginary-time/frequency 
    Monte Carlo data.}.
    In this regime, the FS is electron-like and very close to that of the non-interacting system.
    The spectral weight is uniform along the FS and relatively large, the self-energy is rather small and quasiparticles are long-lived. This regime is found at large doping levels or weak interactions, 
    a representative point being $W$ in Fig.~\ref{fig_PD}.

    For stronger interactions and intermediate doping levels, a different metallic regime (green regions) 
    is found in which the topology of the FS (as defined above) is hole-like. 
The system undergoes an interaction-driven Lifshitz transition when crossing into this regime from the weakly correlated metal, as indicated by the green triangles in Fig.~\ref{fig_PD}.
    The self-energy has become relatively large and the quasiparticle lifetime has decreased significantly. We refer to this regime as a strongly correlated metal (representative point: $S$).
    
    The red regions of Fig.~\ref{fig_PD} are characterized by a pseudogap (PG)
    at the antinode, which we detect by multiple criteria based on the spectral function, self-energy and uniform susceptibility, as discussed in details in the SI. Inside the pseudogap region, the lighter red region describes a regime
    where the electronic self-energy undergoes a series of
    momentum-selective crossovers, as discussed in more details below.
    We note that, in agreement with earlier studies~\cite{wu2018pseudogap}, a pseudogap only appears at strong and intermediate coupling when the FS topology is hole-like. 
Comparing the different panels in Fig.~\ref{fig_PD}, the PG regime is seen to become more extended as temperature is lowered.

    In order to address the important question of the interplay between the 
    spatial range of magnetic correlations and the 
    formation of the PG, we depict in Fig.~\ref{fig_PD} contour lines of equal spin correlation length $\xi$
    (dashed black lines). It is seen that at weak interactions the PG is associated with fairly long-ranged
    spin correlations~\cite{vilk_tremblay_1997,chubukov2010}. 
    In contrast, at stronger coupling, a PG is already found at 
    a high temperature when the correlation length is only a couple of lattice sites. 
    This is a key qualitative difference between the nature of the PG regime at weaker 
    (representative point $P_1$) and stronger coupling ($P_2, P_3$). Other 
    differences between these two regimes of the PG are further described below. 

    The nature of spin correlations actually undergoes a qualitative change from commensurate 
    ($\mathbf{q}=(\pi,\pi)$) at low doping 
    and higher temperature, to incommensurate ($\mathbf{q}=(q,\pi)$) at higher doping and lower temperature, 
    in agreement with the finding of previous studies~\cite{vsimkovic2021weak,huang2018stripe,maier_2021_fluctuating}.
    For $T=0.2$ the crossover happens around $10\%$ doping, while at $T=0.1$ roughly $7.5\%$ is sufficient (and very weakly dependent on $U$). 
    At these intermediate temperatures, the onset of the pseudogap is not directly sensitive
    to the commensurate or incommensurate nature of magnetic correlations.
    
\subsection{Fingerprints of crossovers between regimes}    
    
\begin{figure}[h]
\centering
\includegraphics[width=\columnwidth]{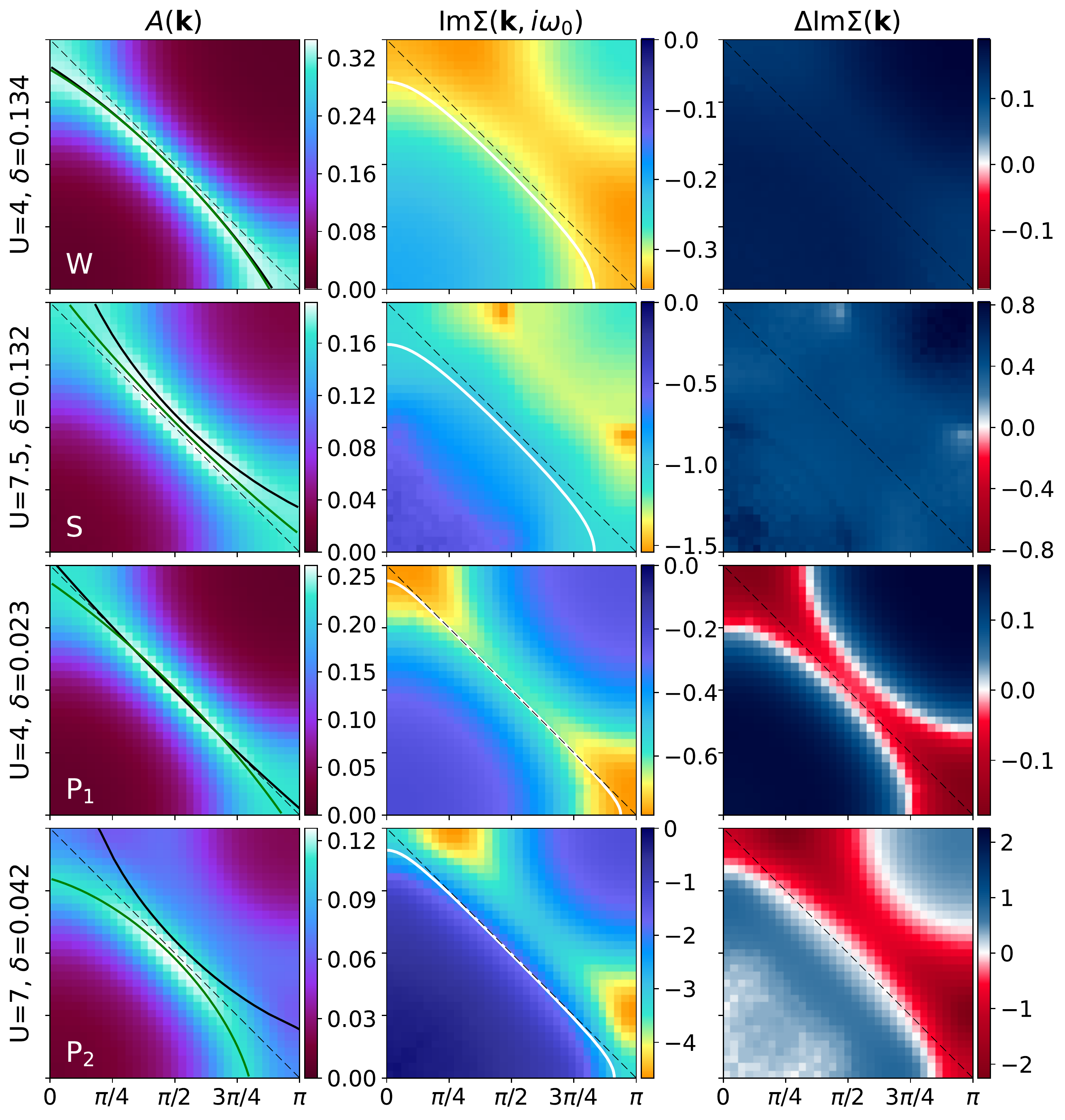}
\caption{
{\it Fingerprints of different regimes.}
The momentum-resolved spectral function $A(\mathbf{k})$, the imaginary part of the self-energy $\text{Im}\,\Sigma(\mathbf{k},i\omega_0)$, and the difference between the imaginary part of the self-energy at the two lowest Matsubara frequencies, $\Delta \,\text{Im}\,\Sigma(\mathbf{k})=\text{Im}\,\Sigma(\mathbf{k},i\omega_0)-\text{Im}\,\Sigma(\mathbf{k},i\omega_1)$, are shown for selected points $W,S,P_1,P_2$ 
in the phase diagram of Fig.~\ref{fig_regions}, 
at a temperature $T=0.2$. 
The white lines indicate the non-interacting Fermi surface. 
The green lines show the maximum of the spectral function. The zero-energy quasiparticle lines are shown
in black (see text).
\label{fig_regions}
}
\end{figure}

    In Fig.~\ref{fig_regions} we present momentum-resolved spectral properties
    over a quarter of the Brillouin zone (BZ) at $T=0.2$, for selected 
    points in the phase diagram of Fig.~\ref{fig_PD} corresponding to the different physical regimes.
    In the left column we display the low-energy spectral function proxy $A(\mathbf{k})$. 
The middle column represents the imaginary part of the self-energy at the lowest Matsubara frequency $\operatorname{Im}\Sigma(\mathbf{k},i\omega_0)$ and the right column shows the slope of the imaginary part of the self-energy obtained from the first two Matsubara frequencies: $\Delta \operatorname{Im}\Sigma(\mathbf{k}) \equiv \operatorname{Im}\Sigma(\mathbf{k},i\omega_0) - \operatorname{Im}\Sigma(\mathbf{k},i\omega_1)$. In a conventional metallic phase, this quantity is positive. We have chosen a momentum resolution of $64\times64$ for all quantities.

    The top row ($W$) of Fig.~\ref{fig_regions} corresponds to the weakly correlated metal ($U=4, n=0.866$).
    The Fermi surface obtained from the maximum of $A(\mathbf{k})$ in the BZ (green line) 
    essentially coincides with that of the non-interacting system (white line in the
    middle column). It is also very close to the
    zero-energy quasiparticle line obtained from $\epsilon_\mathbf{k} - \mu + \mathrm{Re}\Sigma(\mathbf{k}, i\omega_0) = 0$ (black line), which indicates the expected FS if lifetime effects coming
    from the imaginary part of the self-energy were neglected. The
    fact that it is close to the interacting FS is consistent with the rather small 
    and mostly uniform self-energy in this regime. 
    The spectral weight along the whole FS is large and essentially uniform in this regime.
    
    The second row ($S$) corresponds to a strongly correlated metal at intermediate doping ($U=7.5$ and $n=0.868$). 
    The momentum dependence of $\mathrm{Re}\,\Sigma(\mathbf{k},i\omega_0)$ induces a reshaping of the zero-energy
    quasiparticle line that becomes hole-like. 
    The FS is also hole-like, in strong contrast to that of the non-interacting system, 
    but its location does not quite coincide with that of the \mbox{zero-energy} quasiparticle line because 
    the imaginary part of the \mbox{self-energy} has prominent features close to $(\pi, \pi/2)$. 
    Lifetime effects suppress the antinodal spectral weight by about $9\%$ with respect to the node. The change of FS topology from electron-like to hole-like can be interpreted as a \mbox{correlation-induced} Lifschitz transition and is shown with green dots in Fig.~\ref{fig_regions}.
    For both regimes $S$ and $W$, $\Delta \mathrm{Im} \Sigma(\mathbf{k})$ is positive over the whole BZ, compatible with a metallic behaviour.
    
    The third row ($P_1$) is characteristic of a weak-coupling pseudogap ($U=4$ and $n=0.977$).
    Similarly to the case above, the real part of the self-energy
    shifts the zero-energy quasiparticle line to a hole-like shape. However, as
    the maxima in the imaginary part have moved closer to $(\pi,0)$,
    the interacting FS actually remains electron-like and the
    antinodal spectral weight is further reduced by roughly $13\%$ as compared to the node. As temperature is decreased,
    the antinodal spectral weight is reduced, confirming the presence of a pseudogap.

    The last row ($P_2$) displays a pseudogap with more pronounced features and located deeper inside the strong-coupling regime ($U=7$ and $n=0.977$). Quasiparticles are short-lived, with a Fermi arc forming around the nodal region, while the antinode 
    spectral intensity is reduced by about $14\%$. As in the
    weak-coupling pseudogap, the zero-energy quasiparticle line is strongly modified and is hole-like. Lifetime effects strongly suppress the spectral weight above the antiferromagnetic Brillouin zone and the maxima of the spectral function define an electron-like FS.
    For both $P_1$ and $P_2$, the slope of the self-energy is negative in a region just above the
    antiferromagnetic BZ. As shown in more
    detail in the SI, this quantity is a good indicator
    of the onset of the pseudogap region, which commences when the slope changes sign close to the antinode
    (full red circles in Fig.~\ref{fig_regions}). As the doping is decreased, the area
    of the Brillouin zone where the slope changes sign extends out of the antinodal region into the nodal region (indicated by open red circles in Fig.~\ref{fig_regions}, see SI for a detailed discussion).
    
    It is interesting to note that, as temperature is decreased in the pseudogap region, the imaginary part of the self-energy increases, although the momentum space region where it is large remains outside the antiferromagnetic BZ (see SI). 
    As a result, the lifetime effects get stronger with decreasing temperature
    at the antinode, while they have a much weaker effect at the node where
    quasiparticles remain quite coherent. This dichotomy between antinodal
    and nodal quasiparticles is also observed in cluster extensions of
    dynamical mean-field theory 
    ~\cite{parcollet2004,tremblay2006,haule2007strongly,ferrero2009,gull2010momentum}. 

\subsection{Insights from a modified spin-fluctuation theory}

\begin{figure}
\centering
\includegraphics[width=\columnwidth]{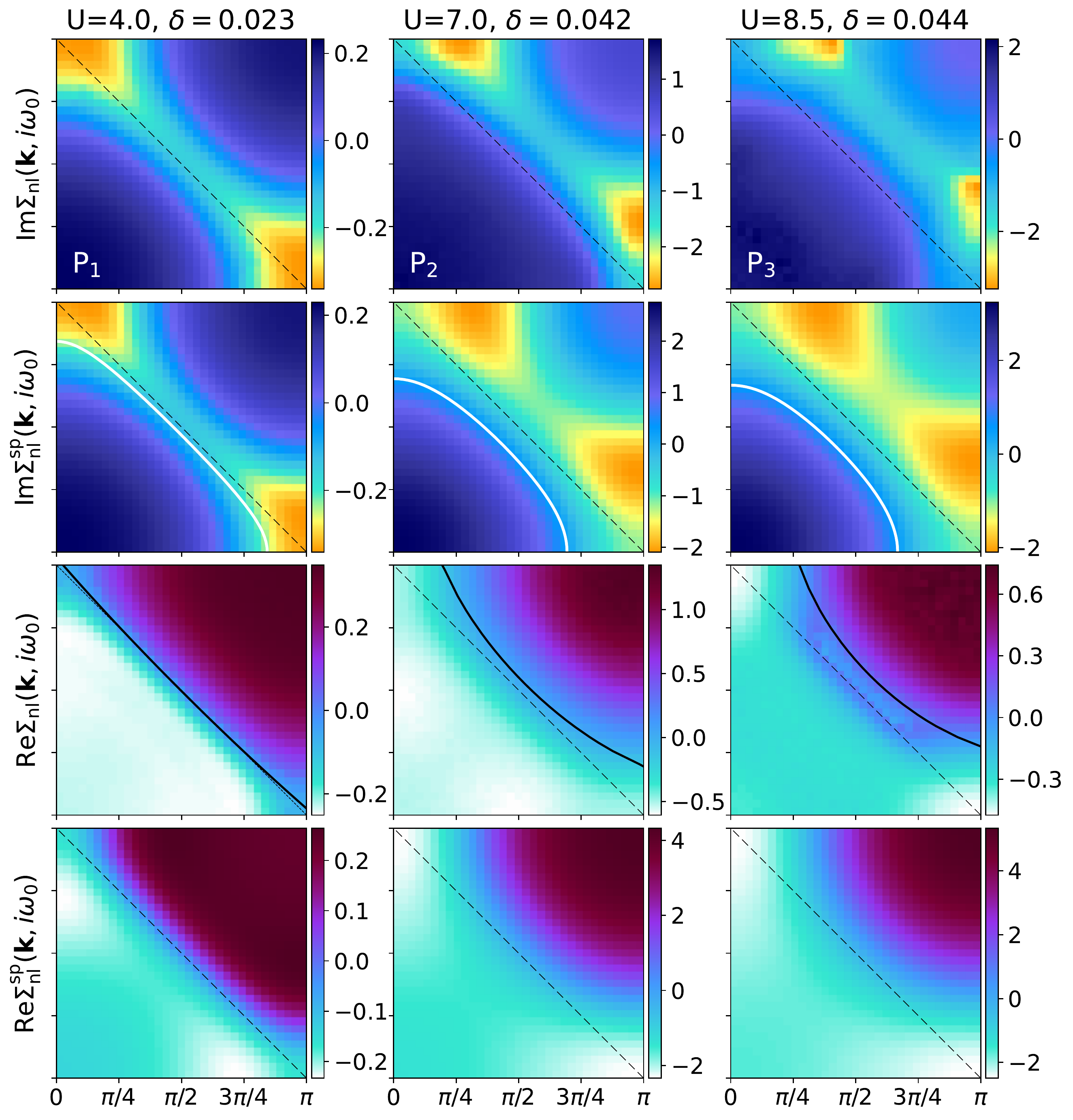}
\caption{{\it Comparison to modified spin-fluctuation theory.} 
Momentum resolved non-local components of the real (bottom rows) and imaginary parts (top rows) of the self-energy are shown for selected points within the pseudogap regime at $T=0.2$.
The numerically exact results (first and third rows) are compared to a spin-fluctuation theory 
fitting procedure (second and fourth rows), as described in the main text. Black lines indicate the 
zero-energy quasiparticle lines. White lines indicate the FS 
associated with the bare Green's function $G_0(\mathbf{k}, i\omega_0=i\pi T, \bar{\mu})$ used in the fits.}
\label{fig_fitting}
\end{figure}

\begin{figure}[h!]
\centering
\includegraphics[width=0.8\columnwidth]{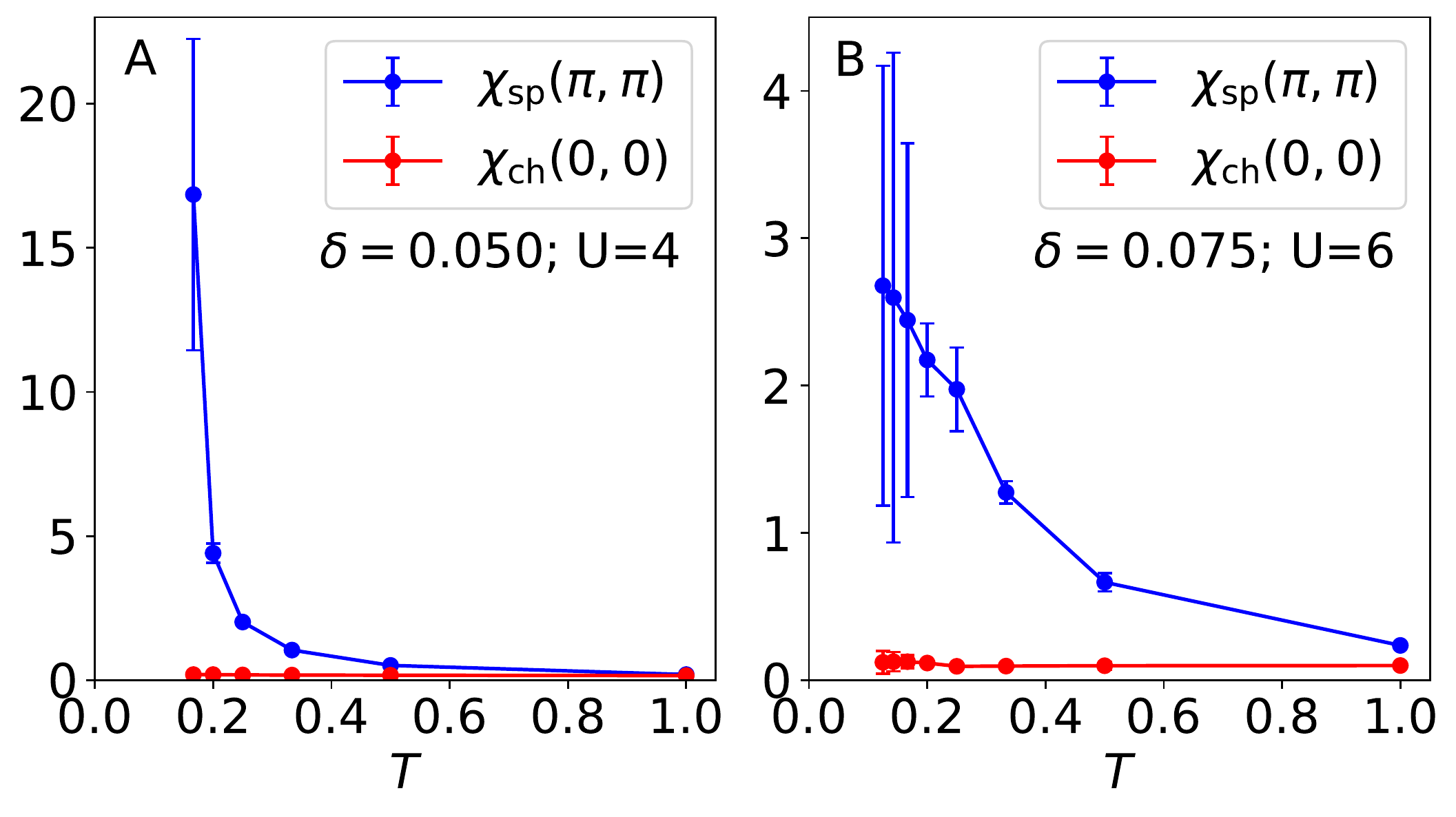}
\caption{
{\it Spin and charge correlations.} 
The zero-frequency spin and charge susceptibilities at their maximum value in momentum space are displayed 
as a function of temperature, for representative examples of the 
weak-coupling (left panel) and strong-coupling (right panel) pseudogap regimes.}
\label{fig_spinvscharge}
\end{figure}

In this section, we ask whether the PG regime can be described by some form of spin-fluctuation theory. 
It is known from past work~\cite{vilk_tremblay_1997,chubukov2010,M7} that this is indeed the case at weak coupling.  
The question is whether such a description is also possible at strong coupling, 
despite the rather short correlation length which invalidates the conditions for a conventional application of spin-fluctuation theory. 
We note that previous work based on a `fluctuation diagnostics'~\cite{schafer2021} in the framework of 
both cluster extensions of DMFT~\cite{gunnarsson2015fluctuation} and diagrammatic MC~\cite{wu_controlling} 
has shown that the spin channel is indeed where the action takes place in relation to the formation of the PG.
This point is further reinforced by a direct evaluation of the spin and charge susceptibilities 
in both the weak- and the strong-coupling PG region. 
These quantities are displayed in Fig.~\ref{fig_spinvscharge} (see
also Ref.~\cite{vsimkovic2021weak}) and indicate that the 
physics is dominated by spin fluctuations in the temperature regime that we investigate 
while the charge response is, in contrast, very weak. 
This clearly points at the pseudogap being of magnetic origin rather than due to the fluctuations 
of a low-$T$ charge order, which is also consistent with the conclusions from cluster extensions of 
DMFT. 
These considerations provide a strong incentive for attempting a spin-fluctuation-inspired description of the PG. 

To this end, we divide the self-energy into a local (uniform in momentum space) and non-local part:
$\Sigma = \Sigma_{\text{loc}} + \Sigma_{\text{nl}}$.  
The local part is quite large, especially in the strong-coupling regime, 
and is not adequately approximated by spin-fluctuation theory. 
In the SI we assess the accuracy of dynamical mean-field theory (DMFT) in computing the local component.
For the non-local component, we draw inspiration from Hedin's equation 
$\Sigma= - G\star W\star \Gamma$ 
involving convolutional products over momenta and frequencies, with $W = U- U^2\chi_{\text{sp}}$. Here $\Gamma$ is the vertex function and $\chi_{\text{sp}}$ is the dynamical spin susceptibility.
We approximate this exact expression by considering the following {\it ansatz} for the non-local part of the self-energy:
\begin{equation}\label{eq:spinfluct} 
    \Sigma^{\text{sp}}_{\text{nl}}(\mathbf{k},i \omega_0) = 
    \bar{\gamma} \, U^2 \, T  \frac{1}{N}\sum_{\mathbf{q}} 
    \frac{G_0(\mathbf{k}+\mathbf{q}, i\omega_0,\bar{\mu})}{(\bm{\pi} - q)^2 + \bar{\xi}^{-2} }.
    \end{equation}
Here, we have replaced the vertex $\Gamma$ by a constant $\bar{\gamma}$ 
and the effective spin interaction $W$ by an Ornstein-Zernike form of the commensurate spin susceptibility 
$\chi_{\text{sp}}$ centered around $\bm{\pi} = (\pi,\pi)$ and with correlation length $\bar{\xi}$. 
In Eq.~\ref{eq:spinfluct}, we use a non-interacting form of the Green's function $G_0$ which, importantly, 
involves an adjustable chemical potential $\bar{\mu}$. Furthermore, we have limited the frequency convolution to the zero bosonic Matsubara frequency only, an approximation which is 
    known to become more accurate at low-$T$ when a pseudogap opens~\cite{vilk_tremblay_1997,M7}.
    We use a fitting procedure on our numerically exact data in order to determine the three parameters $\bar{\gamma},\bar{\mu},\bar{\xi}$, 
    and consider only the imaginary part of the self-energy in the optimization process. 
    In Fig.~\ref{fig_fitting}  we present the real and imaginary parts of the non-local self-energy for three different points 
    within the pseudogap regime, comparing our numerically exact results to the optimized spin-fluctuation expression.
    
     The first column of Fig.~\ref{fig_fitting} shows an example of the weak-coupling pseudogap regime ($P_1$: $U=4$ and $n=0.977$). Here we find remarkable agreement between the self-energy fit and the original data, both for the real and the imaginary part. The momentum dependence as well as overall magnitude of the fit is close to perfect. The parameter $\bar{\mu}=-0.26$ is somewhat lower than the non-interacting chemical potential corresponding to the density ($\mu_0 = -0.10$). The parameter $\bar{\xi}=5.0$ is close to the actual (commensurate) value of $\xi=4.3$ obtained numerically. Finally, $\bar{\gamma}=0.5$, which hints to the fact that the $\Gamma$ vertex is relatively uniform and not very large.
     
     The middle column corresponds to a point in the strong-coupling pseudogap regime ($P_2$: $U=7$, $n=0.958$). 
    Our spin-fluctuation {\it ansatz} still produces a qualitatively correct picture, 
    but differences are apparent at the quantitative level. 
    The extrema in the imaginary part are in the correct location, although somewhat broader than in the data. 
    Let us emphasize that adjusting $\bar{\mu}$ is essential to
    correctly place the extrema of $\mathrm{Im}\Sigma$ (see the
    white lines in Fig.~\ref{fig_fitting}). Using such a
    freedom for the non-interacting starting point is indeed
    often used to improve perturbative 
    expansions~\cite{rubtsov2005continuous,olivier,wu_controlling,shifted_action,rdet,vsimkovic2021weak}.
    The fitting procedure yields $\bar{\mu}=-0.89$, $\bar{\xi}=1.60$ (we expect the exact value to be $\xi \lesssim 2$) and $\bar{\gamma}=4.90$, which points to the fact that the $\Gamma$ vertex becomes large in this regime. 
    Remarkably, the real part has the correct momentum structure, but since our fitting procedure for $\bar{\gamma}$ only takes into account the imaginary part, the overall magnitude of the real part is roughly four times too large. 
    The fact that a single consistent value of $\bar{\gamma}$ cannot be found to fit both the imaginary and real parts of the 
    self-energy points to a strong momentum dependence of the $\Gamma$ vertex.
     The right column of Fig.~\ref{fig_fitting} has the same density and an even larger coupling strength ($P_3$: $U=8.5$, $n=0.956$). 
    From fitting the self-energy we observe a continuation of the trend found at lower $U$, where $\bar{\mu}=-1.03$, $\bar{\xi}=1.25$ and $\bar{\gamma}=4.84$.

    We conclude that a properly modified spin-fluctuation theory provides an excellent theoretical description of the {\it non-local} 
    part of the self-energy in the weak-coupling pseudogap regime and still does qualitatively well in its strongly-coupling counterpart. 
    In the latter regime, however, a quantitative fit of the imaginary and real parts cannot be simultaneously achieved. 
    Amending our {\it ansatz} by the possibility of an incommensurate $\chi_{\text{sp}}$ with maxima at $\mathbf{q}=(\pi \pm \bar{\delta},\pi)$ 
    does not improve the fitting procedure. 
    The authors of Ref.~\cite{krien2021spin} made the interesting observation that the vertex becomes complex at strong coupling 
    and proposed that this may be a key to understanding the PG at strong coupling in a spin fluctuation framework. 
    However, we found that allowing for a complex phase $\bar{\gamma} = \bar{\gamma_0} e^{i \bar{\kappa}}$ with the idea of mixing 
    contributions from the real and imaginary self-energies does not actually lead to a better fit. 
    These observations point to the importance of the momentum and frequency dependence of the vertex function in the strong coupling regime. 
We also note that using the interacting Green's function $G$ within our {\it ansatz} (in the spirit of self-consistent, or bold perturbation theory) instead of a non-interacting $G_0$ (with an adjustable $\overline{\mu}$) yields much poorer fits~\cite{vilk_tremblay_1997,M7}. 


\subsection{The fate of the pseudogap at low temperature: handshake with ground state methods}

\begin{figure}[h!]
\centering
\includegraphics[width=0.81\columnwidth]{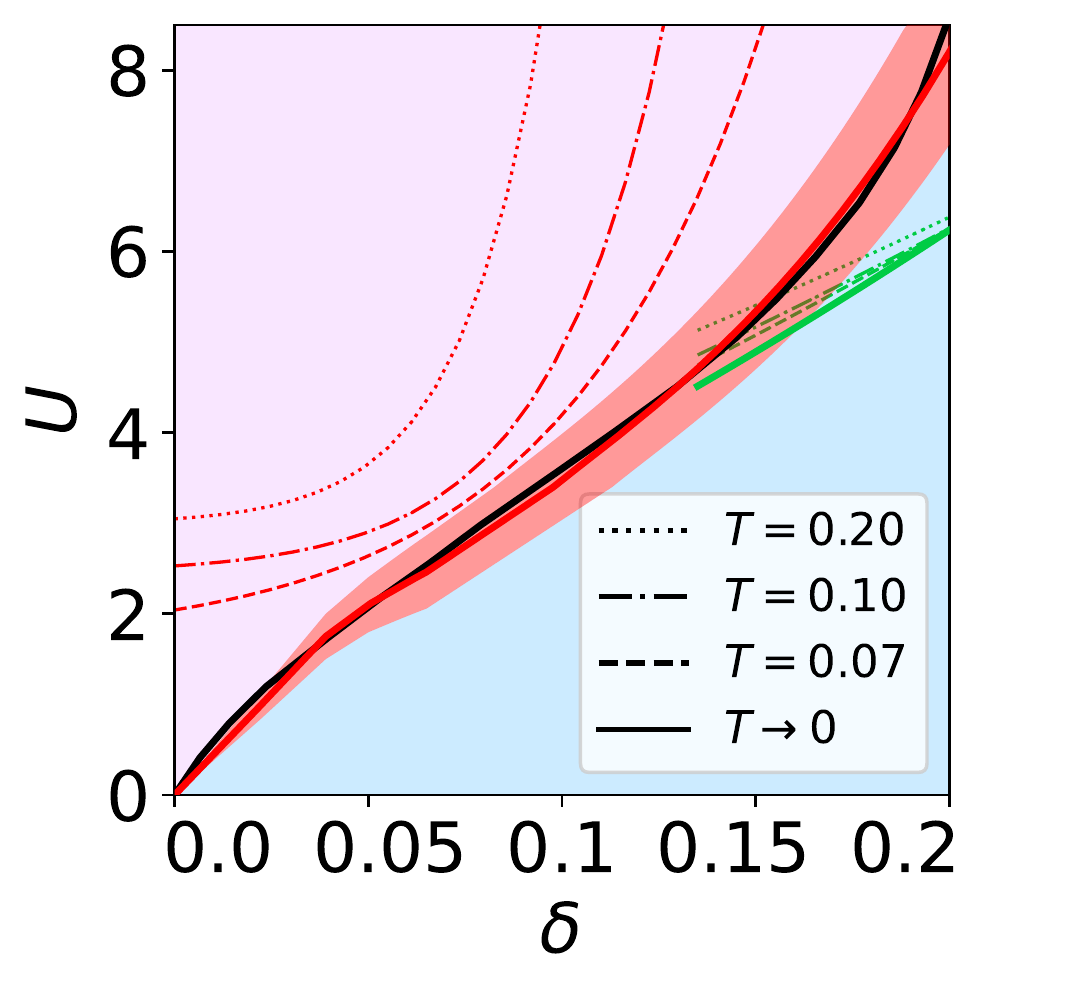}
\caption{
{\it Fate of the pseudogap at low temperature.} 
By extrapolating to $T=0$ the pseudogap crossover at different 
temperatures (dotted, dash-dotted and dashed red lines), we estimate the 
boundary (plain red line) separating the low-doping regime with a pseudogap 
to the higher doping one without. 
This boundary is found to coincide within error bars to the phase transition line 
reported in Ref.~\cite{xu2021stripes} (black line) which separates 
ground-states with (pink) and without (light blue) 
spin/charge stripe ordering. 
The green line indicates the low-$T$ extrapolation of the Lifschitz transition line.
}
\label{fig_extr}
\end{figure}

A major open issue in relation to the doped Hubbard model is the connection between the physical nature of the ground state and 
that of finite temperature crossovers. Distinct sets of computational methods have been successfully used in investigating 
separately these questions~\cite{arovas2021hubbard,qin2021hubbard},
but a handshake between these approaches is still mostly lacking. In the present context, an outstanding question is 
what happens to the pseudogap regime upon cooling towards $T=0$. Does charge and/or spin ordering take place? Do the Fermi arcs
observed at high-$T$ eventually evolve into a reconstruction of the Fermi surface at low-$T$? 
These questions have also been the subject of intense debate and experimental investigations in the context of cuprates \cite{proust2019remarkable}.  
Here, we make progress towards such a handshake by performing an extrapolation towards $T=0$ 
of the crossovers found above and comparing to a recent ground-state study~\cite{xu2021stripes}. 


     In Fig.~\ref{fig_extr} we show the results of an extrapolation down to $T=0$ of our numerically exact finite-$T$ results 
     for the position of the various crossovers (details are provided in SM). 
     The plain red line on Fig.~\ref{fig_extr} indicates the extrapolated $T=0$ boundary between the (pink) region 
     with a PG and that without a PG (light blue). 
     The $T=0$ extrapolation of the FS topology (Lifschitz) crossover coincides with the 
     PG boundary up to a doping level of around $13.5\%$, and deviates from it at higher doping. 
     The full black line on Fig.~\ref{fig_extr} is adapted from Ref.~\cite{xu2021stripes}. 
     It represents the ground state phase transition between a phase with long-range spin 
     and charge stripe 
     order~\cite{zaanen1989charged,machida1989magnetism,schulz1990incommensurate,white1998density,jianground2020} 
     and a phase at higher doping levels with only short-range spin and/or charge correlations. 
     This boundary was computed by auxiliary field quantum Monte Carlo (AFQMC), and the results are in good agreement with a variational Monte Carlo study~\cite{sorella2021phase}.
     Remarkably, our result for the extrapolated pseudogap boundary is in near-perfect agreement with 
     this phase transition line. 
     This provides striking evidence that the pseudogap regime eventually becomes stripe-ordered at zero $T$.  
     This is one of the major conclusions of our work, which answers the long-standing question of the 
     fate of the pseudogap regime as temperature is lowered towards the ground state. 
     

\subsection{Piecing together a unifying picture}

\begin{figure}[h!]
\centering
\includegraphics[width=0.7\columnwidth]{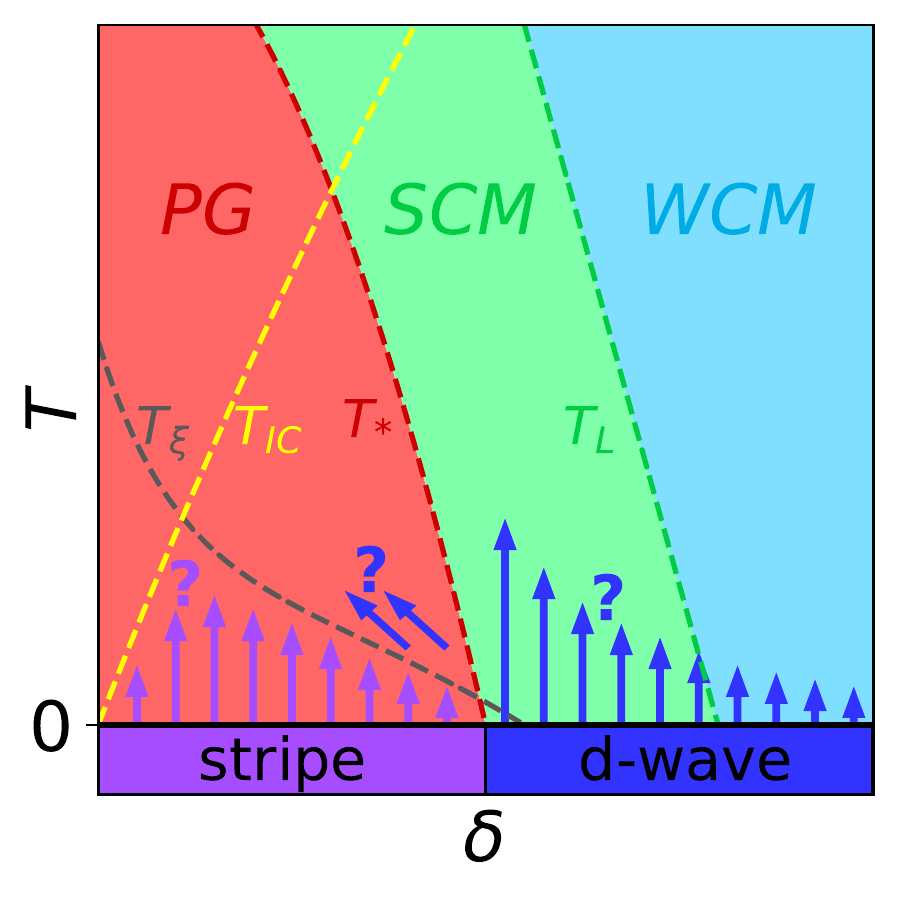}
\caption{
{\it Proposed unifying picture at strong coupling.} 
This schematic strong coupling phase diagram as a function of temperature and doping 
indicates the pseudogap (red), strongly correlated (green) and weakly correlated (blue) metallic regimes 
discussed in the text. 
The dashed gray and yellow lines refer to spin physics: 
below the former ($T_\xi$) the magnetic correlation length exceeds a specified value, 
while below the latter ($T_{IC}$) the magnetic correlations become incommensurate. 
The ground-state is a spin/charge ordered stripe state at low doping (purple region) 
and a superconductor at higher doping (blue region). 
These two zero-temperature phases are likely to extend to finite temperature 
in a manner which is not yet fully understood, 
as represented by arrows and question marks. 
\label{fig_GS}
}
\end{figure}

We conclude this work by attempting to provide a unifying qualitative picture of the physical regimes of the
doped two-dimensional Hubbard model, also emphasizing the questions that are still open.

     In Fig.~\ref{fig_GS} we present a sketch of 
     the proposed strong-coupling phase diagram as a function of temperature and doping level. 
     The pseudogap and Lifschitz crossovers from Fig.~\ref{fig_extr} are indicated by $T^{*}$ and $T_{\text{L}}$.
    Additionally, we display the commensurate to incommensurate spin fluctuation crossover $T_{\text{IC}}$, 
    and the crossover from a short to a long spin correlation length $T_{\xi}$
    which were identified in Ref.~\cite{vsimkovic2021weak}. 
    As established in previous work~\cite{wietekstripes2021,vsimkovic2021weak} and also shown above, 
    charge correlations only pick up at much lower temperatures. 
    This implies that the formation of the pseudogap is driven by spin correlations, consistently 
    with the conclusions from cluster extensions of DMFT~\cite{parcollet2004,tremblay2006,haule2007strongly,ferrero2009,gull2010momentum}.
    Charge correlations are, however, a necessary ingredient for the stripe ordering  
    which was established to exist in the ground state \cite{qinabsence2020,xu2021stripes,sorella2021phase}. 
    We postulate that charge correlations develop only once incommensurate spin correlations have 
    grown to be sufficiently long-ranged
    (as seen in Fig.~\ref{fig_extr}). 
    Very recently, strong indications that charge long-range (or quasi long-range) order indeed takes place through a phase transition 
    at a low non-zero temperature were obtained~\cite{xiao2022temperature}.
    From our data we identify the ideal region of parameters to further investigate this question to be $n\sim0.9$ and $U\sim4$. 
    Incidentally, this is where we experience most difficulties with the resummation of perturbative series from diagrammatic Monte Carlo.
    
    As doping is further increased, the stripe order eventually ceases to
    exist in the ground state. In the weak-to-intermediate coupling
    regime and in the absence of other instabilities,
    the Hubbard model will eventually turn superconducting because of
    the Kohn-Luttinger effect~\cite{chubukov1993}, albeit at
    possibly very low temperatures. For $U\lesssim 4$, it has been established that this instability is of the $d_{x^2-y^2}$ type up to $40\%$ doping~\cite{deng,simkovicsuperfluid2021}.
    At stronger coupling, it has been shown that stripe ordering wins over
    superconductivity over a significant range of parameter space 
    in the absence of next-nearest neighbor hopping~\cite{qinabsence2020}. The situation
    at doping levels just above the critical value where stripe order disappears
    is still under investigation but recent results seem to suggest that
    strong coupling superconductivity exists over some range
    of doping~\cite{sorella2021phase}.
    Finite-temperature studies using approximate methods have also found $d$-wave superconductivity to exist in the vicinity of the pseudogap crossover~\cite{gull2013super,kitatani2019why}.
    However, more work is needed to provide conclusive results about
    the critical temperature and doping extension of this phase. Further, it would be insightful to study changes in entropy within this phase diagram as it could indicate the vicinity to phase separation or favor high-temperature superconductivity ~\cite{lenihan2020entropy}.
    
    In conclusion, we have investigated the two-dimensional Hubbard model using
    a numerically exact diagrammatic Monte Carlo algorithm. We have established
    the finite temperature crossover diagram with a particular focus on the
    pseudogap regime. We have shown that the latter originates 
    in antiferromagnetic spin correlations which are longer ranged 
    at weak coupling and shorter ranged at strong coupling. A 
    suitably modified spin-fluctuation theory was found to successfully reproduce some 
    of the salient qualitative features of the pseudogap regime. 
    A central result of our work is that the pseudogap regime eventually turns into a
    stripe-ordered phase at zero-temperature. 
    Extending the present study to lower temperatures and to a non-zero 
    next nearest-neighbor hopping is, without doubt, highly desirable and will require 
    overcoming further computational challenges.

\acknowledgements
We are grateful to Subir Sachdev, Sandro Sorella, Steven White, Bo Xiao and Shiwei Zhang for insightful discussions.
This work was granted access to the HPC resources of TGCC and IDRIS under the allocations A0090510609 and
A0110510609 attributed by GENCI (Grand Equipement National de Calcul Intensif).
F.S. and M.F. acknowledge the support of the Simons Foundation within the Many Electron Collaboration framework.
The Flatiron Institute is a division of the Simons Foundation. 

\bibliography{main_biblio}

\begin{thebibliography}{64}%
\makeatletter
\providecommand \@ifxundefined [1]{%
 \@ifx{#1\undefined}
}%
\providecommand \@ifnum [1]{%
 \ifnum #1\expandafter \@firstoftwo
 \else \expandafter \@secondoftwo
 \fi
}%
\providecommand \@ifx [1]{%
 \ifx #1\expandafter \@firstoftwo
 \else \expandafter \@secondoftwo
 \fi
}%
\providecommand \natexlab [1]{#1}%
\providecommand \enquote  [1]{``#1''}%
\providecommand \bibnamefont  [1]{#1}%
\providecommand \bibfnamefont [1]{#1}%
\providecommand \citenamefont [1]{#1}%
\providecommand \href@noop [0]{\@secondoftwo}%
\providecommand \href [0]{\begingroup \@sanitize@url \@href}%
\providecommand \@href[1]{\@@startlink{#1}\@@href}%
\providecommand \@@href[1]{\endgroup#1\@@endlink}%
\providecommand \@sanitize@url [0]{\catcode `\\12\catcode `\$12\catcode
  `\&12\catcode `\#12\catcode `\^12\catcode `\_12\catcode `\%12\relax}%
\providecommand \@@startlink[1]{}%
\providecommand \@@endlink[0]{}%
\providecommand \url  [0]{\begingroup\@sanitize@url \@url }%
\providecommand \@url [1]{\endgroup\@href {#1}{\urlprefix }}%
\providecommand \urlprefix  [0]{URL }%
\providecommand \Eprint [0]{\href }%
\providecommand \doibase [0]{https://doi.org/}%
\providecommand \selectlanguage [0]{\@gobble}%
\providecommand \bibinfo  [0]{\@secondoftwo}%
\providecommand \bibfield  [0]{\@secondoftwo}%
\providecommand \translation [1]{[#1]}%
\providecommand \BibitemOpen [0]{}%
\providecommand \bibitemStop [0]{}%
\providecommand \bibitemNoStop [0]{.\EOS\space}%
\providecommand \EOS [0]{\spacefactor3000\relax}%
\providecommand \BibitemShut  [1]{\csname bibitem#1\endcsname}%
\let\auto@bib@innerbib\@empty
\bibitem [{\citenamefont {Bednorz}\ and\ \citenamefont
  {M{\"u}ller}(1986)}]{bednorz1986possible}%
  \BibitemOpen
  \bibfield  {author} {\bibinfo {author} {\bibfnamefont {J.~G.}\ \bibnamefont
  {Bednorz}}\ and\ \bibinfo {author} {\bibfnamefont {K.~A.}\ \bibnamefont
  {M{\"u}ller}},\ }\href@noop {} {\bibfield  {journal} {\bibinfo  {journal}
  {Zeitschrift f{\"u}r Physik B Condensed Matter}\ }\textbf {\bibinfo {volume}
  {64}},\ \bibinfo {pages} {189} (\bibinfo {year} {1986})}\BibitemShut
  {NoStop}%
\bibitem [{\citenamefont {Dirac}(1929)}]{dirac1929prsl}%
  \BibitemOpen
  \bibfield  {author} {\bibinfo {author} {\bibfnamefont {P.~A.~M.}\
  \bibnamefont {Dirac}},\ }\href
  {https://doi.org/http://doi.org/10.1098/rspa.1929.0094} {\bibfield  {journal}
  {\bibinfo  {journal} {Proc. R. Soc. Lond. A}\ }\textbf {\bibinfo {volume}
  {123}},\ \bibinfo {pages} {714–733} (\bibinfo {year} {1929})}\BibitemShut
  {NoStop}%
\bibitem [{\citenamefont {Qin}\ \emph {et~al.}(2021)\citenamefont {Qin},
  \citenamefont {Sch{\"a}fer}, \citenamefont {Andergassen}, \citenamefont
  {Corboz},\ and\ \citenamefont {Gull}}]{qin2021hubbard}%
  \BibitemOpen
  \bibfield  {author} {\bibinfo {author} {\bibfnamefont {M.}~\bibnamefont
  {Qin}}, \bibinfo {author} {\bibfnamefont {T.}~\bibnamefont {Sch{\"a}fer}},
  \bibinfo {author} {\bibfnamefont {S.}~\bibnamefont {Andergassen}}, \bibinfo
  {author} {\bibfnamefont {P.}~\bibnamefont {Corboz}},\ and\ \bibinfo {author}
  {\bibfnamefont {E.}~\bibnamefont {Gull}},\ }\href@noop {} {\bibfield
  {journal} {\bibinfo  {journal} {arXiv preprint arXiv:2104.00064}\ } (\bibinfo
  {year} {2021})}\BibitemShut {NoStop}%
\bibitem [{\citenamefont {Arovas}\ \emph {et~al.}(2021)\citenamefont {Arovas},
  \citenamefont {Berg}, \citenamefont {Kivelson},\ and\ \citenamefont
  {Raghu}}]{arovas2021hubbard}%
  \BibitemOpen
  \bibfield  {author} {\bibinfo {author} {\bibfnamefont {D.~P.}\ \bibnamefont
  {Arovas}}, \bibinfo {author} {\bibfnamefont {E.}~\bibnamefont {Berg}},
  \bibinfo {author} {\bibfnamefont {S.}~\bibnamefont {Kivelson}},\ and\
  \bibinfo {author} {\bibfnamefont {S.}~\bibnamefont {Raghu}},\ }\href@noop {}
  {\bibfield  {journal} {\bibinfo  {journal} {arXiv preprint arXiv:2103.12097}\
  } (\bibinfo {year} {2021})}\BibitemShut {NoStop}%
\bibitem [{\citenamefont {Anderson}(1987)}]{anderson1987resonating}%
  \BibitemOpen
  \bibfield  {author} {\bibinfo {author} {\bibfnamefont {P.~W.}\ \bibnamefont
  {Anderson}},\ }\href {https://doi.org/10.1126/science.235.4793.1196}
  {\bibfield  {journal} {\bibinfo  {journal} {Science}\ }\textbf {\bibinfo
  {volume} {235}},\ \bibinfo {pages} {1196} (\bibinfo {year} {1987})},\ \Eprint
  {https://arxiv.org/abs/https://www.science.org/doi/pdf/10.1126/science.235.4793.1196}
  {https://www.science.org/doi/pdf/10.1126/science.235.4793.1196} \BibitemShut
  {NoStop}%
\bibitem [{\citenamefont {Imada}\ \emph {et~al.}(1998)\citenamefont {Imada},
  \citenamefont {Fujimori},\ and\ \citenamefont {Tokura}}]{imada_rmp_1998}%
  \BibitemOpen
  \bibfield  {author} {\bibinfo {author} {\bibfnamefont {M.}~\bibnamefont
  {Imada}}, \bibinfo {author} {\bibfnamefont {A.}~\bibnamefont {Fujimori}},\
  and\ \bibinfo {author} {\bibfnamefont {Y.}~\bibnamefont {Tokura}},\ }\href
  {https://doi.org/10.1103/RevModPhys.70.1039} {\bibfield  {journal} {\bibinfo
  {journal} {Rev. Mod. Phys.}\ }\textbf {\bibinfo {volume} {70}},\ \bibinfo
  {pages} {1039} (\bibinfo {year} {1998})}\BibitemShut {NoStop}%
\bibitem [{\citenamefont {White}\ and\ \citenamefont
  {Scalapino}(1998)}]{white1998density}%
  \BibitemOpen
  \bibfield  {author} {\bibinfo {author} {\bibfnamefont {S.~R.}\ \bibnamefont
  {White}}\ and\ \bibinfo {author} {\bibfnamefont {D.~J.}\ \bibnamefont
  {Scalapino}},\ }\href {https://doi.org/10.1103/PhysRevLett.80.1272}
  {\bibfield  {journal} {\bibinfo  {journal} {Phys. Rev. Lett.}\ }\textbf
  {\bibinfo {volume} {80}},\ \bibinfo {pages} {1272} (\bibinfo {year}
  {1998})}\BibitemShut {NoStop}%
\bibitem [{\citenamefont {Jiang}\ \emph {et~al.}(2020)\citenamefont {Jiang},
  \citenamefont {Zaanen}, \citenamefont {Devereaux},\ and\ \citenamefont
  {Jiang}}]{jianground2020}%
  \BibitemOpen
  \bibfield  {author} {\bibinfo {author} {\bibfnamefont {Y.-F.}\ \bibnamefont
  {Jiang}}, \bibinfo {author} {\bibfnamefont {J.}~\bibnamefont {Zaanen}},
  \bibinfo {author} {\bibfnamefont {T.~P.}\ \bibnamefont {Devereaux}},\ and\
  \bibinfo {author} {\bibfnamefont {H.-C.}\ \bibnamefont {Jiang}},\ }\href
  {https://doi.org/10.1103/PhysRevResearch.2.033073} {\bibfield  {journal}
  {\bibinfo  {journal} {Phys. Rev. Research}\ }\textbf {\bibinfo {volume}
  {2}},\ \bibinfo {pages} {033073} (\bibinfo {year} {2020})}\BibitemShut
  {NoStop}%
\bibitem [{\citenamefont {Wietek}\ \emph {et~al.}(2021)\citenamefont {Wietek},
  \citenamefont {He}, \citenamefont {White}, \citenamefont {Georges},\ and\
  \citenamefont {Stoudenmire}}]{wietekstripes2021}%
  \BibitemOpen
  \bibfield  {author} {\bibinfo {author} {\bibfnamefont {A.}~\bibnamefont
  {Wietek}}, \bibinfo {author} {\bibfnamefont {Y.-Y.}\ \bibnamefont {He}},
  \bibinfo {author} {\bibfnamefont {S.~R.}\ \bibnamefont {White}}, \bibinfo
  {author} {\bibfnamefont {A.}~\bibnamefont {Georges}},\ and\ \bibinfo {author}
  {\bibfnamefont {E.~M.}\ \bibnamefont {Stoudenmire}},\ }\href
  {https://doi.org/10.1103/PhysRevX.11.031007} {\bibfield  {journal} {\bibinfo
  {journal} {Phys. Rev. X}\ }\textbf {\bibinfo {volume} {11}},\ \bibinfo
  {pages} {031007} (\bibinfo {year} {2021})}\BibitemShut {NoStop}%
\bibitem [{\citenamefont {Zaanen}\ and\ \citenamefont
  {Gunnarsson}(1989{\natexlab{a}})}]{zaanencharged1989}%
  \BibitemOpen
  \bibfield  {author} {\bibinfo {author} {\bibfnamefont {J.}~\bibnamefont
  {Zaanen}}\ and\ \bibinfo {author} {\bibfnamefont {O.}~\bibnamefont
  {Gunnarsson}},\ }\href {https://doi.org/10.1103/PhysRevB.40.7391} {\bibfield
  {journal} {\bibinfo  {journal} {Phys. Rev. B}\ }\textbf {\bibinfo {volume}
  {40}},\ \bibinfo {pages} {7391} (\bibinfo {year}
  {1989}{\natexlab{a}})}\BibitemShut {NoStop}%
\bibitem [{\citenamefont {Schulz}(1990)}]{schulz1990incommensurate}%
  \BibitemOpen
  \bibfield  {author} {\bibinfo {author} {\bibfnamefont {H.~J.}\ \bibnamefont
  {Schulz}},\ }\href {https://doi.org/10.1103/PhysRevLett.64.1445} {\bibfield
  {journal} {\bibinfo  {journal} {Phys. Rev. Lett.}\ }\textbf {\bibinfo
  {volume} {64}},\ \bibinfo {pages} {1445} (\bibinfo {year}
  {1990})}\BibitemShut {NoStop}%
\bibitem [{\citenamefont {Machida}(1989)}]{machida1989magnetism}%
  \BibitemOpen
  \bibfield  {author} {\bibinfo {author} {\bibfnamefont {K.}~\bibnamefont
  {Machida}},\ }\href@noop {} {\bibfield  {journal} {\bibinfo  {journal}
  {Physica C: Superconductivity}\ }\textbf {\bibinfo {volume} {158}},\ \bibinfo
  {pages} {192} (\bibinfo {year} {1989})}\BibitemShut {NoStop}%
\bibitem [{\citenamefont {Maier}\ \emph {et~al.}(2005)\citenamefont {Maier},
  \citenamefont {Jarrell}, \citenamefont {Pruschke},\ and\ \citenamefont
  {Hettler}}]{maier2005review}%
  \BibitemOpen
  \bibfield  {author} {\bibinfo {author} {\bibfnamefont {T.}~\bibnamefont
  {Maier}}, \bibinfo {author} {\bibfnamefont {M.}~\bibnamefont {Jarrell}},
  \bibinfo {author} {\bibfnamefont {T.}~\bibnamefont {Pruschke}},\ and\
  \bibinfo {author} {\bibfnamefont {M.~H.}\ \bibnamefont {Hettler}},\ }\href
  {https://doi.org/10.1103/RevModPhys.77.1027} {\bibfield  {journal} {\bibinfo
  {journal} {Rev. Mod. Phys.}\ }\textbf {\bibinfo {volume} {77}},\ \bibinfo
  {pages} {1027} (\bibinfo {year} {2005})}\BibitemShut {NoStop}%
\bibitem [{\citenamefont {Tremblay}\ \emph {et~al.}(2006)\citenamefont
  {Tremblay}, \citenamefont {Kyung},\ and\ \citenamefont
  {Sénéchal}}]{tremblay2006}%
  \BibitemOpen
  \bibfield  {author} {\bibinfo {author} {\bibfnamefont {A.-M.}\ \bibnamefont
  {Tremblay}}, \bibinfo {author} {\bibfnamefont {B.}~\bibnamefont {Kyung}},\
  and\ \bibinfo {author} {\bibfnamefont {D.}~\bibnamefont {Sénéchal}},\
  }\href {https://doi.org/10.1063/1.2199446} {\bibfield  {journal} {\bibinfo
  {journal} {Low Temperature Physics}\ }\textbf {\bibinfo {volume} {32}},\
  \bibinfo {pages} {424} (\bibinfo {year} {2006})},\ \Eprint
  {https://arxiv.org/abs/http://dx.doi.org/10.1063/1.2199446}
  {http://dx.doi.org/10.1063/1.2199446} \BibitemShut {NoStop}%
\bibitem [{\citenamefont {Kotliar}\ \emph {et~al.}(2006)\citenamefont
  {Kotliar}, \citenamefont {Savrasov}, \citenamefont {Haule}, \citenamefont
  {Oudovenko}, \citenamefont {Parcollet},\ and\ \citenamefont
  {Marianetti}}]{kotliar2006rmp}%
  \BibitemOpen
  \bibfield  {author} {\bibinfo {author} {\bibfnamefont {G.}~\bibnamefont
  {Kotliar}}, \bibinfo {author} {\bibfnamefont {S.~Y.}\ \bibnamefont
  {Savrasov}}, \bibinfo {author} {\bibfnamefont {K.}~\bibnamefont {Haule}},
  \bibinfo {author} {\bibfnamefont {V.~S.}\ \bibnamefont {Oudovenko}}, \bibinfo
  {author} {\bibfnamefont {O.}~\bibnamefont {Parcollet}},\ and\ \bibinfo
  {author} {\bibfnamefont {C.~A.}\ \bibnamefont {Marianetti}},\ }\href
  {https://doi.org/10.1103/RevModPhys.78.865} {\bibfield  {journal} {\bibinfo
  {journal} {Rev. Mod. Phys.}\ }\textbf {\bibinfo {volume} {78}},\ \bibinfo
  {pages} {865} (\bibinfo {year} {2006})}\BibitemShut {NoStop}%
\bibitem [{\citenamefont {Macridin}\ \emph {et~al.}(2006)\citenamefont
  {Macridin}, \citenamefont {Jarrell}, \citenamefont {Maier}, \citenamefont
  {Kent},\ and\ \citenamefont {D'Azevedo}}]{macridin_prl_2006}%
  \BibitemOpen
  \bibfield  {author} {\bibinfo {author} {\bibfnamefont {A.}~\bibnamefont
  {Macridin}}, \bibinfo {author} {\bibfnamefont {M.}~\bibnamefont {Jarrell}},
  \bibinfo {author} {\bibfnamefont {T.}~\bibnamefont {Maier}}, \bibinfo
  {author} {\bibfnamefont {P.~R.~C.}\ \bibnamefont {Kent}},\ and\ \bibinfo
  {author} {\bibfnamefont {E.}~\bibnamefont {D'Azevedo}},\ }\href
  {https://doi.org/10.1103/PhysRevLett.97.036401} {\bibfield  {journal}
  {\bibinfo  {journal} {Phys. Rev. Lett.}\ }\textbf {\bibinfo {volume} {97}},\
  \bibinfo {pages} {036401} (\bibinfo {year} {2006})}\BibitemShut {NoStop}%
\bibitem [{\citenamefont {Gull}\ \emph {et~al.}(2010)\citenamefont {Gull},
  \citenamefont {Ferrero}, \citenamefont {Parcollet}, \citenamefont {Georges},\
  and\ \citenamefont {Millis}}]{gull2010momentum}%
  \BibitemOpen
  \bibfield  {author} {\bibinfo {author} {\bibfnamefont {E.}~\bibnamefont
  {Gull}}, \bibinfo {author} {\bibfnamefont {M.}~\bibnamefont {Ferrero}},
  \bibinfo {author} {\bibfnamefont {O.}~\bibnamefont {Parcollet}}, \bibinfo
  {author} {\bibfnamefont {A.}~\bibnamefont {Georges}},\ and\ \bibinfo {author}
  {\bibfnamefont {A.~J.}\ \bibnamefont {Millis}},\ }\href
  {https://doi.org/10.1103/PhysRevB.82.155101} {\bibfield  {journal} {\bibinfo
  {journal} {Phys. Rev. B}\ }\textbf {\bibinfo {volume} {82}},\ \bibinfo
  {pages} {155101} (\bibinfo {year} {2010})}\BibitemShut {NoStop}%
\bibitem [{\citenamefont {Gunnarsson}\ \emph {et~al.}(2015)\citenamefont
  {Gunnarsson}, \citenamefont {Sch\"afer}, \citenamefont {LeBlanc},
  \citenamefont {Gull}, \citenamefont {Merino}, \citenamefont {Sangiovanni},
  \citenamefont {Rohringer},\ and\ \citenamefont
  {Toschi}}]{gunnarsson2015fluctuation}%
  \BibitemOpen
  \bibfield  {author} {\bibinfo {author} {\bibfnamefont {O.}~\bibnamefont
  {Gunnarsson}}, \bibinfo {author} {\bibfnamefont {T.}~\bibnamefont
  {Sch\"afer}}, \bibinfo {author} {\bibfnamefont {J.~P.~F.}\ \bibnamefont
  {LeBlanc}}, \bibinfo {author} {\bibfnamefont {E.}~\bibnamefont {Gull}},
  \bibinfo {author} {\bibfnamefont {J.}~\bibnamefont {Merino}}, \bibinfo
  {author} {\bibfnamefont {G.}~\bibnamefont {Sangiovanni}}, \bibinfo {author}
  {\bibfnamefont {G.}~\bibnamefont {Rohringer}},\ and\ \bibinfo {author}
  {\bibfnamefont {A.}~\bibnamefont {Toschi}},\ }\href
  {https://doi.org/10.1103/PhysRevLett.114.236402} {\bibfield  {journal}
  {\bibinfo  {journal} {Phys. Rev. Lett.}\ }\textbf {\bibinfo {volume} {114}},\
  \bibinfo {pages} {236402} (\bibinfo {year} {2015})}\BibitemShut {NoStop}%
\bibitem [{\citenamefont {Wu}\ \emph {et~al.}(2018)\citenamefont {Wu},
  \citenamefont {Scheurer}, \citenamefont {Chatterjee}, \citenamefont
  {Sachdev}, \citenamefont {Georges},\ and\ \citenamefont
  {Ferrero}}]{wu2018pseudogap}%
  \BibitemOpen
  \bibfield  {author} {\bibinfo {author} {\bibfnamefont {W.}~\bibnamefont
  {Wu}}, \bibinfo {author} {\bibfnamefont {M.~S.}\ \bibnamefont {Scheurer}},
  \bibinfo {author} {\bibfnamefont {S.}~\bibnamefont {Chatterjee}}, \bibinfo
  {author} {\bibfnamefont {S.}~\bibnamefont {Sachdev}}, \bibinfo {author}
  {\bibfnamefont {A.}~\bibnamefont {Georges}},\ and\ \bibinfo {author}
  {\bibfnamefont {M.}~\bibnamefont {Ferrero}},\ }\href
  {https://doi.org/10.1103/PhysRevX.8.021048} {\bibfield  {journal} {\bibinfo
  {journal} {Phys. Rev. X}\ }\textbf {\bibinfo {volume} {8}},\ \bibinfo {pages}
  {021048} (\bibinfo {year} {2018})}\BibitemShut {NoStop}%
\bibitem [{\citenamefont {Krien}\ \emph {et~al.}(2021)\citenamefont {Krien},
  \citenamefont {Worm}, \citenamefont {Chalupa}, \citenamefont {Toschi},\ and\
  \citenamefont {Held}}]{krien2021spin}%
  \BibitemOpen
  \bibfield  {author} {\bibinfo {author} {\bibfnamefont {F.}~\bibnamefont
  {Krien}}, \bibinfo {author} {\bibfnamefont {P.}~\bibnamefont {Worm}},
  \bibinfo {author} {\bibfnamefont {P.}~\bibnamefont {Chalupa}}, \bibinfo
  {author} {\bibfnamefont {A.}~\bibnamefont {Toschi}},\ and\ \bibinfo {author}
  {\bibfnamefont {K.}~\bibnamefont {Held}},\ }\href@noop {} {\bibfield
  {journal} {\bibinfo  {journal} {arXiv preprint arXiv:2107.06529}\ } (\bibinfo
  {year} {2021})}\BibitemShut {NoStop}%
\bibitem [{\citenamefont {Prokof’ev}\ and\ \citenamefont
  {Svistunov}(2008)}]{prokof2008fermi}%
  \BibitemOpen
  \bibfield  {author} {\bibinfo {author} {\bibfnamefont {N.}~\bibnamefont
  {Prokof’ev}}\ and\ \bibinfo {author} {\bibfnamefont {B.}~\bibnamefont
  {Svistunov}},\ }\href@noop {} {\bibfield  {journal} {\bibinfo  {journal}
  {Physical Review B}\ }\textbf {\bibinfo {volume} {77}},\ \bibinfo {pages}
  {020408} (\bibinfo {year} {2008})}\BibitemShut {NoStop}%
\bibitem [{\citenamefont {Kozik}\ \emph {et~al.}(2010)\citenamefont {Kozik},
  \citenamefont {Van~Houcke}, \citenamefont {Gull}, \citenamefont {Pollet},
  \citenamefont {Prokof'ev}, \citenamefont {Svistunov},\ and\ \citenamefont
  {Troyer}}]{kozik2010diagrammatic}%
  \BibitemOpen
  \bibfield  {author} {\bibinfo {author} {\bibfnamefont {E.}~\bibnamefont
  {Kozik}}, \bibinfo {author} {\bibfnamefont {K.}~\bibnamefont {Van~Houcke}},
  \bibinfo {author} {\bibfnamefont {E.}~\bibnamefont {Gull}}, \bibinfo {author}
  {\bibfnamefont {L.}~\bibnamefont {Pollet}}, \bibinfo {author} {\bibfnamefont
  {N.}~\bibnamefont {Prokof'ev}}, \bibinfo {author} {\bibfnamefont
  {B.}~\bibnamefont {Svistunov}},\ and\ \bibinfo {author} {\bibfnamefont
  {M.}~\bibnamefont {Troyer}},\ }\href@noop {} {\bibfield  {journal} {\bibinfo
  {journal} {EPL (Europhysics Letters)}\ }\textbf {\bibinfo {volume} {90}},\
  \bibinfo {pages} {10004} (\bibinfo {year} {2010})}\BibitemShut {NoStop}%
\bibitem [{\citenamefont {Van~Houcke}\ \emph {et~al.}(2012)\citenamefont
  {Van~Houcke}, \citenamefont {Werner}, \citenamefont {Kozik}, \citenamefont
  {Prokof’ev}, \citenamefont {Svistunov}, \citenamefont {Ku}, \citenamefont
  {Sommer}, \citenamefont {Cheuk}, \citenamefont {Schirotzek},\ and\
  \citenamefont {Zwierlein}}]{van2012feynman}%
  \BibitemOpen
  \bibfield  {author} {\bibinfo {author} {\bibfnamefont {K.}~\bibnamefont
  {Van~Houcke}}, \bibinfo {author} {\bibfnamefont {F.}~\bibnamefont {Werner}},
  \bibinfo {author} {\bibfnamefont {E.}~\bibnamefont {Kozik}}, \bibinfo
  {author} {\bibfnamefont {N.}~\bibnamefont {Prokof’ev}}, \bibinfo {author}
  {\bibfnamefont {B.}~\bibnamefont {Svistunov}}, \bibinfo {author}
  {\bibfnamefont {M.}~\bibnamefont {Ku}}, \bibinfo {author} {\bibfnamefont
  {A.}~\bibnamefont {Sommer}}, \bibinfo {author} {\bibfnamefont
  {L.}~\bibnamefont {Cheuk}}, \bibinfo {author} {\bibfnamefont
  {A.}~\bibnamefont {Schirotzek}},\ and\ \bibinfo {author} {\bibfnamefont
  {M.}~\bibnamefont {Zwierlein}},\ }\href@noop {} {\bibfield  {journal}
  {\bibinfo  {journal} {Nature Physics}\ }\textbf {\bibinfo {volume} {8}},\
  \bibinfo {pages} {366} (\bibinfo {year} {2012})}\BibitemShut {NoStop}%
\bibitem [{\citenamefont {Wu}\ \emph {et~al.}(2017)\citenamefont {Wu},
  \citenamefont {Ferrero}, \citenamefont {Georges},\ and\ \citenamefont
  {Kozik}}]{wu_controlling}%
  \BibitemOpen
  \bibfield  {author} {\bibinfo {author} {\bibfnamefont {W.}~\bibnamefont
  {Wu}}, \bibinfo {author} {\bibfnamefont {M.}~\bibnamefont {Ferrero}},
  \bibinfo {author} {\bibfnamefont {A.}~\bibnamefont {Georges}},\ and\ \bibinfo
  {author} {\bibfnamefont {E.}~\bibnamefont {Kozik}},\ }\href
  {https://doi.org/10.1103/PhysRevB.96.041105} {\bibfield  {journal} {\bibinfo
  {journal} {Phys. Rev. B}\ }\textbf {\bibinfo {volume} {96}},\ \bibinfo
  {pages} {041105(R)} (\bibinfo {year} {2017})}\BibitemShut {NoStop}%
\bibitem [{\citenamefont {Rossi}(2017)}]{cdet}%
  \BibitemOpen
  \bibfield  {author} {\bibinfo {author} {\bibfnamefont {R.}~\bibnamefont
  {Rossi}},\ }\href {https://doi.org/10.1103/PhysRevLett.119.045701} {\bibfield
   {journal} {\bibinfo  {journal} {Phys. Rev. Lett.}\ }\textbf {\bibinfo
  {volume} {119}},\ \bibinfo {pages} {045701} (\bibinfo {year}
  {2017})}\BibitemShut {NoStop}%
\bibitem [{\citenamefont {Chen}\ and\ \citenamefont
  {Haule}(2019)}]{chen2019combined}%
  \BibitemOpen
  \bibfield  {author} {\bibinfo {author} {\bibfnamefont {K.}~\bibnamefont
  {Chen}}\ and\ \bibinfo {author} {\bibfnamefont {K.}~\bibnamefont {Haule}},\
  }\href@noop {} {\bibfield  {journal} {\bibinfo  {journal} {Nature
  communications}\ }\textbf {\bibinfo {volume} {10}},\ \bibinfo {pages} {1}
  (\bibinfo {year} {2019})}\BibitemShut {NoStop}%
\bibitem [{\citenamefont {{\v{S}}imkovic~IV}\ \emph {et~al.}(2020)\citenamefont
  {{\v{S}}imkovic~IV}, \citenamefont {LeBlanc}, \citenamefont {Kim},
  \citenamefont {Deng}, \citenamefont {Prokof’ev}, \citenamefont
  {Svistunov},\ and\ \citenamefont {Kozik}}]{vsimkovic2020extended}%
  \BibitemOpen
  \bibfield  {author} {\bibinfo {author} {\bibfnamefont {F.}~\bibnamefont
  {{\v{S}}imkovic~IV}}, \bibinfo {author} {\bibfnamefont {J.}~\bibnamefont
  {LeBlanc}}, \bibinfo {author} {\bibfnamefont {A.~J.}\ \bibnamefont {Kim}},
  \bibinfo {author} {\bibfnamefont {Y.}~\bibnamefont {Deng}}, \bibinfo {author}
  {\bibfnamefont {N.}~\bibnamefont {Prokof’ev}}, \bibinfo {author}
  {\bibfnamefont {B.}~\bibnamefont {Svistunov}},\ and\ \bibinfo {author}
  {\bibfnamefont {E.}~\bibnamefont {Kozik}},\ }\href@noop {} {\bibfield
  {journal} {\bibinfo  {journal} {Physical Review Letters}\ }\textbf {\bibinfo
  {volume} {124}},\ \bibinfo {pages} {017003} (\bibinfo {year}
  {2020})}\BibitemShut {NoStop}%
\bibitem [{\citenamefont {Georges}\ \emph {et~al.}(1996)\citenamefont
  {Georges}, \citenamefont {Kotliar}, \citenamefont {Krauth},\ and\
  \citenamefont {Rozenberg}}]{antoine_dmft}%
  \BibitemOpen
  \bibfield  {author} {\bibinfo {author} {\bibfnamefont {A.}~\bibnamefont
  {Georges}}, \bibinfo {author} {\bibfnamefont {G.}~\bibnamefont {Kotliar}},
  \bibinfo {author} {\bibfnamefont {W.}~\bibnamefont {Krauth}},\ and\ \bibinfo
  {author} {\bibfnamefont {M.~J.}\ \bibnamefont {Rozenberg}},\ }\href
  {https://doi.org/10.1103/RevModPhys.68.13} {\bibfield  {journal} {\bibinfo
  {journal} {Rev. Mod. Phys.}\ }\textbf {\bibinfo {volume} {68}},\ \bibinfo
  {pages} {13} (\bibinfo {year} {1996})}\BibitemShut {NoStop}%
\bibitem [{Note1()}]{Note1}%
  \BibitemOpen
  \bibinfo {note} {Let us emphasize that, throughout this article, we {\protect
  \it do not} perform numerical analytic continuations but rather base our
  study on a direct analysis of imaginary-time/frequency Monte Carlo
  data.}\BibitemShut {Stop}%
\bibitem [{\citenamefont {Xu}\ \emph {et~al.}(2021)\citenamefont {Xu},
  \citenamefont {Shi}, \citenamefont {Vitali}, \citenamefont {Qin},\ and\
  \citenamefont {Zhang}}]{xu2021stripes}%
  \BibitemOpen
  \bibfield  {author} {\bibinfo {author} {\bibfnamefont {H.}~\bibnamefont
  {Xu}}, \bibinfo {author} {\bibfnamefont {H.}~\bibnamefont {Shi}}, \bibinfo
  {author} {\bibfnamefont {E.}~\bibnamefont {Vitali}}, \bibinfo {author}
  {\bibfnamefont {M.}~\bibnamefont {Qin}},\ and\ \bibinfo {author}
  {\bibfnamefont {S.}~\bibnamefont {Zhang}},\ }\href@noop {} {\bibfield
  {journal} {\bibinfo  {journal} {arXiv preprint arXiv:2112.02187}\ } (\bibinfo
  {year} {2021})}\BibitemShut {NoStop}%
\bibitem [{\citenamefont {{Y.M. Vilk}}\ and\ \citenamefont {{A.-M.S.
  Tremblay}}(1997)}]{vilk_tremblay_1997}%
  \BibitemOpen
  \bibfield  {author} {\bibinfo {author} {\bibnamefont {{Y.M. Vilk}}}\ and\
  \bibinfo {author} {\bibnamefont {{A.-M.S. Tremblay}}},\ }\href
  {https://doi.org/10.1051/jp1:1997135} {\bibfield  {journal} {\bibinfo
  {journal} {J. Phys. I France}\ }\textbf {\bibinfo {volume} {7}},\ \bibinfo
  {pages} {1309} (\bibinfo {year} {1997})}\BibitemShut {NoStop}%
\bibitem [{\citenamefont {Sedrakyan}\ and\ \citenamefont
  {Chubukov}(2010)}]{chubukov2010}%
  \BibitemOpen
  \bibfield  {author} {\bibinfo {author} {\bibfnamefont {T.~A.}\ \bibnamefont
  {Sedrakyan}}\ and\ \bibinfo {author} {\bibfnamefont {A.~V.}\ \bibnamefont
  {Chubukov}},\ }\href {https://doi.org/10.1103/PhysRevB.81.174536} {\bibfield
  {journal} {\bibinfo  {journal} {Phys. Rev. B}\ }\textbf {\bibinfo {volume}
  {81}},\ \bibinfo {pages} {174536} (\bibinfo {year} {2010})}\BibitemShut
  {NoStop}%
\bibitem [{\citenamefont {{\v{S}}imkovic~IV}\ \emph {et~al.}(2021)\citenamefont
  {{\v{S}}imkovic~IV}, \citenamefont {Rossi},\ and\ \citenamefont
  {Ferrero}}]{vsimkovic2021weak}%
  \BibitemOpen
  \bibfield  {author} {\bibinfo {author} {\bibfnamefont {F.}~\bibnamefont
  {{\v{S}}imkovic~IV}}, \bibinfo {author} {\bibfnamefont {R.}~\bibnamefont
  {Rossi}},\ and\ \bibinfo {author} {\bibfnamefont {M.}~\bibnamefont
  {Ferrero}},\ }\href@noop {} {\bibfield  {journal} {\bibinfo  {journal} {arXiv
  preprint arXiv:2110.05863}\ } (\bibinfo {year} {2021})}\BibitemShut {NoStop}%
\bibitem [{\citenamefont {Huang}\ \emph {et~al.}(2018)\citenamefont {Huang},
  \citenamefont {Mendl}, \citenamefont {Jiang}, \citenamefont {Moritz},\ and\
  \citenamefont {Devereaux}}]{huang2018stripe}%
  \BibitemOpen
  \bibfield  {author} {\bibinfo {author} {\bibfnamefont {E.~W.}\ \bibnamefont
  {Huang}}, \bibinfo {author} {\bibfnamefont {C.~B.}\ \bibnamefont {Mendl}},
  \bibinfo {author} {\bibfnamefont {H.-C.}\ \bibnamefont {Jiang}}, \bibinfo
  {author} {\bibfnamefont {B.}~\bibnamefont {Moritz}},\ and\ \bibinfo {author}
  {\bibfnamefont {T.~P.}\ \bibnamefont {Devereaux}},\ }\href@noop {} {\bibfield
   {journal} {\bibinfo  {journal} {npj Quantum Materials}\ }\textbf {\bibinfo
  {volume} {3}},\ \bibinfo {pages} {1} (\bibinfo {year} {2018})}\BibitemShut
  {NoStop}%
\bibitem [{\citenamefont {Mai}\ \emph {et~al.}(2022)\citenamefont {Mai},
  \citenamefont {Karakuzu}, \citenamefont {Balduzzi}, \citenamefont
  {Johnston},\ and\ \citenamefont {Maier}}]{maier_2021_fluctuating}%
  \BibitemOpen
  \bibfield  {author} {\bibinfo {author} {\bibfnamefont {P.}~\bibnamefont
  {Mai}}, \bibinfo {author} {\bibfnamefont {S.}~\bibnamefont {Karakuzu}},
  \bibinfo {author} {\bibfnamefont {G.}~\bibnamefont {Balduzzi}}, \bibinfo
  {author} {\bibfnamefont {S.}~\bibnamefont {Johnston}},\ and\ \bibinfo
  {author} {\bibfnamefont {T.~A.}\ \bibnamefont {Maier}},\ }\href
  {https://doi.org/10.1073/pnas.2112806119} {\bibfield  {journal} {\bibinfo
  {journal} {Proceedings of the National Academy of Sciences}\ }\textbf
  {\bibinfo {volume} {119}},\ \bibinfo {pages} {e2112806119} (\bibinfo {year}
  {2022})}\BibitemShut {NoStop}%
\bibitem [{\citenamefont {Parcollet}\ \emph {et~al.}(2004)\citenamefont
  {Parcollet}, \citenamefont {Biroli},\ and\ \citenamefont
  {Kotliar}}]{parcollet2004}%
  \BibitemOpen
  \bibfield  {author} {\bibinfo {author} {\bibfnamefont {O.}~\bibnamefont
  {Parcollet}}, \bibinfo {author} {\bibfnamefont {G.}~\bibnamefont {Biroli}},\
  and\ \bibinfo {author} {\bibfnamefont {G.}~\bibnamefont {Kotliar}},\ }\href
  {https://doi.org/10.1103/PhysRevLett.92.226402} {\bibfield  {journal}
  {\bibinfo  {journal} {Phys. Rev. Lett.}\ }\textbf {\bibinfo {volume} {92}},\
  \bibinfo {pages} {226402} (\bibinfo {year} {2004})}\BibitemShut {NoStop}%
\bibitem [{\citenamefont {Haule}\ and\ \citenamefont
  {Kotliar}(2007)}]{haule2007strongly}%
  \BibitemOpen
  \bibfield  {author} {\bibinfo {author} {\bibfnamefont {K.}~\bibnamefont
  {Haule}}\ and\ \bibinfo {author} {\bibfnamefont {G.}~\bibnamefont
  {Kotliar}},\ }\href {https://doi.org/10.1103/PhysRevB.76.104509} {\bibfield
  {journal} {\bibinfo  {journal} {Phys. Rev. B}\ }\textbf {\bibinfo {volume}
  {76}},\ \bibinfo {pages} {104509} (\bibinfo {year} {2007})}\BibitemShut
  {NoStop}%
\bibitem [{\citenamefont {Ferrero}\ \emph {et~al.}(2009)\citenamefont
  {Ferrero}, \citenamefont {Cornaglia}, \citenamefont {De~Leo}, \citenamefont
  {Parcollet}, \citenamefont {Kotliar},\ and\ \citenamefont
  {Georges}}]{ferrero2009}%
  \BibitemOpen
  \bibfield  {author} {\bibinfo {author} {\bibfnamefont {M.}~\bibnamefont
  {Ferrero}}, \bibinfo {author} {\bibfnamefont {P.~S.}\ \bibnamefont
  {Cornaglia}}, \bibinfo {author} {\bibfnamefont {L.}~\bibnamefont {De~Leo}},
  \bibinfo {author} {\bibfnamefont {O.}~\bibnamefont {Parcollet}}, \bibinfo
  {author} {\bibfnamefont {G.}~\bibnamefont {Kotliar}},\ and\ \bibinfo {author}
  {\bibfnamefont {A.}~\bibnamefont {Georges}},\ }\href
  {https://doi.org/10.1103/PhysRevB.80.064501} {\bibfield  {journal} {\bibinfo
  {journal} {Phys. Rev. B}\ }\textbf {\bibinfo {volume} {80}},\ \bibinfo
  {pages} {064501} (\bibinfo {year} {2009})}\BibitemShut {NoStop}%
\bibitem [{\citenamefont {Sch\"afer}\ \emph {et~al.}(2021)\citenamefont
  {Sch\"afer}, \citenamefont {Wentzell}, \citenamefont {\ifmmode~\check{S}\else
  \v{S}\fi{}imkovic}, \citenamefont {He}, \citenamefont {Hille}, \citenamefont
  {Klett}, \citenamefont {Eckhardt}, \citenamefont {Arzhang}, \citenamefont
  {Harkov}, \citenamefont {Le~R\'egent}, \citenamefont {Kirsch}, \citenamefont
  {Wang}, \citenamefont {Kim}, \citenamefont {Kozik}, \citenamefont {Stepanov},
  \citenamefont {Kauch}, \citenamefont {Andergassen}, \citenamefont {Hansmann},
  \citenamefont {Rohe}, \citenamefont {Vilk}, \citenamefont {LeBlanc},
  \citenamefont {Zhang}, \citenamefont {Tremblay}, \citenamefont {Ferrero},
  \citenamefont {Parcollet},\ and\ \citenamefont {Georges}}]{M7}%
  \BibitemOpen
  \bibfield  {author} {\bibinfo {author} {\bibfnamefont {T.}~\bibnamefont
  {Sch\"afer}}, \bibinfo {author} {\bibfnamefont {N.}~\bibnamefont {Wentzell}},
  \bibinfo {author} {\bibfnamefont {F.}~\bibnamefont {\ifmmode~\check{S}\else
  \v{S}\fi{}imkovic}}, \bibinfo {author} {\bibfnamefont {Y.-Y.}\ \bibnamefont
  {He}}, \bibinfo {author} {\bibfnamefont {C.}~\bibnamefont {Hille}}, \bibinfo
  {author} {\bibfnamefont {M.}~\bibnamefont {Klett}}, \bibinfo {author}
  {\bibfnamefont {C.~J.}\ \bibnamefont {Eckhardt}}, \bibinfo {author}
  {\bibfnamefont {B.}~\bibnamefont {Arzhang}}, \bibinfo {author} {\bibfnamefont
  {V.}~\bibnamefont {Harkov}}, \bibinfo {author} {\bibfnamefont {F.~m. c.-M.}\
  \bibnamefont {Le~R\'egent}}, \bibinfo {author} {\bibfnamefont
  {A.}~\bibnamefont {Kirsch}}, \bibinfo {author} {\bibfnamefont
  {Y.}~\bibnamefont {Wang}}, \bibinfo {author} {\bibfnamefont {A.~J.}\
  \bibnamefont {Kim}}, \bibinfo {author} {\bibfnamefont {E.}~\bibnamefont
  {Kozik}}, \bibinfo {author} {\bibfnamefont {E.~A.}\ \bibnamefont {Stepanov}},
  \bibinfo {author} {\bibfnamefont {A.}~\bibnamefont {Kauch}}, \bibinfo
  {author} {\bibfnamefont {S.}~\bibnamefont {Andergassen}}, \bibinfo {author}
  {\bibfnamefont {P.}~\bibnamefont {Hansmann}}, \bibinfo {author}
  {\bibfnamefont {D.}~\bibnamefont {Rohe}}, \bibinfo {author} {\bibfnamefont
  {Y.~M.}\ \bibnamefont {Vilk}}, \bibinfo {author} {\bibfnamefont {J.~P.~F.}\
  \bibnamefont {LeBlanc}}, \bibinfo {author} {\bibfnamefont {S.}~\bibnamefont
  {Zhang}}, \bibinfo {author} {\bibfnamefont {A.-M.~S.}\ \bibnamefont
  {Tremblay}}, \bibinfo {author} {\bibfnamefont {M.}~\bibnamefont {Ferrero}},
  \bibinfo {author} {\bibfnamefont {O.}~\bibnamefont {Parcollet}},\ and\
  \bibinfo {author} {\bibfnamefont {A.}~\bibnamefont {Georges}},\ }\href
  {https://doi.org/10.1103/PhysRevX.11.011058} {\bibfield  {journal} {\bibinfo
  {journal} {Phys. Rev. X}\ }\textbf {\bibinfo {volume} {11}},\ \bibinfo
  {pages} {011058} (\bibinfo {year} {2021})}\BibitemShut {NoStop}%
\bibitem [{\citenamefont {Schäfer}\ and\ \citenamefont
  {Toschi}(2021)}]{schafer2021}%
  \BibitemOpen
  \bibfield  {author} {\bibinfo {author} {\bibfnamefont {T.}~\bibnamefont
  {Schäfer}}\ and\ \bibinfo {author} {\bibfnamefont {A.}~\bibnamefont
  {Toschi}},\ }\href {https://doi.org/10.1088/1361-648x/abeb44} {\bibfield
  {journal} {\bibinfo  {journal} {Journal of Physics: Condensed Matter}\
  }\textbf {\bibinfo {volume} {33}},\ \bibinfo {pages} {214001} (\bibinfo
  {year} {2021})}\BibitemShut {NoStop}%
\bibitem [{\citenamefont {Rubtsov}\ \emph {et~al.}(2005)\citenamefont
  {Rubtsov}, \citenamefont {Savkin},\ and\ \citenamefont
  {Lichtenstein}}]{rubtsov2005continuous}%
  \BibitemOpen
  \bibfield  {author} {\bibinfo {author} {\bibfnamefont {A.~N.}\ \bibnamefont
  {Rubtsov}}, \bibinfo {author} {\bibfnamefont {V.~V.}\ \bibnamefont
  {Savkin}},\ and\ \bibinfo {author} {\bibfnamefont {A.~I.}\ \bibnamefont
  {Lichtenstein}},\ }\href@noop {} {\bibfield  {journal} {\bibinfo  {journal}
  {Physical Review B}\ }\textbf {\bibinfo {volume} {72}},\ \bibinfo {pages}
  {035122} (\bibinfo {year} {2005})}\BibitemShut {NoStop}%
\bibitem [{\citenamefont {Profumo}\ \emph {et~al.}(2015)\citenamefont
  {Profumo}, \citenamefont {Groth}, \citenamefont {Messio}, \citenamefont
  {Parcollet},\ and\ \citenamefont {Waintal}}]{olivier}%
  \BibitemOpen
  \bibfield  {author} {\bibinfo {author} {\bibfnamefont {R.~E.~V.}\
  \bibnamefont {Profumo}}, \bibinfo {author} {\bibfnamefont {C.}~\bibnamefont
  {Groth}}, \bibinfo {author} {\bibfnamefont {L.}~\bibnamefont {Messio}},
  \bibinfo {author} {\bibfnamefont {O.}~\bibnamefont {Parcollet}},\ and\
  \bibinfo {author} {\bibfnamefont {X.}~\bibnamefont {Waintal}},\ }\href
  {https://doi.org/10.1103/PhysRevB.91.245154} {\bibfield  {journal} {\bibinfo
  {journal} {Phys. Rev. B}\ }\textbf {\bibinfo {volume} {91}},\ \bibinfo
  {pages} {245154} (\bibinfo {year} {2015})}\BibitemShut {NoStop}%
\bibitem [{\citenamefont {Rossi}\ \emph {et~al.}(2016)\citenamefont {Rossi},
  \citenamefont {Werner}, \citenamefont {Prokof'ev},\ and\ \citenamefont
  {Svistunov}}]{shifted_action}%
  \BibitemOpen
  \bibfield  {author} {\bibinfo {author} {\bibfnamefont {R.}~\bibnamefont
  {Rossi}}, \bibinfo {author} {\bibfnamefont {F.}~\bibnamefont {Werner}},
  \bibinfo {author} {\bibfnamefont {N.}~\bibnamefont {Prokof'ev}},\ and\
  \bibinfo {author} {\bibfnamefont {B.}~\bibnamefont {Svistunov}},\ }\href
  {https://doi.org/10.1103/PhysRevB.93.161102} {\bibfield  {journal} {\bibinfo
  {journal} {Phys. Rev. B}\ }\textbf {\bibinfo {volume} {93}},\ \bibinfo
  {pages} {161102} (\bibinfo {year} {2016})}\BibitemShut {NoStop}%
\bibitem [{\citenamefont {Rossi}\ \emph {et~al.}(2020)\citenamefont {Rossi},
  \citenamefont {{\v{S}}imkovic},\ and\ \citenamefont {Ferrero}}]{rdet}%
  \BibitemOpen
  \bibfield  {author} {\bibinfo {author} {\bibfnamefont {R.}~\bibnamefont
  {Rossi}}, \bibinfo {author} {\bibfnamefont {F.}~\bibnamefont
  {{\v{S}}imkovic}},\ and\ \bibinfo {author} {\bibfnamefont {M.}~\bibnamefont
  {Ferrero}},\ }\href {https://doi.org/10.1209/0295-5075/132/11001} {\bibfield
  {journal} {\bibinfo  {journal} {{EPL} (Europhysics Letters)}\ }\textbf
  {\bibinfo {volume} {132}},\ \bibinfo {pages} {11001} (\bibinfo {year}
  {2020})}\BibitemShut {NoStop}%
\bibitem [{\citenamefont {Proust}\ and\ \citenamefont
  {Taillefer}(2019)}]{proust2019remarkable}%
  \BibitemOpen
  \bibfield  {author} {\bibinfo {author} {\bibfnamefont {C.}~\bibnamefont
  {Proust}}\ and\ \bibinfo {author} {\bibfnamefont {L.}~\bibnamefont
  {Taillefer}},\ }\href@noop {} {\bibfield  {journal} {\bibinfo  {journal}
  {Annual Review of Condensed Matter Physics}\ }\textbf {\bibinfo {volume}
  {10}},\ \bibinfo {pages} {409} (\bibinfo {year} {2019})}\BibitemShut
  {NoStop}%
\bibitem [{\citenamefont {Zaanen}\ and\ \citenamefont
  {Gunnarsson}(1989{\natexlab{b}})}]{zaanen1989charged}%
  \BibitemOpen
  \bibfield  {author} {\bibinfo {author} {\bibfnamefont {J.}~\bibnamefont
  {Zaanen}}\ and\ \bibinfo {author} {\bibfnamefont {O.}~\bibnamefont
  {Gunnarsson}},\ }\href {https://doi.org/10.1103/PhysRevB.40.7391} {\bibfield
  {journal} {\bibinfo  {journal} {Phys. Rev. B}\ }\textbf {\bibinfo {volume}
  {40}},\ \bibinfo {pages} {7391} (\bibinfo {year}
  {1989}{\natexlab{b}})}\BibitemShut {NoStop}%
\bibitem [{\citenamefont {Sorella}(2021)}]{sorella2021phase}%
  \BibitemOpen
  \bibfield  {author} {\bibinfo {author} {\bibfnamefont {S.}~\bibnamefont
  {Sorella}},\ }\href@noop {} {\bibfield  {journal} {\bibinfo  {journal} {arXiv
  preprint arXiv:2101.07045}\ } (\bibinfo {year} {2021})}\BibitemShut {NoStop}%
\bibitem [{\citenamefont {Qin}\ \emph {et~al.}(2020)\citenamefont {Qin},
  \citenamefont {Chung}, \citenamefont {Shi}, \citenamefont {Vitali},
  \citenamefont {Hubig}, \citenamefont {Schollw\"ock}, \citenamefont {White},\
  and\ \citenamefont {Zhang}}]{qinabsence2020}%
  \BibitemOpen
  \bibfield  {author} {\bibinfo {author} {\bibfnamefont {M.}~\bibnamefont
  {Qin}}, \bibinfo {author} {\bibfnamefont {C.-M.}\ \bibnamefont {Chung}},
  \bibinfo {author} {\bibfnamefont {H.}~\bibnamefont {Shi}}, \bibinfo {author}
  {\bibfnamefont {E.}~\bibnamefont {Vitali}}, \bibinfo {author} {\bibfnamefont
  {C.}~\bibnamefont {Hubig}}, \bibinfo {author} {\bibfnamefont
  {U.}~\bibnamefont {Schollw\"ock}}, \bibinfo {author} {\bibfnamefont {S.~R.}\
  \bibnamefont {White}},\ and\ \bibinfo {author} {\bibfnamefont
  {S.}~\bibnamefont {Zhang}} (\bibinfo {collaboration} {Simons Collaboration on
  the Many-Electron Problem}),\ }\href
  {https://doi.org/10.1103/PhysRevX.10.031016} {\bibfield  {journal} {\bibinfo
  {journal} {Phys. Rev. X}\ }\textbf {\bibinfo {volume} {10}},\ \bibinfo
  {pages} {031016} (\bibinfo {year} {2020})}\BibitemShut {NoStop}%
\bibitem [{\citenamefont {Xiao}\ \emph {et~al.}(2022)\citenamefont {Xiao},
  \citenamefont {He}, \citenamefont {Georges},\ and\ \citenamefont
  {Zhang}}]{xiao2022temperature}%
  \BibitemOpen
  \bibfield  {author} {\bibinfo {author} {\bibfnamefont {B.}~\bibnamefont
  {Xiao}}, \bibinfo {author} {\bibfnamefont {Y.-Y.}\ \bibnamefont {He}},
  \bibinfo {author} {\bibfnamefont {A.}~\bibnamefont {Georges}},\ and\ \bibinfo
  {author} {\bibfnamefont {S.}~\bibnamefont {Zhang}},\ }\href@noop {}
  {\bibfield  {journal} {\bibinfo  {journal} {arXiv preprint arXiv:2202.11741}\
  } (\bibinfo {year} {2022})}\BibitemShut {NoStop}%
\bibitem [{\citenamefont {Chubukov}(1993)}]{chubukov1993}%
  \BibitemOpen
  \bibfield  {author} {\bibinfo {author} {\bibfnamefont {A.~V.}\ \bibnamefont
  {Chubukov}},\ }\href {https://doi.org/10.1103/PhysRevB.48.1097} {\bibfield
  {journal} {\bibinfo  {journal} {Phys. Rev. B}\ }\textbf {\bibinfo {volume}
  {48}},\ \bibinfo {pages} {1097} (\bibinfo {year} {1993})}\BibitemShut
  {NoStop}%
\bibitem [{\citenamefont {Deng}\ \emph {et~al.}(2015)\citenamefont {Deng},
  \citenamefont {Kozik}, \citenamefont {Prokof'ev},\ and\ \citenamefont
  {Svistunov}}]{deng}%
  \BibitemOpen
  \bibfield  {author} {\bibinfo {author} {\bibfnamefont {Y.}~\bibnamefont
  {Deng}}, \bibinfo {author} {\bibfnamefont {E.}~\bibnamefont {Kozik}},
  \bibinfo {author} {\bibfnamefont {N.~V.}\ \bibnamefont {Prokof'ev}},\ and\
  \bibinfo {author} {\bibfnamefont {B.~V.}\ \bibnamefont {Svistunov}},\
  }\href@noop {} {\bibfield  {journal} {\bibinfo  {journal} {EPL}\ }\textbf
  {\bibinfo {volume} {110}} (\bibinfo {year} {2015})}\BibitemShut {NoStop}%
\bibitem [{\citenamefont {\ifmmode~\check{S}\else \v{S}\fi{}imkovic}\ \emph
  {et~al.}(2021)\citenamefont {\ifmmode~\check{S}\else \v{S}\fi{}imkovic},
  \citenamefont {Deng},\ and\ \citenamefont {Kozik}}]{simkovicsuperfluid2021}%
  \BibitemOpen
  \bibfield  {author} {\bibinfo {author} {\bibfnamefont {F.}~\bibnamefont
  {\ifmmode~\check{S}\else \v{S}\fi{}imkovic}}, \bibinfo {author}
  {\bibfnamefont {Y.}~\bibnamefont {Deng}},\ and\ \bibinfo {author}
  {\bibfnamefont {E.}~\bibnamefont {Kozik}},\ }\href
  {https://doi.org/10.1103/PhysRevB.104.L020507} {\bibfield  {journal}
  {\bibinfo  {journal} {Phys. Rev. B}\ }\textbf {\bibinfo {volume} {104}},\
  \bibinfo {pages} {L020507} (\bibinfo {year} {2021})}\BibitemShut {NoStop}%
\bibitem [{\citenamefont {Gull}\ \emph {et~al.}(2013)\citenamefont {Gull},
  \citenamefont {Parcollet},\ and\ \citenamefont {Millis}}]{gull2013super}%
  \BibitemOpen
  \bibfield  {author} {\bibinfo {author} {\bibfnamefont {E.}~\bibnamefont
  {Gull}}, \bibinfo {author} {\bibfnamefont {O.}~\bibnamefont {Parcollet}},\
  and\ \bibinfo {author} {\bibfnamefont {A.~J.}\ \bibnamefont {Millis}},\
  }\href {https://doi.org/10.1103/PhysRevLett.110.216405} {\bibfield  {journal}
  {\bibinfo  {journal} {Phys. Rev. Lett.}\ }\textbf {\bibinfo {volume} {110}},\
  \bibinfo {pages} {216405} (\bibinfo {year} {2013})}\BibitemShut {NoStop}%
\bibitem [{\citenamefont {Kitatani}\ \emph {et~al.}(2019)\citenamefont
  {Kitatani}, \citenamefont {Sch\"afer}, \citenamefont {Aoki},\ and\
  \citenamefont {Held}}]{kitatani2019why}%
  \BibitemOpen
  \bibfield  {author} {\bibinfo {author} {\bibfnamefont {M.}~\bibnamefont
  {Kitatani}}, \bibinfo {author} {\bibfnamefont {T.}~\bibnamefont {Sch\"afer}},
  \bibinfo {author} {\bibfnamefont {H.}~\bibnamefont {Aoki}},\ and\ \bibinfo
  {author} {\bibfnamefont {K.}~\bibnamefont {Held}},\ }\href
  {https://doi.org/10.1103/PhysRevB.99.041115} {\bibfield  {journal} {\bibinfo
  {journal} {Phys. Rev. B}\ }\textbf {\bibinfo {volume} {99}},\ \bibinfo
  {pages} {041115} (\bibinfo {year} {2019})}\BibitemShut {NoStop}%
\bibitem [{\citenamefont {Lenihan}\ \emph {et~al.}(2021)\citenamefont
  {Lenihan}, \citenamefont {Kim}, \citenamefont {\ifmmode
  \check{S}\else~\v{S}\fi{}imkovic IV.},\ and\ \citenamefont
  {Kozik}}]{lenihan2020entropy}%
  \BibitemOpen
  \bibfield  {author} {\bibinfo {author} {\bibfnamefont {C.}~\bibnamefont
  {Lenihan}}, \bibinfo {author} {\bibfnamefont {A.~J.}\ \bibnamefont {Kim}},
  \bibinfo {author} {\bibfnamefont {F.}~\bibnamefont {\ifmmode
  \check{S}\else~\v{S}\fi{}imkovic IV.}},\ and\ \bibinfo {author}
  {\bibfnamefont {E.}~\bibnamefont {Kozik}},\ }\href
  {https://doi.org/10.1103/PhysRevLett.126.105701} {\bibfield  {journal}
  {\bibinfo  {journal} {Phys. Rev. Lett.}\ }\textbf {\bibinfo {volume} {126}},\
  \bibinfo {pages} {105701} (\bibinfo {year} {2021})}\BibitemShut {NoStop}%
\bibitem [{\citenamefont {Hubbard}(1963)}]{hubbard63}%
  \BibitemOpen
  \bibfield  {author} {\bibinfo {author} {\bibfnamefont {J.}~\bibnamefont
  {Hubbard}},\ }\href@noop {} {\bibfield  {journal} {\bibinfo  {journal} {Proc.
  R. Soc. Lond. A}\ }\textbf {\bibinfo {volume} {276}},\ \bibinfo {pages} {238}
  (\bibinfo {year} {1963})}\BibitemShut {NoStop}%
\bibitem [{\citenamefont {Kanamori}(1963)}]{kanamori}%
  \BibitemOpen
  \bibfield  {author} {\bibinfo {author} {\bibfnamefont {J.}~\bibnamefont
  {Kanamori}},\ }\href@noop {} {\bibfield  {journal} {\bibinfo  {journal}
  {Progress of Theoretical Physics}\ }\textbf {\bibinfo {volume} {30}},\
  \bibinfo {pages} {275} (\bibinfo {year} {1963})}\BibitemShut {NoStop}%
\bibitem [{\citenamefont {Anderson}(1963)}]{anderson1963theory}%
  \BibitemOpen
  \bibfield  {author} {\bibinfo {author} {\bibfnamefont {P.~W.}\ \bibnamefont
  {Anderson}},\ }\href@noop {} {\bibfield  {journal} {\bibinfo  {journal}
  {Solid state physics}\ }\textbf {\bibinfo {volume} {14}},\ \bibinfo {pages}
  {99} (\bibinfo {year} {1963})}\BibitemShut {NoStop}%
\bibitem [{\citenamefont {Moutenet}\ \emph {et~al.}(2018)\citenamefont
  {Moutenet}, \citenamefont {Wu},\ and\ \citenamefont
  {Ferrero}}]{alice_michel}%
  \BibitemOpen
  \bibfield  {author} {\bibinfo {author} {\bibfnamefont {A.}~\bibnamefont
  {Moutenet}}, \bibinfo {author} {\bibfnamefont {W.}~\bibnamefont {Wu}},\ and\
  \bibinfo {author} {\bibfnamefont {M.}~\bibnamefont {Ferrero}},\ }\href
  {https://doi.org/10.1103/PhysRevB.97.085117} {\bibfield  {journal} {\bibinfo
  {journal} {Phys. Rev. B}\ }\textbf {\bibinfo {volume} {97}},\ \bibinfo
  {pages} {085117} (\bibinfo {year} {2018})}\BibitemShut {NoStop}%
\bibitem [{\citenamefont {\ifmmode~\check{S}\else \v{S}\fi{}imkovic}\ and\
  \citenamefont {Kozik}(2019)}]{fedor_sigma}%
  \BibitemOpen
  \bibfield  {author} {\bibinfo {author} {\bibfnamefont {F.}~\bibnamefont
  {\ifmmode~\check{S}\else \v{S}\fi{}imkovic}}\ and\ \bibinfo {author}
  {\bibfnamefont {E.}~\bibnamefont {Kozik}},\ }\href
  {https://doi.org/10.1103/PhysRevB.100.121102} {\bibfield  {journal} {\bibinfo
   {journal} {Phys. Rev. B}\ }\textbf {\bibinfo {volume} {100}},\ \bibinfo
  {pages} {121102(R)} (\bibinfo {year} {2019})}\BibitemShut {NoStop}%
\bibitem [{\citenamefont {Rossi}(2018)}]{rr_sigma}%
  \BibitemOpen
  \bibfield  {author} {\bibinfo {author} {\bibfnamefont {R.}~\bibnamefont
  {Rossi}},\ }\href@noop {} {\bibfield  {journal} {\bibinfo  {journal}
  {arXiv:1802.04743}\ } (\bibinfo {year} {2018})}\BibitemShut {NoStop}%
\bibitem [{\citenamefont {IV}\ and\ \citenamefont
  {Ferrero}(2022)}]{vsimkovic2021fast}%
  \BibitemOpen
  \bibfield  {author} {\bibinfo {author} {\bibfnamefont {F.~i. c.~v.}\
  \bibnamefont {IV}}\ and\ \bibinfo {author} {\bibfnamefont {M.}~\bibnamefont
  {Ferrero}},\ }\href {https://doi.org/10.1103/PhysRevB.105.125104} {\bibfield
  {journal} {\bibinfo  {journal} {Phys. Rev. B}\ }\textbf {\bibinfo {volume}
  {105}},\ \bibinfo {pages} {125104} (\bibinfo {year} {2022})}\BibitemShut
  {NoStop}%
\bibitem [{\citenamefont {{\v{S}}imkovic}\ and\ \citenamefont
  {Rossi}(2021)}]{MCMCMC}%
  \BibitemOpen
  \bibfield  {author} {\bibinfo {author} {\bibfnamefont {F.}~\bibnamefont
  {{\v{S}}imkovic}}\ and\ \bibinfo {author} {\bibfnamefont {R.}~\bibnamefont
  {Rossi}},\ }\href@noop {} {\bibfield  {journal} {\bibinfo  {journal} {arXiv
  preprint arXiv:2102.05613}\ } (\bibinfo {year} {2021})}\BibitemShut {NoStop}%
\bibitem [{\citenamefont {Benfatto}\ \emph {et~al.}(2006)\citenamefont
  {Benfatto}, \citenamefont {Giuliani},\ and\ \citenamefont
  {Mastropietro}}]{benfatto_convergence}%
  \BibitemOpen
  \bibfield  {author} {\bibinfo {author} {\bibfnamefont {G.}~\bibnamefont
  {Benfatto}}, \bibinfo {author} {\bibfnamefont {A.}~\bibnamefont {Giuliani}},\
  and\ \bibinfo {author} {\bibfnamefont {V.}~\bibnamefont {Mastropietro}},\
  }\href@noop {} {\bibfield  {journal} {\bibinfo  {journal} {Annales Henri
  Poincaré}\ }\textbf {\bibinfo {volume} {7}},\ \bibinfo {pages} {809}
  (\bibinfo {year} {2006})}\BibitemShut {NoStop}%
\end{thebibliography}%
\onecolumngrid
\newpage
\clearpage
\appendix

\section{Model} 
    In this work we study thermal-equibrium properties of the Fermi-Hubbard model~\cite{hubbard63,kanamori,anderson1963theory,qin2021hubbard,arovas2021hubbard} on the square lattice, 
     defined by the hamiltonian:
    \begin{align}
    \hat{H} = \sum_{\mathbf{k}, \sigma} \epsilon_{\mathbf{k}} \,
    \hat{c}_{\mathbf{k} \sigma}^\dagger
    \hat{c}_{\mathbf{k} \sigma}^{\phantom{\dagger}}+U\sum_{\mathbf{r}} \hat{n}_{\mathbf{r}\uparrow}\,
    \hat{n}_{\mathbf{r} \downarrow} ,
    \label{H}
    \end{align}
where
    $\mathbf{k}\in [-\pi,\pi]^2$ is the reciprocal lattice momentum,
    $\sigma \in \{ \uparrow, \downarrow \} $ is the fermionic spin, $\hat{c}^\dagger_{{\mathbf{ k}}\sigma}$ and $\hat{c}_{\mathbf{k}\sigma}$ denote the fermionic creation and annihilation operators, $\mathbf{r}\in \{0,1,\dots,L-1\}^2$ labels lattice sites and $L$ is the linear system size (in this work we use $L=64$), $U$ is the onsite
    repulsion strength, $\mu$ the chemical potential, the square lattice dispersion relation is given by $\epsilon_{\mathbf{k}}=-2\,t\left(\cos k_x+\cos k_y\right)$, where $t$ is the nearest-neighbor hopping amplitude (we set $t=1$ in our units), and $\hat{n}_{\mathbf{r}\sigma}$ counts the number of particles with spin $\sigma$ at site $\mathbf{r}$.
    
    We consider thermal-equilibrium properties at temperature $T$ in the grand-canonical ensemble with chemical potential $\mu$, and we denote by $\langle \hat{O}\rangle$ the thermal average of an operator $\hat{O}$:
\begin{equation}
        \langle\hat{O}\rangle\equiv \frac{\text{Tr}\;\hat{O}\;
e^{-\hat{H}'/T}
        }{
               \text{Tr}\, e^{ -\hat{H}'/T}
        },
    \end{equation}
    where $\hat{H}'=\hat{H}-\mu\sum_{\mathbf{r},\sigma}\hat{n}_{\mathbf{r}\sigma}$ is the grand-canonical hamiltonian. To compute dynamical quantities, we use the Matsubara formalism and the imaginary-time Heisenberg representation $\hat{O}(\tau)\equiv e^{\tau\hat{H}'}\,\hat{O}\,e^{-\tau\hat{H}'}$ of an operator $\hat{O}$.

The one-particle Green's function $G$ in the Matsubara representation is
\begin{equation}
    G(\mathbf{k},i\omega_n)\equiv -\int_0^{1/T} d\tau\; e^{i\omega_n\tau}\;\left\langle \hat{c}_{\mathbf{k}\uparrow}(\tau)\;\hat{c}_{\mathbf{k}\uparrow}^\dagger\right\rangle,
\end{equation}
where $\omega_n=(2n+1)\,\pi\,T$ is a fermionic Matsubara frequency. It is possible to express $G$ in terms of the spectral function $A(\mathbf{k},\omega)$ by
\begin{equation}
    G(\mathbf{k},i\omega_n)=\int_{-\infty}^\infty d\omega\;\frac{A(\mathbf{k},\omega)}{i\omega_n-\omega}
\end{equation}
We define the zero-energy spectral function proxy $A(\mathbf{k})$ as
\begin{equation}
A(\mathbf{k})\equiv -\frac{1}{\pi}\,\text{Im}\;G(\mathbf{k},i \pi T),
\end{equation}
which is a valid approximation of $A(\mathbf{k},\omega=0)$ at low-enough temperature as
\begin{equation}
    A(\mathbf{k})\approx -\frac{1}{\pi}\,\text{Im}\;G(\mathbf{k},i 0^+)=A(\mathbf{k},\omega=0).
\end{equation}

The self-energy $\Sigma$ can be introduced from the Dyson's equation
\begin{equation}
    \Sigma(\mathbf{k},i\omega_n)\equiv \left[G_0(\mathbf{k},i\omega_n)\right]^{-1}-\left[G(\mathbf{k},i\omega_n)\right]^{-1}.
\end{equation}
where $G_0$ is the non-interacting Green's function. We also define 
\begin{equation}
    \Delta \text{Im}\,\Sigma(\mathbf{k})\equiv \text{Im}\,\Sigma(\mathbf{k},i\omega_0)-\text{Im}\,\Sigma(\mathbf{k},i\omega_1),
\end{equation}
whose sign, in combination with the value of $\text{Im}\,\Sigma(\mathbf{k},i\omega_0)$, is used to distinguish between metallic and insulating behavior of the self-energy as $\text{Im}\;\Sigma(\mathbf{k},i0^+)= 0$ for momentum $\mathbf{k}$ belonging to the Fermi surface in a Fermi liquid at zero temperature.

The static real-space spin and charge susceptibilities are defined as
\begin{equation}
    \chi_{\text{sp}}(\mathbf{r})=\int_0^{1/T}d\tau\;\langle\hat{S}_z(\mathbf{r},\tau)\,\hat{S}_z(\mathbf{r}, 0)\rangle, 
\end{equation}
\begin{equation}
    \chi_{\text{ch}}(\mathbf{r})=\int_0^{1/T}d\tau\;\langle\delta\hat{n}(\mathbf{r},\tau)\,\delta\hat{n}(\mathbf{r}, 0)\rangle, 
\end{equation}
where $\hat{S}_z(\mathbf{r})\equiv \frac{1}{2}\left(\hat{n}_{\mathbf{r}\uparrow} -\hat{n}_{\mathbf{r}\downarrow}\right)$, $\delta\hat{n}(\mathbf{r})\equiv \sum_\sigma\left(\hat{n}_{\mathbf{r}\sigma}-\langle\hat{n}_{\mathbf{r}\sigma}\rangle\right)$.

\section{Methods}
We perform a numerical study of the Hubbard model by means of a version of Connected Determinant Diagrammatic Monte Carlo (CDet) algorithm~\cite{cdet}, which allows to calculate high-order diagrammatic contributions to any (imaginary-time) physical quantity for arbitrary system sizes. This gives access to numerically controlled results and fine momentum-space resolution.

We compute bare (corresponding to Feynman diagrams containing non-interacting Green's functions) double interaction-chemical-potential expansions in terms of the interaction strength $U$ and a chemical-potential shift $\alpha\,U$, as introduced in Ref.~\cite{vsimkovic2021weak}, which we summarize below. We write any quantity of interest, $\mathcal{O}$, as an explicit function of the chemical potential $\mu$ and the interaction strength $\mathcal{O}(\mu,U)$. From $\mathcal{O}(\mu,U)$, we introduce two auxiliary mathematical quantities: the ``Hartree'' expansion
\begin{equation}
    \mathcal{O}_{\text{Hartree}}(\mu_0,U)\equiv\mathcal{O}(\mu_0+U\,n_0/2,U)=\sum_{k=0}^\infty U^k\;\mathcal{O}_{\text{Hartree};k}(\mu_0),
\end{equation}
where $n_0$ is the number of particle per site at $U=0$, and the ``double'' expansion:
\begin{equation}
        \mathcal{O}_{\text{Double}}(\mu_0,\alpha, U)\equiv\mathcal{O}(\mu_0+\alpha\,U,U)=\sum_{k=0}^\infty\sum_{j=0}^k U^k\;\alpha^j\;\mathcal{O}_{\text{Double};k,j}(\mu_0).
\end{equation}
Diagrammatically, with respect to the bare series, the Hartree series does not contain any tadpole insertions, while the double series contains arbitrary chemical potential insertions. The calculation of $\mathcal{O}_{\text{Hartree};k}$ is performed using the algorithm of Ref.~\cite{cdet}, while for obtaining $\mathcal{O}_{\text{Double};k,j}$ we use the method of Ref.~\cite{vsimkovic2021weak}. For the self-energy, we make use of a slightly modified algorithm compared to previous realisations~\cite{alice_michel,fedor_sigma,rr_sigma}, which we detail for completeness in Sec.~\ref{sm-sec-self-energy}. We use a fast principal minor algorithm for the simultaneous evaluation of an exponential number of determinants~\cite{vsimkovic2021fast} and the Many Configuration Markov Chain Monte Carlo for numerical integration~\cite{MCMCMC}. 

 We choose the chemical potential  shift $\alpha(U)$ that fixes the average particle number to a constant as a function of $U$. Jointly using these expansions provides us a way to double-check the final result. 

It is well known~\cite{benfatto_convergence} that for fermions on a lattice at finite temperature the perturbative series has a nonzero radius of convergence. When outside the radius of convergence, we resum the series by means of Pad\'e approximants~\cite{fedor_sigma}.

The correlation length $\xi$ is obtained from fitting the spin susceptibility with a double-Lorentzian Ornstein-Zernike form (with constant offset):
\begin{equation}
    \chi_{\text{sp}}(\mathbf{q}) \approx A\left(\frac{1}{|\mathbf{q}-(Q_x,Q_y)|^2+\xi^2}+\frac{1}{|\mathbf{q}-(Q_y,Q_x)|^2+\xi^2}\right).
\end{equation}
     
     \subsection{Algorithm for the self-energy}\label{sm-sec-self-energy}
     For a set of spacetime vertices $\{X_1,\dots,X_n\}$ representating the Hubbard on-site interactions of an order-$n$ Feynman diagram, we define the $n\times n$ matrix
     \begin{equation}        [\mathbb{G}_{\sigma}]_{jk} = G_{0;\sigma}(X_j,X_k)(1-\delta_{j,k})-\alpha \delta_{j,k},
     \end{equation}
     where $G_{0;\sigma}$ is the bare one-particle propagator.
     For a subset $S\subseteq \{X_1,\dots,X_n\}$, we define
     \begin{equation}
         Z(S):=\det_S \mathbb{G}_\uparrow \;\det_S \mathbb{G}_\downarrow ,
     \end{equation}
     where $\det_S$ means that only the subset of indices $S$ is retained when computing the determinant.
     $Z(S)$ is the sum of all diagrams built on the set of vertices $S$ contributing to the partition function. $Z(S)$ is a polynomial in $\alpha$ of degree $2\,|S|$. 
     For $S\subseteq\{X_1,\dots,X_n\}$, and for $X_j,X_k\in S$, we define the $|S|\times |S|$ matrix $\mathbb{Z}(S)$ by
     \begin{equation}
     \mathbb{Z}_{jk}(S)=\frac{\partial  }{\partial {G_{0;\uparrow}(X_k,X_j)}} Z(S)=\left[\mathbb{G}_{\uparrow}^{-1}\right]_{jk} Z(S).
     \end{equation}
     Let $\Xi_{jk}$ be the connected part of $\mathbb{Z}_{jk}$, recursively defined from
\begin{equation}
    \Xi_{jk}(S) = \mathbb{Z}_{jk}(S) -\sum_{S'\subsetneq S}\Xi_{jl}(S')\;Z(S\setminus S'),
\end{equation}
with $\Xi_{jk}(\emptyset) = 0$.
  We introduce the matrix $\rho$ by
  \begin{equation}
      \rho_{jk}(S)=\sum_{l}[\mathbb{G}_\uparrow]_{jl} \;\Xi_{lk}(S).
  \end{equation}
We can finally compute the contribution to the self-energy $\Sigma_{X_j,X_k}(S)=:\Sigma_{jk}(S)$ coming from the $S$ vertices, for all pairs of vertices $X_j,X_k$, as
\begin{equation}\label{eq-self-energy-appendix-final}
    \Sigma_{jk}(S)=\Xi_{jk}(S)-\sum_{S'\subsetneq S}\sum_{l}\Sigma_{jl}(S')\,\rho_{lk}(S\setminus S')-\sum_{S'\subsetneq S}\Xi_{jj}(S')\,\rho_{jk}(S\setminus S').
\end{equation}
The diagrammatic interpretation of this algebraic procedure is the elimination of one-particle-reducible diagrams (second term in the r.h.s. of Eq.~\eqref{eq-self-energy-appendix-final}) and bold tadpole contributions (third term in the r.h.s. of Eq.~\eqref{eq-self-energy-appendix-final}) from the sum of all connected diagrams. 
These equations are solved in the field of truncated polynomials of degree $n$ in $\alpha$ with an overhead of $O(n^2)$ for multiplication and division. In order to directly obtain quantities at momentum-frequency $K$, the Fourier transform of $\Sigma$ is used as Monte Carlo weight at each step
\begin{equation}
    \Sigma_{K} =
    \left\langle\frac{1}{n(n-1)}\sum_{j,k}e^{iK\cdot (X_j-X_k)}\;
    \Sigma_{j,k}\right\rangle.
\end{equation}

\section{Criteria to identify the pseudogap region}

\begin{figure}[h!]
\centering
\includegraphics[width=0.5\columnwidth]{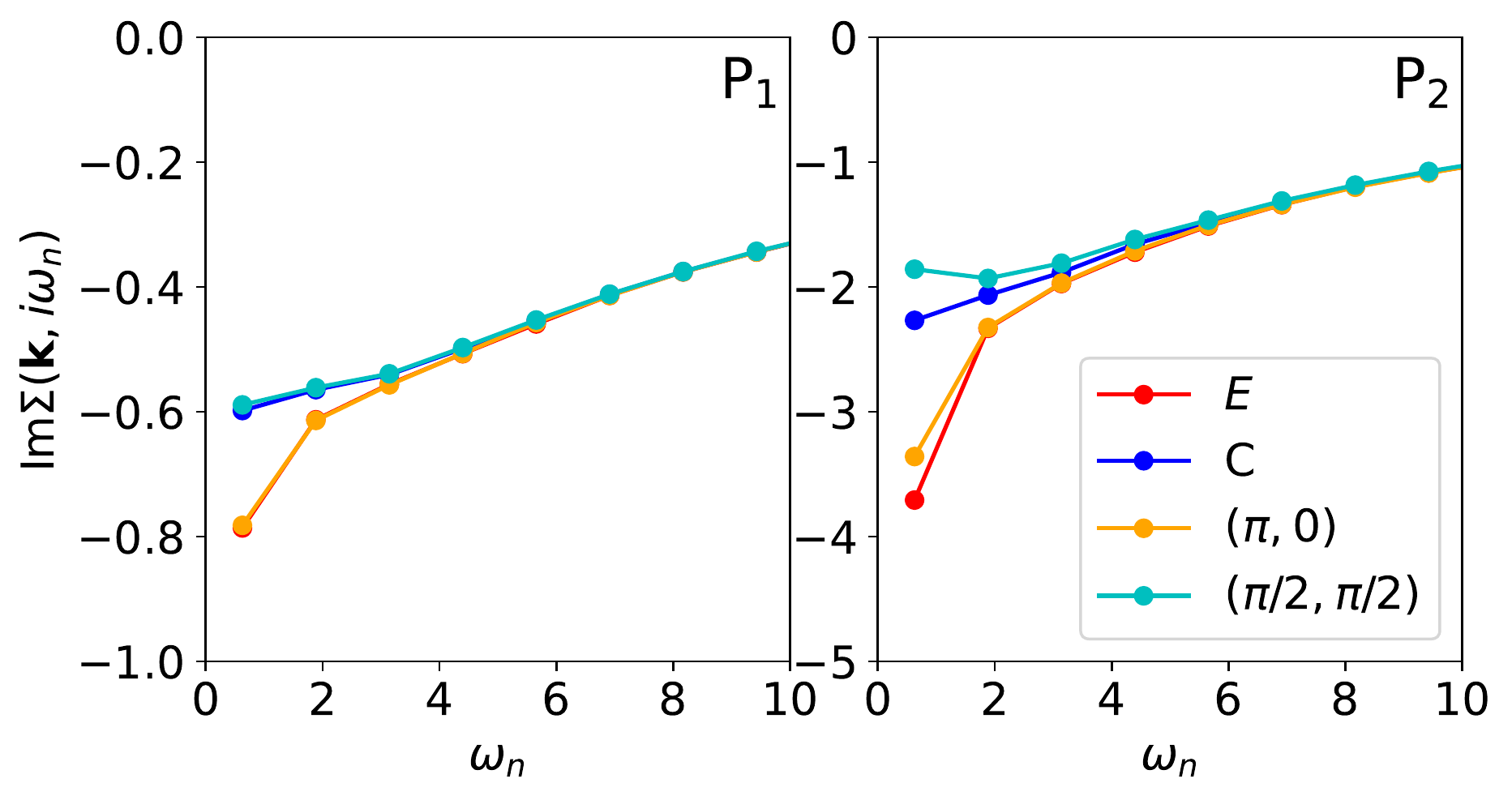}
\caption{The frequency dependence of the imaginary part of the self-energy is shown for the weak-coupling pseudogap ($P_1$, left) and the strong-coupling pseudogap ($P_2$, right). The displayed momenta correspond to the center (C) and the edge (E) as well as $(\pi,0)$ and $(\pi/2,\pi/2)$.}
\label{fig_center_edge}
\end{figure}  

In this section we discuss multiple distinct numerical criteria for the identification of the pseudogap regime. 

We have seen in Fig.~\ref{fig_regions} of the main text, that one pseudogap criterion is the change of slope  of the imaginary part of the self-energy, $\Delta \operatorname{Im} \Sigma \equiv \operatorname{Im}\Sigma(\mathbf{k},i\omega_0) - \operatorname{Im}\Sigma(\mathbf{k},i\omega_1)$. As we have access to the full momentum resolution of the self-energy, we can pinpoint the first $\mathbf{k}$-point at which the slope changes sign as well as study the evolution of this criterion as a function of temperature. To this end, we introduce two momenta called \emph{center} (C) and \emph{edge} (E) defined as the points in reciprocal space where
the imaginary part of the self-energy reaches its maximum along the lines $(\mathbf{q}, \mathbf{q})$ and $(\mathbf{q}, \pi)$, respectively. 

We found in all examined parameter regimes that the slope first changes at the edge point and then this (red) region of negative slope quickly develops towards the center point. We note that, when the system is doped, both of these momenta move away from the $(\pi/2,\pi/2)$ and $(\pi,0)$ points, which we will denote by \emph{nodal} (N) and \emph{antinodal} (AN) momenta in the following. In Fig.~\ref{fig_center_edge}, the frequency dependence of $\operatorname{Im}\Sigma$ is compared for the center and the edge as well as the nodal $(\pi,0)$ and antinodal $(\pi/2,\pi/2)$ momenta. We see that while in the weak-coupling pseudogap regime (P$_1$) there is barely any noticeable difference between E (C) and AN (N). In the strong-coupling pseudogap regime (P$_2$) we see significant differences between these momenta. In particular, the slope for both E and C is more negative than their antinodal and nodal counterparts. Throughout this paper we identify the pseudogap crossover with the change of the slope at the first, edge  momentum, which is very well defined. Note that this possibly yields slightly higher crossover temperatures as compared to other criteria in literature and might be considered as a precursor. However, we find that this difference is not very significant and this crossover clearly signals the onset of severe deformations of the self-energy and spectral function due to antiferromagnetic fluctuations. 

\begin{figure}[h!]
\centering
\includegraphics[width=0.49\columnwidth]{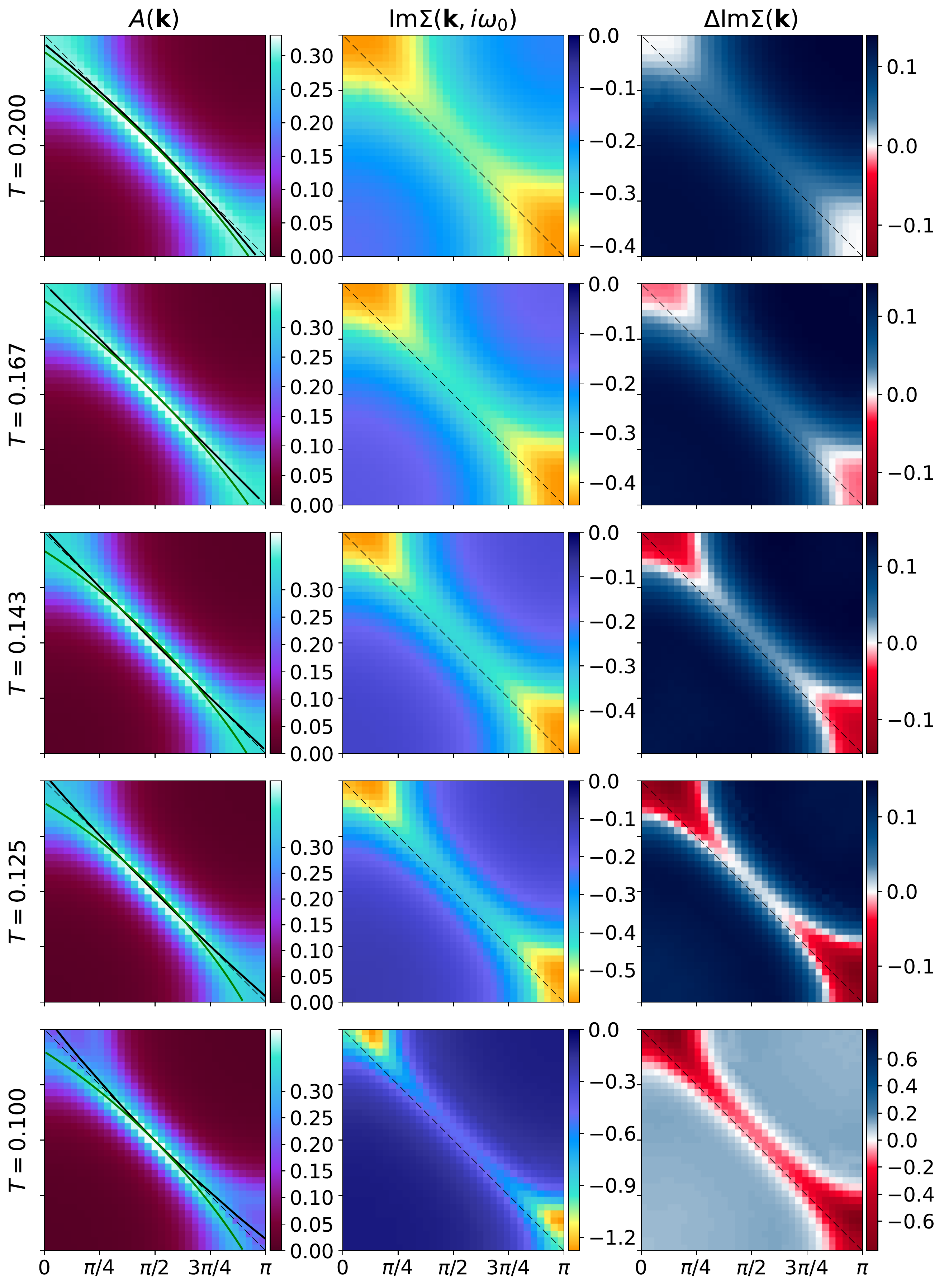}
\includegraphics[width=0.49\columnwidth]{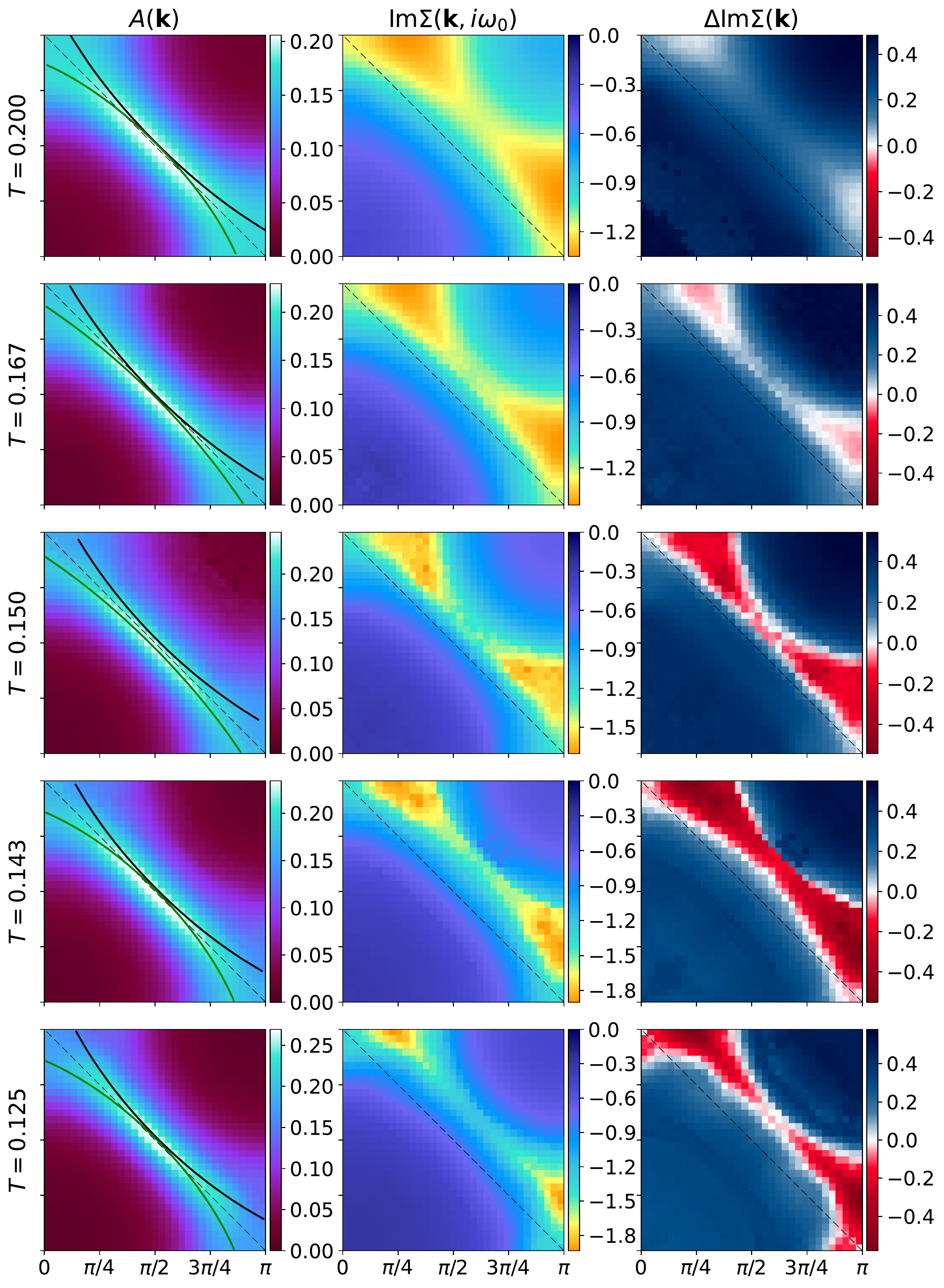}
\caption{The momentum-resolved zero-frequency spectral function proxy $A(\bf k)$, the imaginary part of the self-energy and the difference between the
imaginary part of the self-energy at the two lowest Matsubara frequencies are shown for various temperatures for examples of a  weak-coupling pseudogap (left, $\delta=0.043$, $U=3.5$) and a strong-coupling pseudogap (right, $\delta=0.087$, $U=6.0$).}
\label{fig_PG_vs_T}
\end{figure}

\begin{figure}[h!]
\centering
\includegraphics[width=0.3\textwidth]{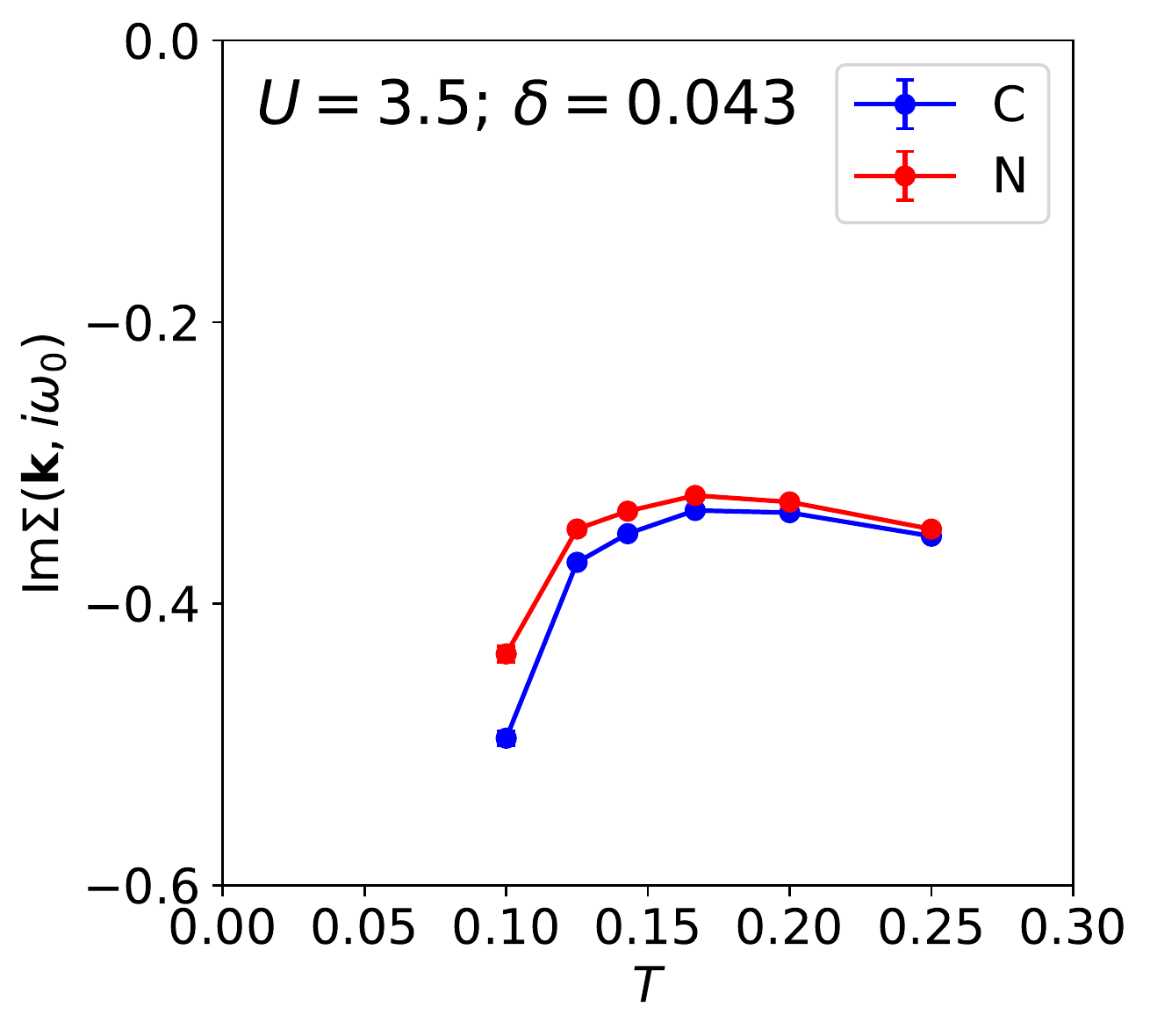}
\includegraphics[width=0.3\textwidth]{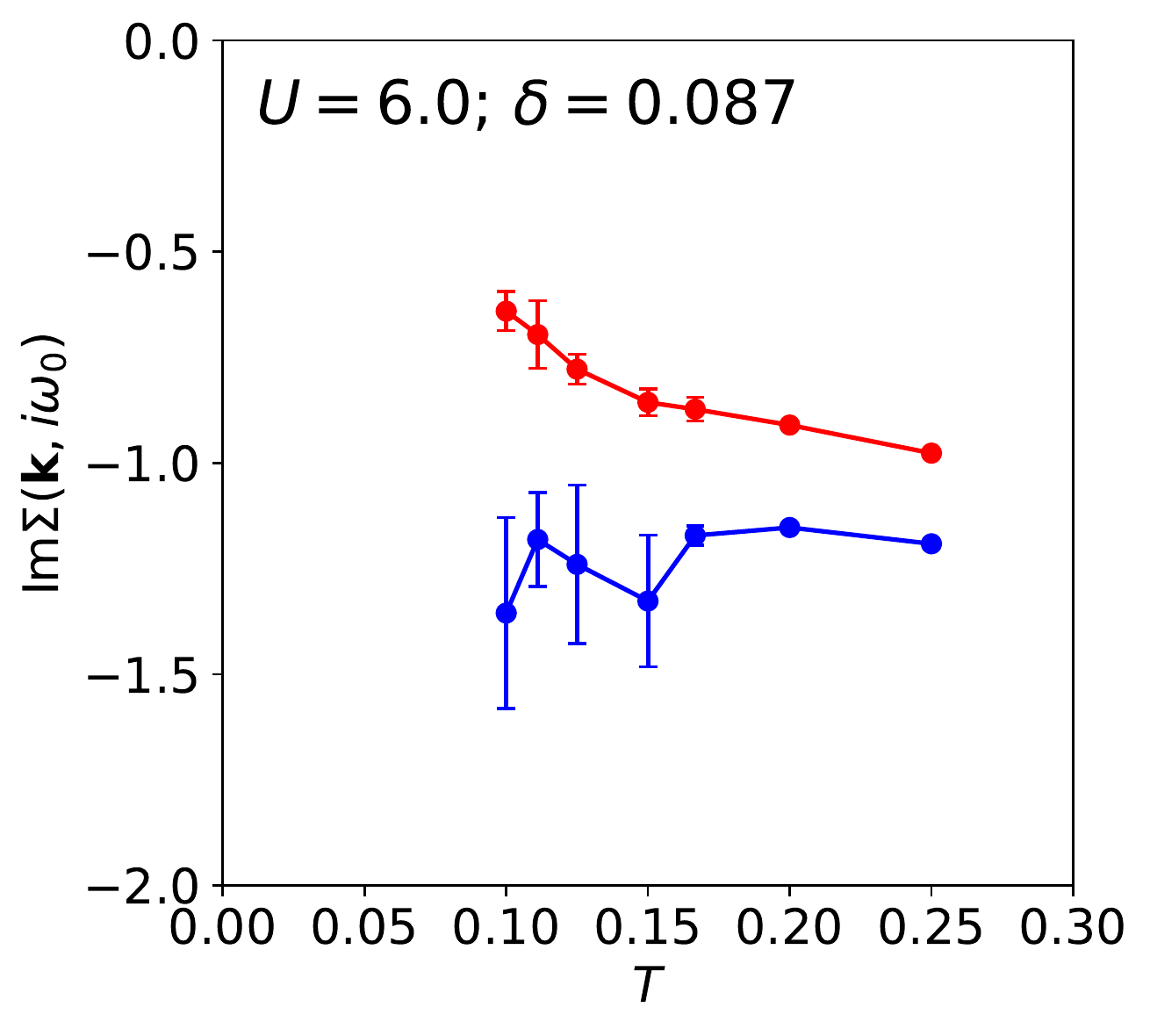}
\caption{The temperature dependence of the self-energy at the node (N) as well as center (C) momenta is shown across examples of a weakly-coupling (left) and strong-coupling (right) pseudogap crossovers.}
\label{fig_CvsN}
\end{figure}

In Fig.~\ref{fig_PG_vs_T} we show the temperature evolution of the spectral function, the imaginary part of the self-energy, as well as its slope $\Delta \operatorname{Im} \Sigma$. In the left panel the weak-coupling pseudogap regime is investigated ($\delta=0.043$ and $U=3.5$). We observe that the slope first changes in the vicinity of the momentum $(0,\pi)$ and then this region gradually grows towards momentum $(\pi/2,\pi/2)$. This case is strongly reminiscent of what is observed at half-filling. In contrast, in the right panel we display the strong-coupling pseudogap regime ($\delta=0.087$ and $U=6.0$). Here the slope first changes at roughly $(\pi/4, \pi)$ and grows inwards towards roughly $(5\pi/8, 5\pi/8)$, closely following the regions where the imaginary part of the self-energy is largest. We would like to stress that this is the first time a numerically unbiased method has been able to observe such behaviour of the self-energy in the doped Hubbard model.

\begin{figure}[h!]
\centering
\includegraphics[width=0.49\columnwidth]{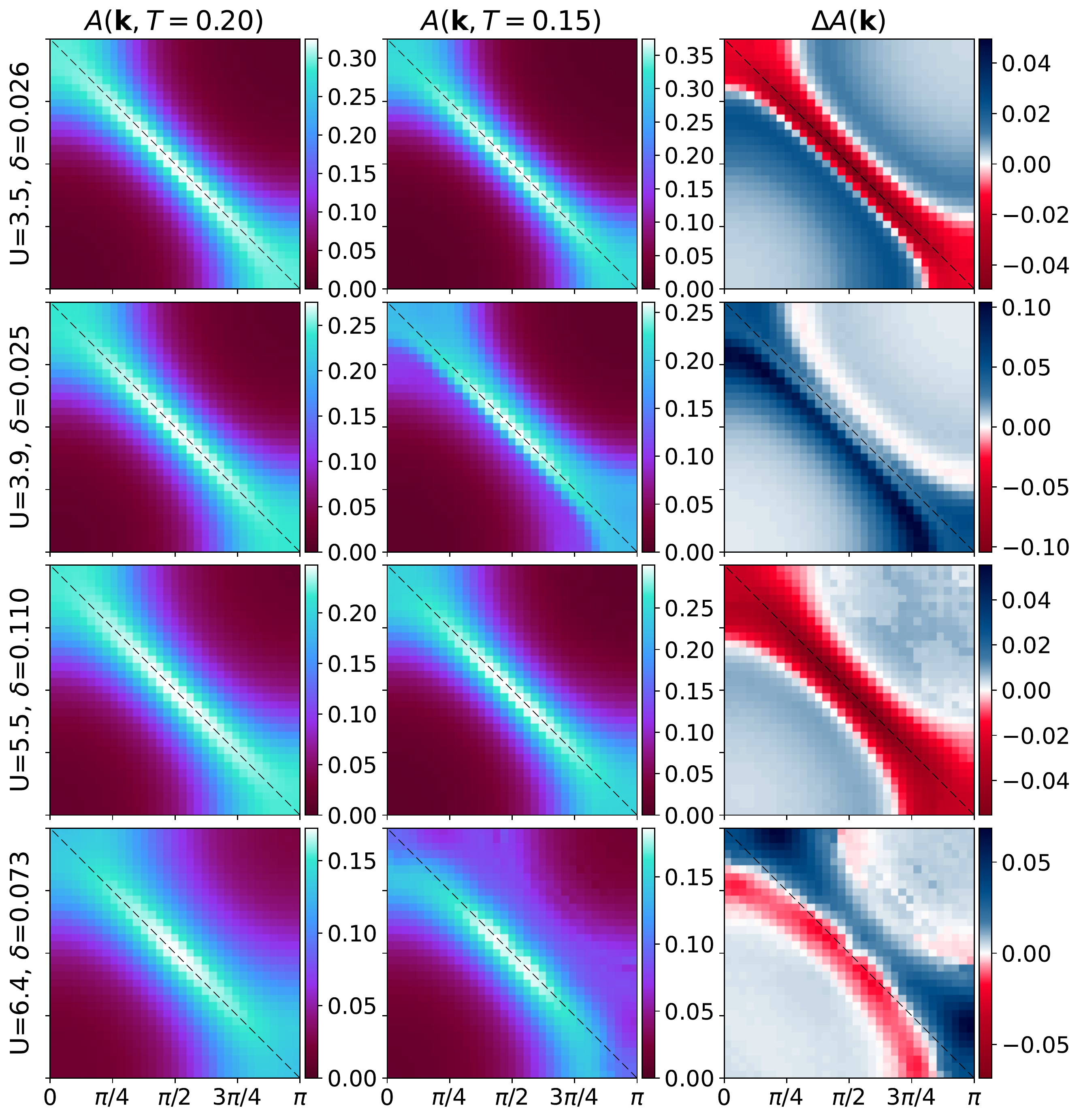}
\caption{The momentum resolved spectral function $A(k)$ is shown for two temperatures ($T=\{0.2,0.15\}$) as well as their relative difference $\Delta A(\mathbf{k}) = A(\mathbf{k}, T=0.2)-A(\mathbf{k}, T=0.15)$}
\label{fig_dA_PG}
\end{figure}

We further study the temperature dependence of the imaginary self-energy at the node (N) as well as center (C) momenta for examples of the weak- and strong-coupling pseudogap regimes in Fig.~\ref{fig_CvsN}. At high temperatures, above the pseudogap regimes, the behaviour is similar in both cases, the imaginary self-energy increases as temperature is lowered. Below the crossover temperature in the weakly-correlated regime we see both a sharp decrease for both momenta. In contrast, for the strongly-correlated case, the self energy actually increases at the node momentum, whilst staying constant within error bars at the center momentum. This represents a stark qualitative difference between the two pseudogap regimes and could in principle be used to identify the boundary between them.

\begin{figure}[h!]
\centering
\includegraphics[width=0.55\textwidth]{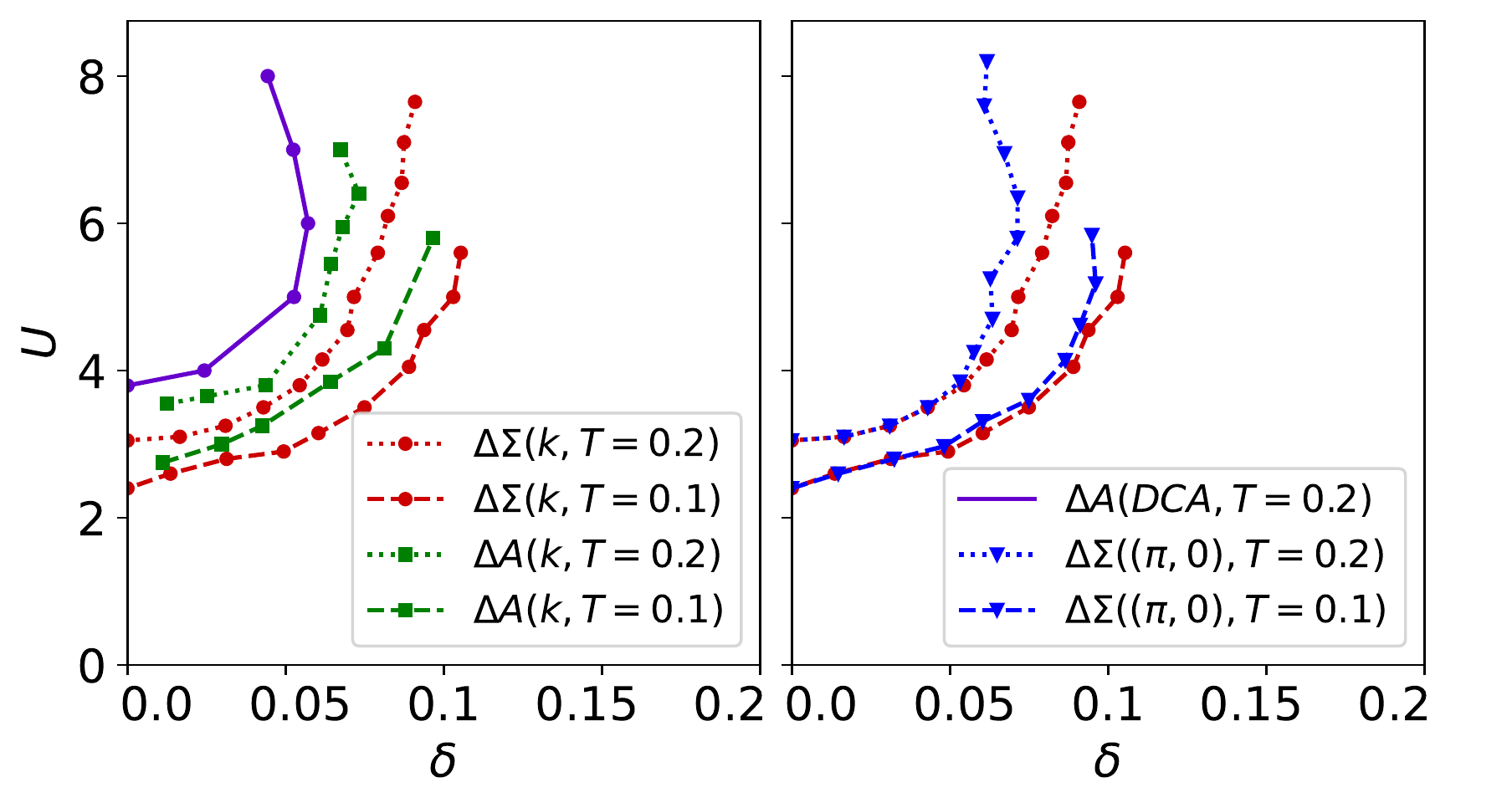}
\caption{The location of the pseudogap crossover at temperatures $T=\{0.1,0.2\}$ as established from different criteria. Results from the criterion of the difference between the imaginary part of the self-energy at the two lowest Matsubara frequencies for the edge (E) momentum is shown in red. The same for momentum $(\pi,0)$ is shown in blue. Results from the temperature difference of the spectral function criterion are shown in green. We used temperatures $T=\{0.2,0.15\}$ to establish the slope around $T=0.2$ and temperatures $T=\{0.125,0.1\}$ for the slope around $T=0.1$.}
\label{fig_PG_A_vs_Sigma}
\end{figure}

\begin{figure}[h!]
\centering
\includegraphics[width=0.3\textwidth]{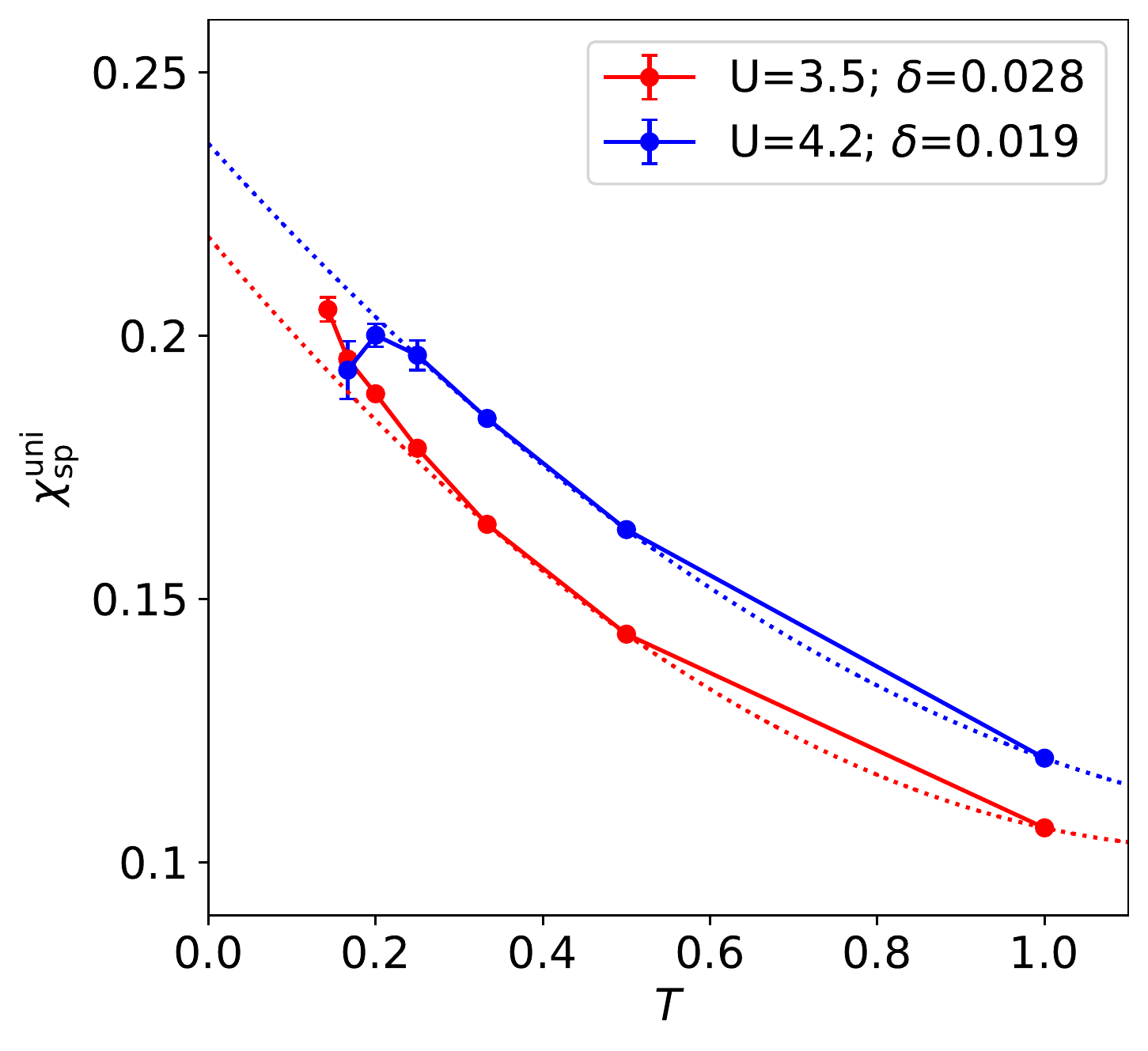}
\caption{The uniform spin susceptibility as a function of temperature. Dashed lines are second order polynomial fits from high temperatures, for an example from the weak-coupling (red) and weakly-coupled pseudogap (blue) regions.}
\label{fig_uniform}
\end{figure}

In Fig.~\ref{fig_dA_PG} we show the momentum-resolved spectral function $A(\mathbf{k})$ for two temperatures $T=\{0.15, 0.20\}$ as well as their relative difference $\Delta A(\mathbf{k}) \equiv A(\mathbf{k},T=0.20) - A(\mathbf{k},T=0.15)$. The first row is representative of the weakly correlated metallic regime and we see that the spectral weight grows everywhere along the Fermi surface as temperature is decreased. Momenta away from the Fermi surface show the opposite behaviour, but since their spectral weight is extremely low they are of little relevance. The second row shows a weak-coupling pseudogap regime and, in contrast to the previous case, the spectral weight has decreased for essentially all momenta. The third row  corresponds to a strongly correlated metal and the situation is similar to the weakly correlated metal. The final row is a strong-coupling pseudogap regime. Here we observe a momentum-dependent change from the strongly correlated metal to the strong-coupling pseudogap.

\begin{figure}[h!]
\centering
\includegraphics[width=0.8\textwidth]{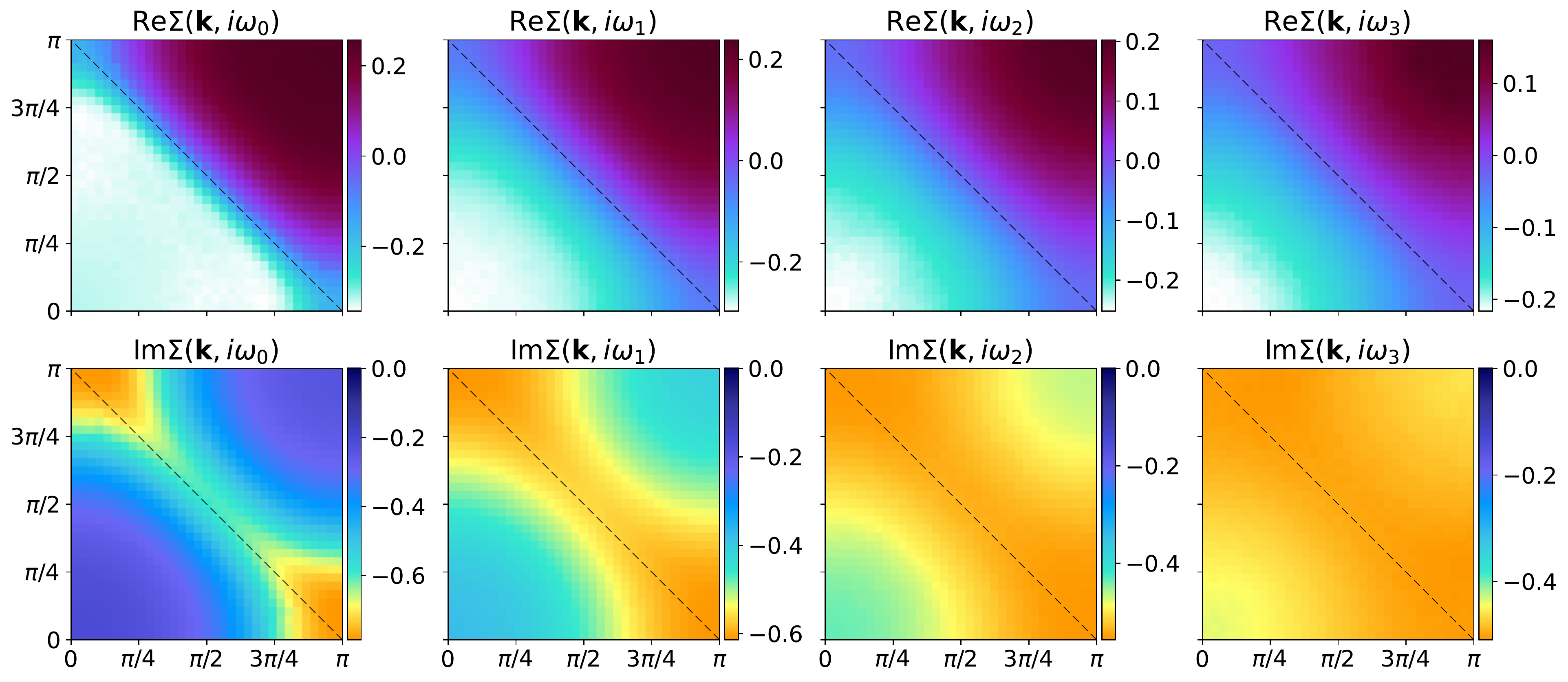}
\caption{The momentum resolved self-energy is shown for the four lowest frequencies for the $P_1$ point of Fig.\ref{fig_PD} in the main text ($U=4$, $T=0.2$ and $\delta=0.023$). }
\label{freq_P1}
\end{figure}

Similar to the self-energy, the temperature dependence of the spectral function first changes around the edge point and away from the Fermi surface. This change then propagates towards the Fermi Surface in particular around the antinodal region. This crossover is less natural to evaluate within diagrammatic Monte Carlo than the self-energy crossover as one needs to ideally compute many temperatures. We identify the crossover as the moment when the spectral function for the edge momentum (defined from the maximum of the spectral function along the line $(0,0)\to(0,\pi)\to(\pi,\pi)$) changes temperature dependence. We show a comparison between the two criteria for two temperatures ($T=\{0.1,0.2\}$) in Fig.~\ref{fig_PG_A_vs_Sigma}. Note that we used temperatures $T=\{0.2,0.15\}$ to establish the behaviour of $\Delta A (\mathbf{k})$ around $T=0.2$ and temperatures $T=\{0.125,0.1\}$ around $T=0.1$. We observe that the pseudogap regions are of similar shape between the two criteria, but our self-energy crossover preempts the spectral function crossover. This is in line with our previous statements about the self-energy crossover being an immediate precursor of the pseudogap regime.

In our self-energy criterion for the pseudogap crossover we look for the first momentum point on the Fermi surface, as defined by the spectral function, to manifest a change of slope. Since the spectral function maximum is relatively spread, especially around the antinode, this first momentum to change slope does not necessarily have to be $k=(\pi,0)$. To elucidate this effect we compare in the right panel of Fig.~\ref{fig_PG_A_vs_Sigma} our criterion with a modified version, which only takes into account the slope at the $k=(\pi,0)$ momentum. We see that whilst the two curves are practically identical for small-to-intermediate interactions $U\lesssim4$, they start to deviate thereafter. This does not come as a surprise, since the Fermi surface itself shifts away from the Fermi surface of the half-filled model and the $k=(\pi,0)$ momentum point is no longer on it. This justifies our choice of looking for any one momentum point to change slope as long as it has enough spectral weight.

Finally, another criterion is the drop of uniform spin susceptibility $\chi_{\text{sp}}^{\text{uni}}$ at low temperature. In Fig.~\ref{fig_uniform} we show $\chi_{\text{sp}}^{\text{uni}}$ for the weakly correlated metal ($U=3.5$, $\delta=0.028$) and weak-coupling pseudogap regime ($U=4.2$, $\delta=0.019$). While we observe a steady increase in the uniform spin susceptibility with decreasing temperature in the first case, we can clearly see a decrease around $T=0.2$ in the second case, clearly marking the onset of the pseudogap regime. In the strong-coupling pseudogap regime the error bars on our results do not allow us to clearly identify this downturn in the uniform spin susceptibility and we thus leave this task to future studies.

\begin{figure}[h!]
\centering
\includegraphics[width=0.8\textwidth]{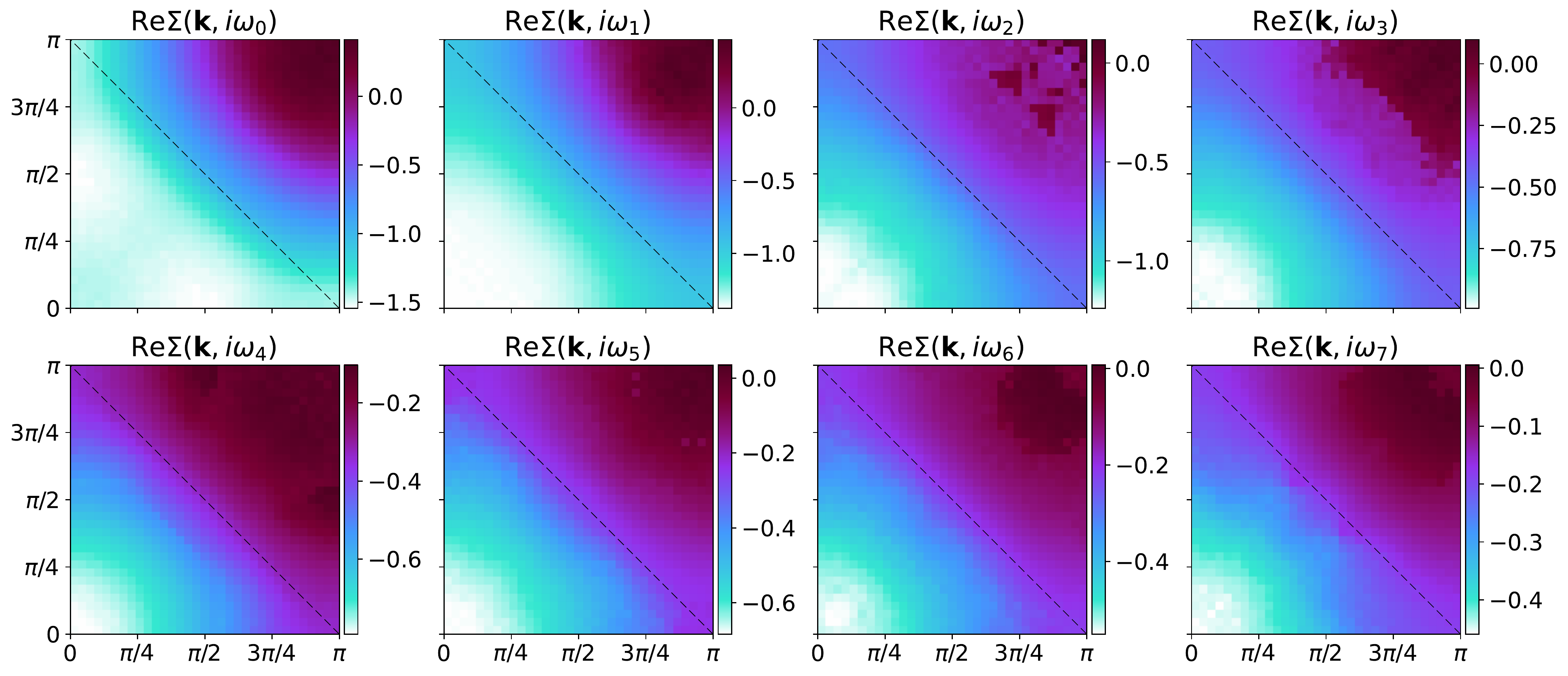}
\includegraphics[width=0.8\textwidth]{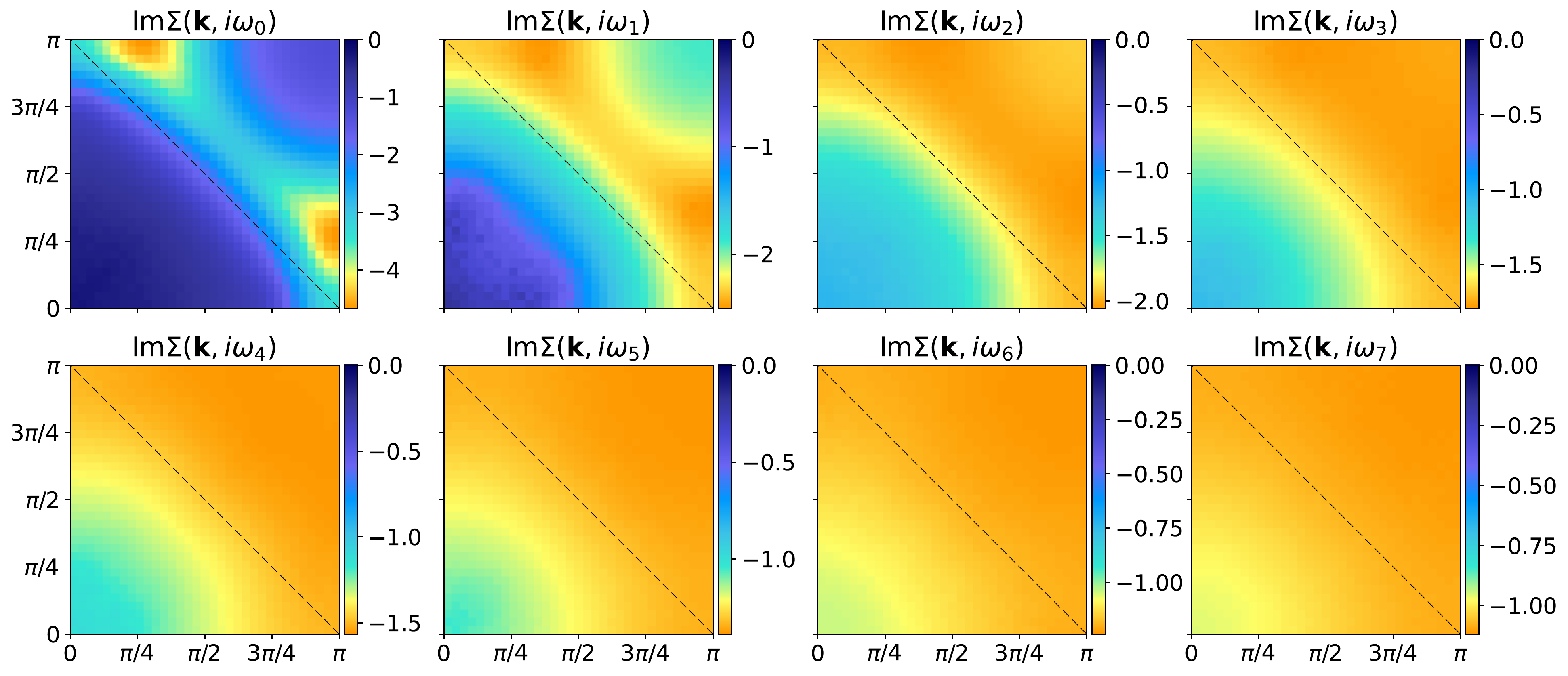}
\caption{The momentum resolved self-energy is shown for the eight lowest frequencies for the $P_2$ point of Fig.~\ref{fig_PD} in the main text ($U=7$, $T=0.2$ and $\delta=0.042$).}
\label{freq_P2}
\end{figure}

\section{Frequency dependence of the self-energy}

In Fig.~\ref{freq_P1} and Fig.~\ref{freq_P2} we study the momentum-resolved frequency dependence of the real and imaginary parts of the self-energy. In Fig.~\ref{freq_P1}, we investigate the weak-coupling pseudogap regime (P$_1$). We observe that the structure in momentum space rapidly becomes more uniform as the imaginary frequency increases. This is
especially true for the imaginary part and is an indication that the self-energy is more
local at higher frequencies. The picture is qualitatively the same in the strong-coupling pseudogap regime (P$_2$, Fig.~\ref{freq_P2}), but the self-energy becomes local at slightly
larger frequencies.

\begin{figure}[h!]
\centering
\includegraphics[width=0.8\textwidth]{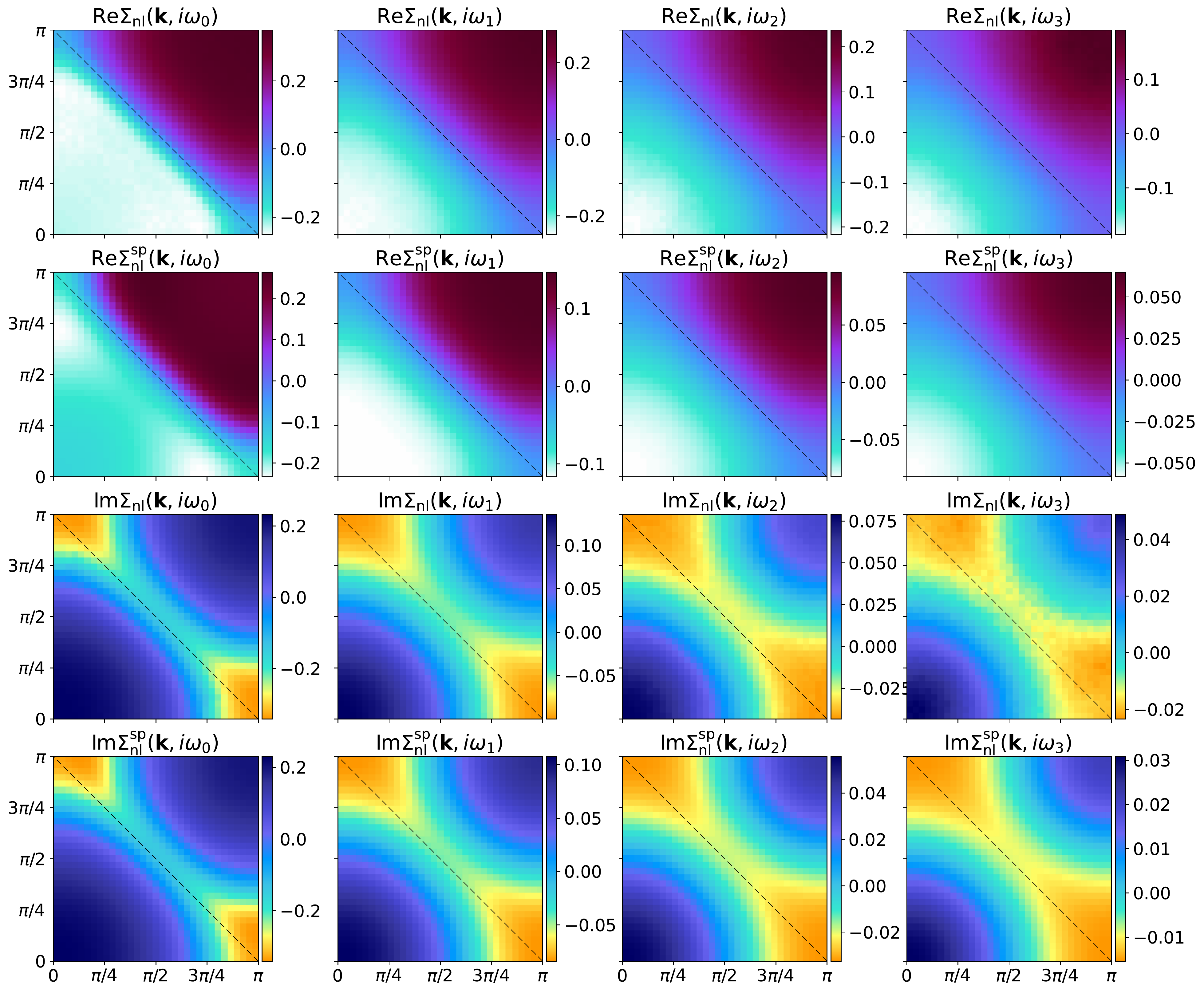}
\caption{The momentum resolved non-local components of the self-energy for the four lowest frequencies at the $P_1$ point of Fig.~\ref{fig_PD} in the main text ($U=4$, $T=0.2$ and $\delta=0.023$). Numerically exact results (first and third rows) are compared to a spin-fluctuation theory fitting procedure (second
and fourth rows), as described in the main text.}
\label{freq_P1_fit}
\end{figure}

\begin{figure}[h!]
\centering
\includegraphics[width=0.8\textwidth]{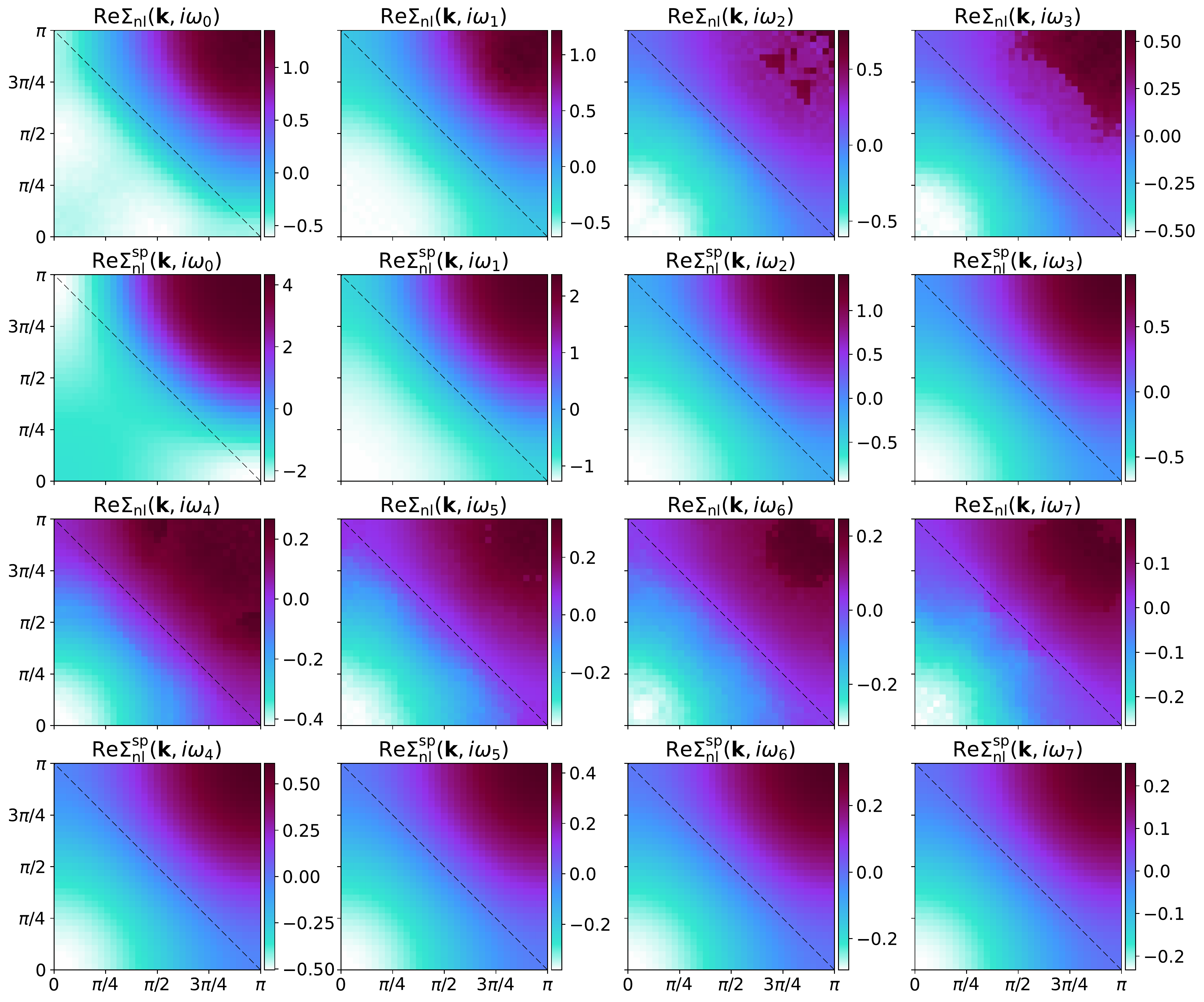}
\caption{The momentum resolved non-local component of the real self-energy for the eight lowest frequencies at the $P_2$ point of Fig.\ref{fig_PD} in the main text ($U=7$, $T=0.2$ and $\delta=0.042$). Numerically exact results (first and third rows) are compared to a spin-fluctuation theory fitting procedure (second
and fourth rows), as described in the main text.}
\label{freq_P2_fit_Re}
\end{figure}

\begin{figure}[h!]
\centering
\includegraphics[width=0.8\textwidth]{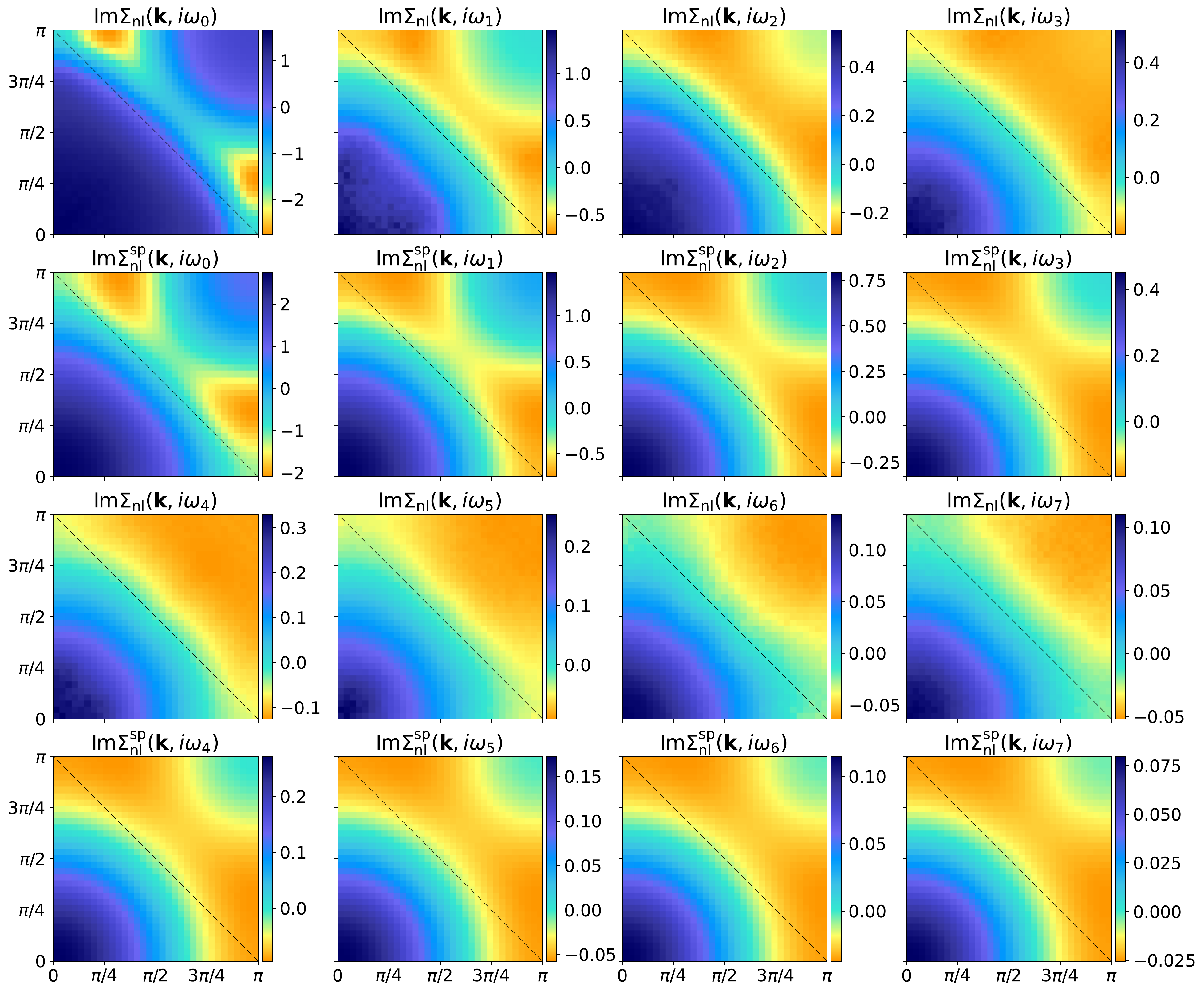}
\caption{The momentum resolved non-local component of the imaginary self-energy for the eight lowest frequencies at the $P_2$ point of Fig.~\ref{fig_PD} in the main text ($U=7$, $T=0.2$ and $\delta=0.042$). Numerically exact results (first and third rows) are compared to a spin-fluctuation theory fitting procedure (second
and fourth rows), as described in the main text.}
\label{freq_P2_fit_Im}
\end{figure}

In Fig.~\ref{freq_P1_fit}, \ref{freq_P2_fit_Re} and \ref{freq_P2_fit_Im} we provide a comparison between numerically exact data and the spin-fluctuation theory fitting procedure, as described in the main text. In should be noted that, as in the main text, the fitting has only been done on the imaginary part of the self-energy for the lowest Matsubara frequency ($i \omega_0$). 

In the case of the weak-coupling pseudogap regime ($P_1$) we observe in \ref{freq_P1_fit} a near-perfect match at all shown frequencies when it comes to momentum-dependency, with only slightly lower absolute self-energy values in the fitted data. 

From Fig.~\ref{freq_P2_fit_Re} and \ref{freq_P2_fit_Im} we also observe a very good correspondence between the exact data and the theoretical fit for the strong-coupling pseudogap regime ($P_2$), albeit slightly worse than in the case of $P_1$, especially when it comes to the imaginary part of the self-energy. All in all, we find that our theoretical fit is performing remarkably well for higher frequencies in both pseudogap regimes studied here. 

\section{Extrapolations to the ground-state}

The zero-temperature curves for the pseudogap region in Fig.~\ref{fig_extr} were obtained from an extrapolation of our finite-temperature data. To this end, we have split our data into two sectors in which the extrapolation was performed with respect to different variables. For $U\lesssim4$ and $\delta\lesssim0.1$ we extrapolated at constant doping $\delta$ and with respect to the interaction $U$. For $U\gtrsim4$ and $\delta\gtrsim.1$ we set $U$ constant and extrapolated with respect to $\delta$. Both extrapolated curves remarkably coincide within error bars at their boundary. For the Lifshitz crossover we only extrapolated at with respect to $\delta$ and at constant $U$.

\begin{figure}[h!]
\centering
\includegraphics[width=0.4\textwidth]{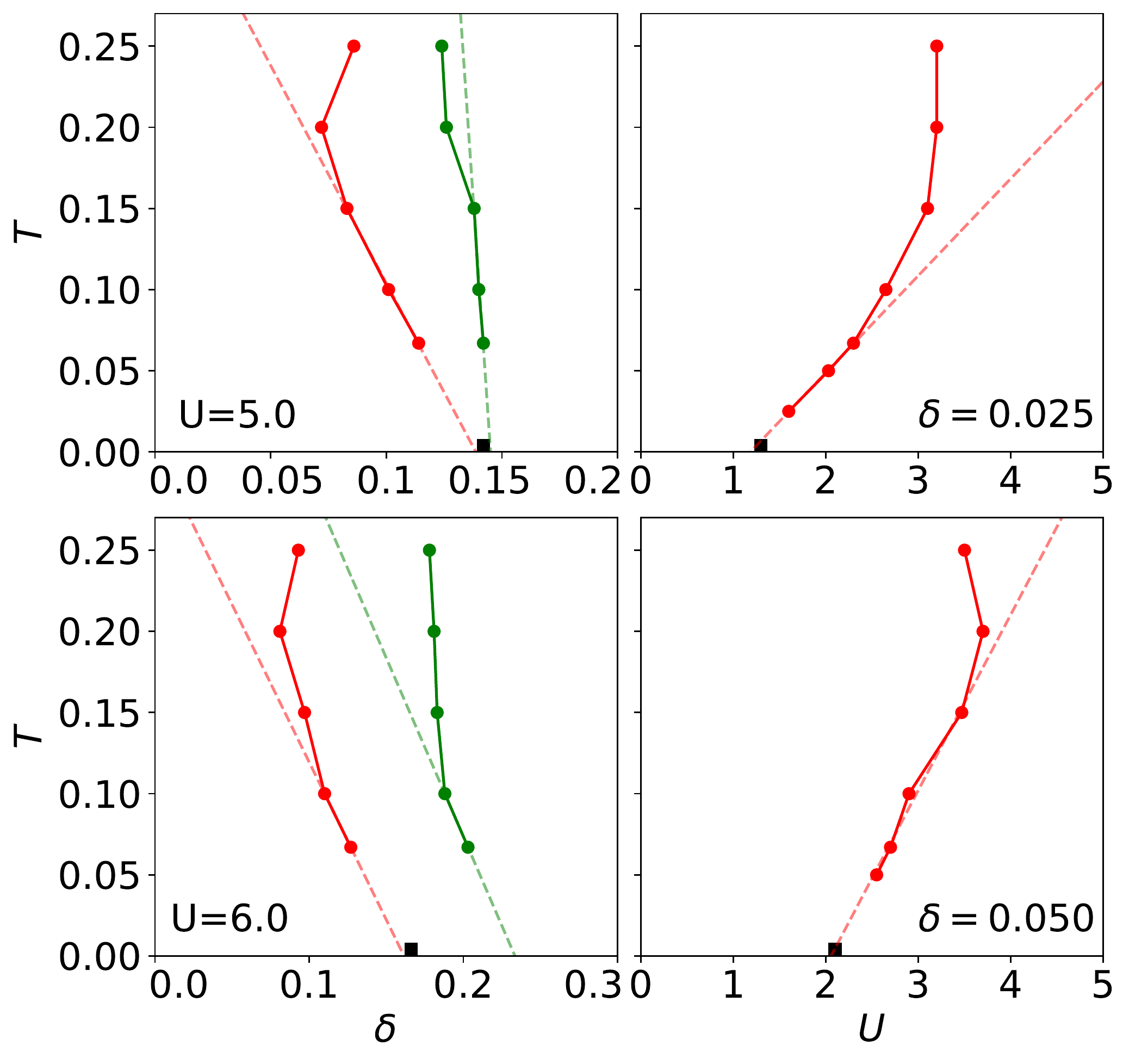}
\caption{The temperature dependence of the pseudogap crossover (red circles) and Lifshitz transition (green circles) are shown for either fixed interaction $U$ (left) or fixed doping $\delta$ (right). AFQMC ground-state results from Ref.~\cite{xu2021stripes} are shown as black squares. Dashed lines correspond to linear extrapolations from the two lowest available temperatures.}
\label{fig_extr_PG}
\end{figure}

In Fig.~\ref{fig_extr_PG} we show the temperature dependence of the pseudogap crossover and Lifshitz crossover for two fixed interaction values $U=\{5,6\}$ (left) and fixed doping values $\delta=\{0.025, 0.05\}$ (right). Circles correspond to our best estimates for the location of the crossover into the pseudogap region (without error bars). AFQMC ground-state results from Ref.~\cite{xu2021stripes} are shown as black squares and dashed lines correspond to linear extrapolations from the two lowest available temperature data points, with relative uncertainties of $\Delta U = 0.05$ and $\Delta \delta =0.002$. For the pseudogap, we see that the extrapolated zero-temperature values match the AFQMC ground-state results extremely well for both the constant doping and interaction examples. Our data for the Lifshitz crossover extrapolates to the AFQMC data point for $U=5$, but starts to deviate at $U=6$ (this effect becomes even more amplified as $U$ is increased), thus indicating a separation of the two crossovers beyond a certain critical value of interaction, as shown in Fig.~\ref{fig_extr_PG}.

\section{Additional insights from the self-energy and spectral function}

\begin{figure}[h!]
\centering
\includegraphics[width=0.23\textwidth]{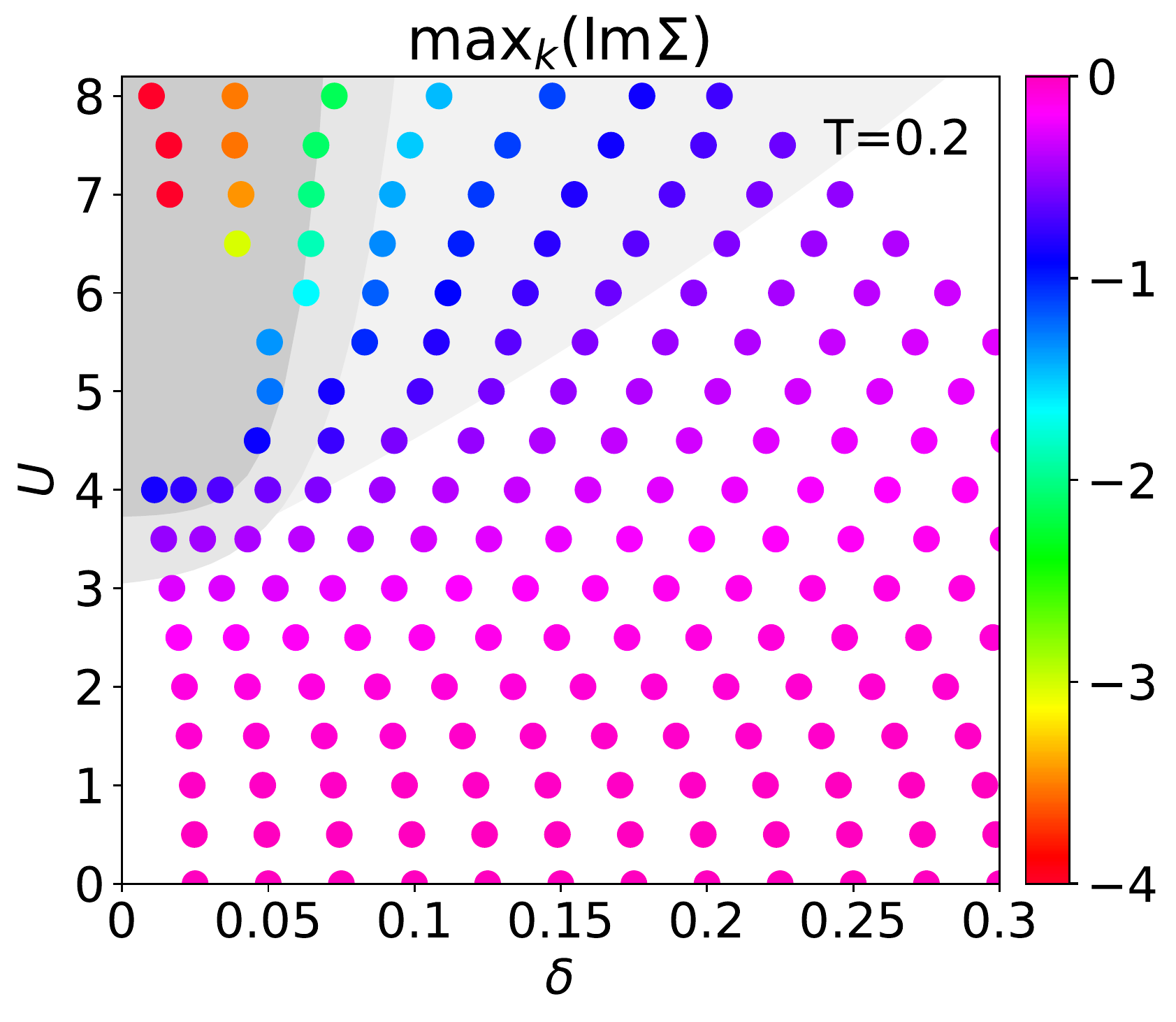}
\includegraphics[width=0.23\textwidth]{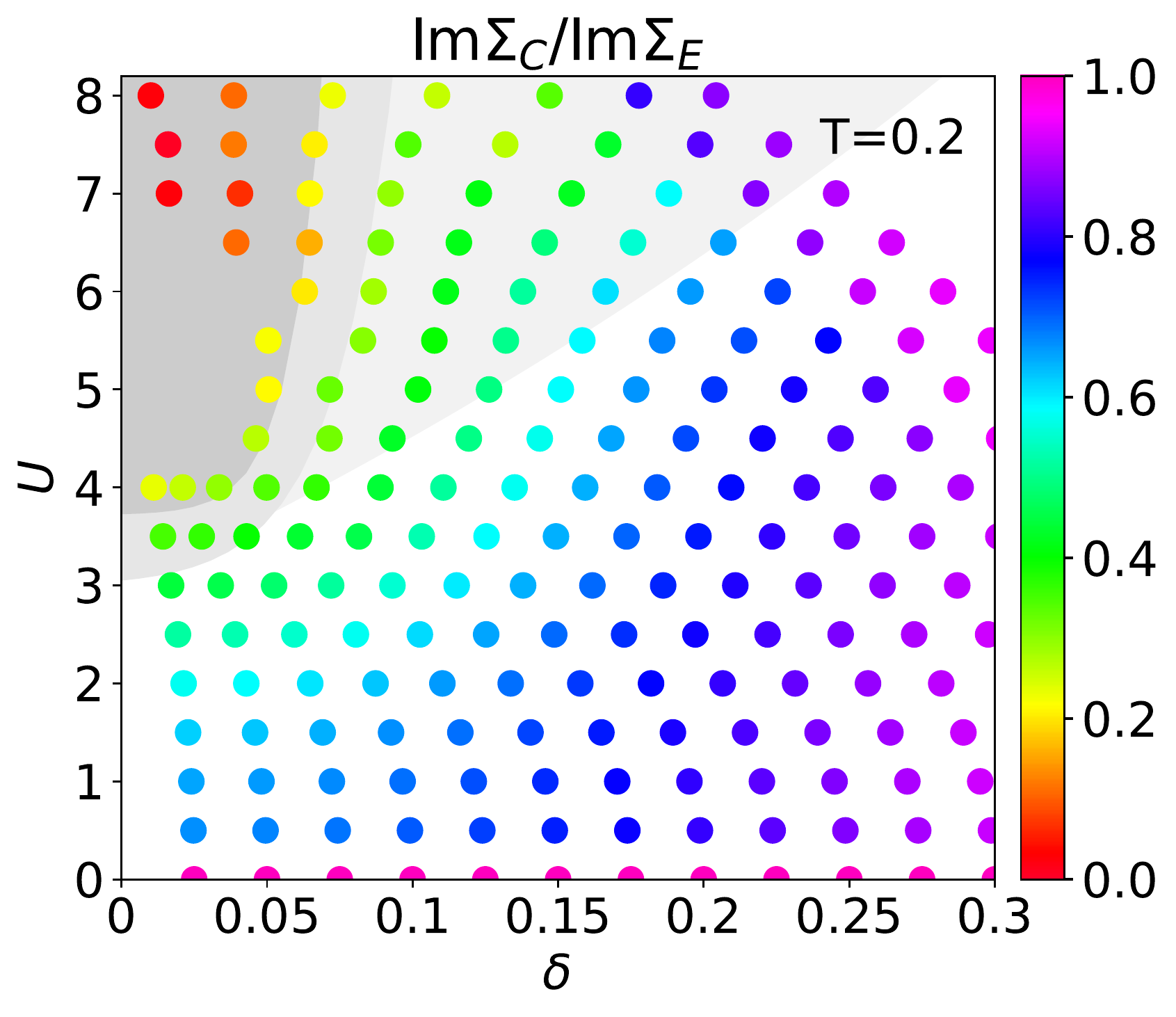}
\includegraphics[width=0.24\textwidth]{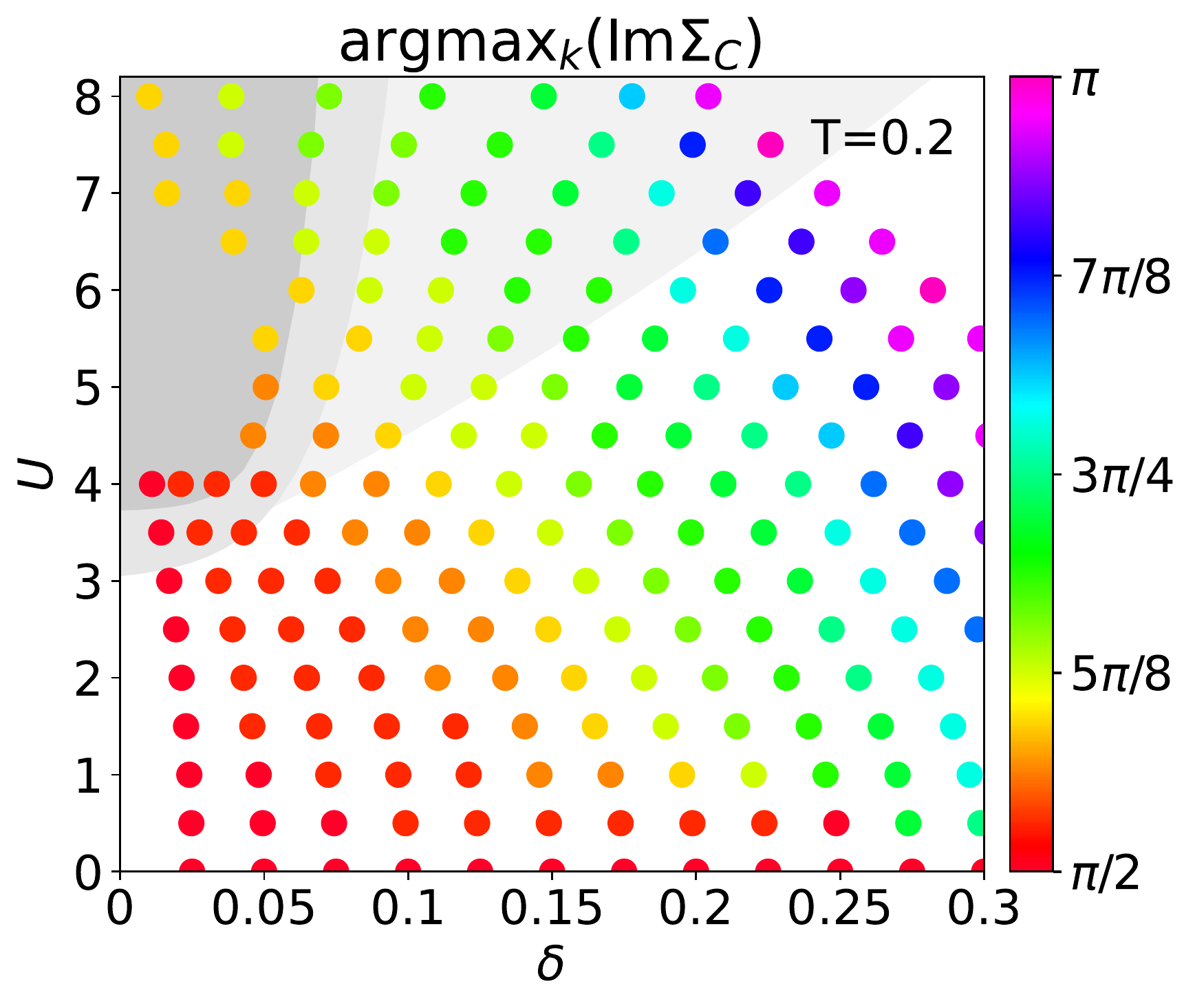}
\includegraphics[width=0.24\textwidth]{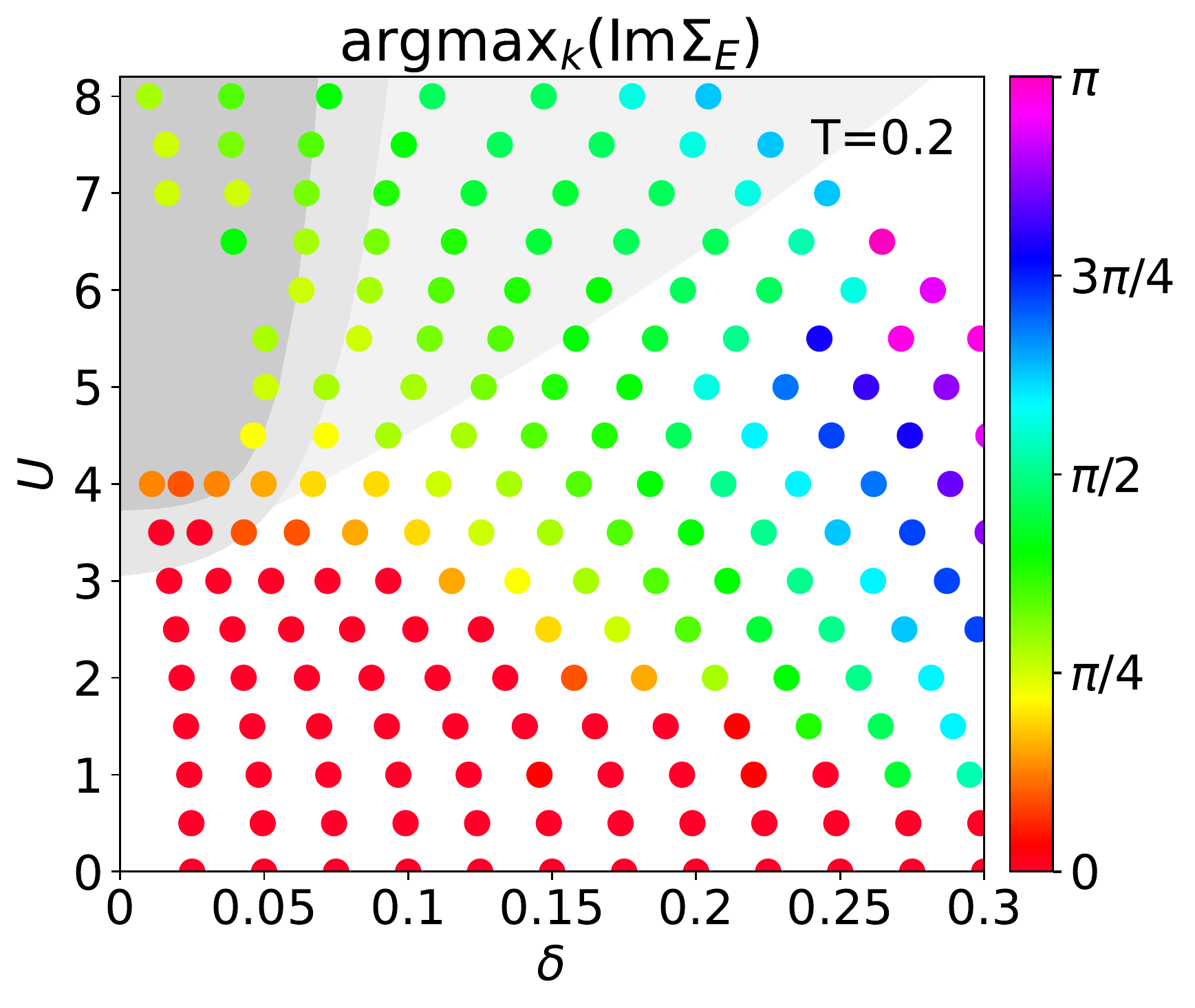}
\caption{Crossover phase diagrams at $T=0.2$. Left to right: The maximum value of the imaginary self-energy in the Brillouin zone; The ratio between the values of imaginary self-energy at the center(C) and edge(E) momenta; The position of the maximum in the imaginary self-energy along the momentum line $(0,0)\to(\pi,\pi)$ corresponding to the center point (C). The maximum along the line $(0,0)\to(0,\pi)\to(\pi,\pi)$ corresponding to the edge point (E). Each circle is a data point. Different shading corresponds to the distinct regions in Fig.\ref{fig_PD}}
\label{fig_Sigma}
\end{figure}

\begin{figure}[h!]
\centering
\includegraphics[width=0.23\textwidth]{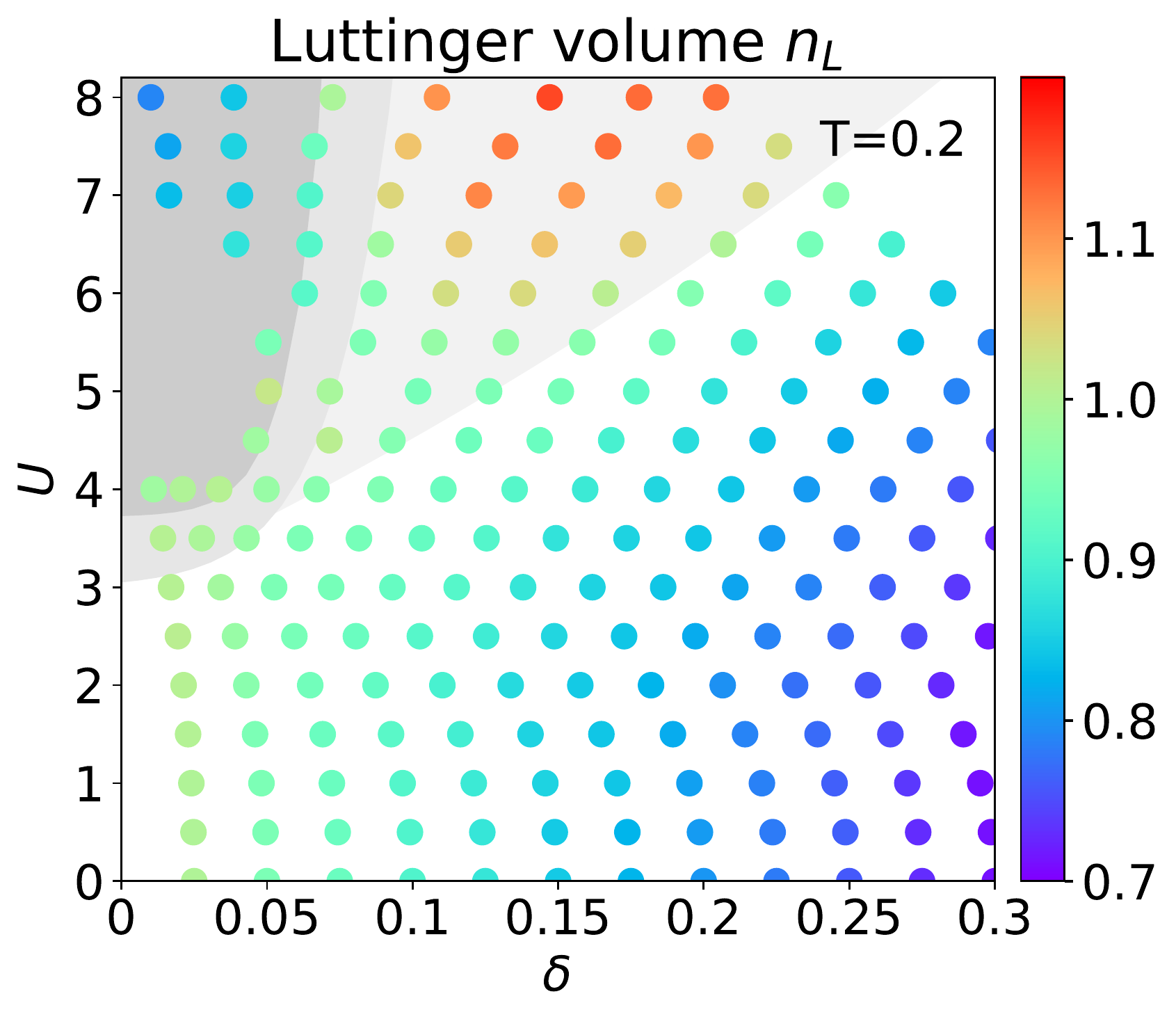}
\includegraphics[width=0.24\textwidth]{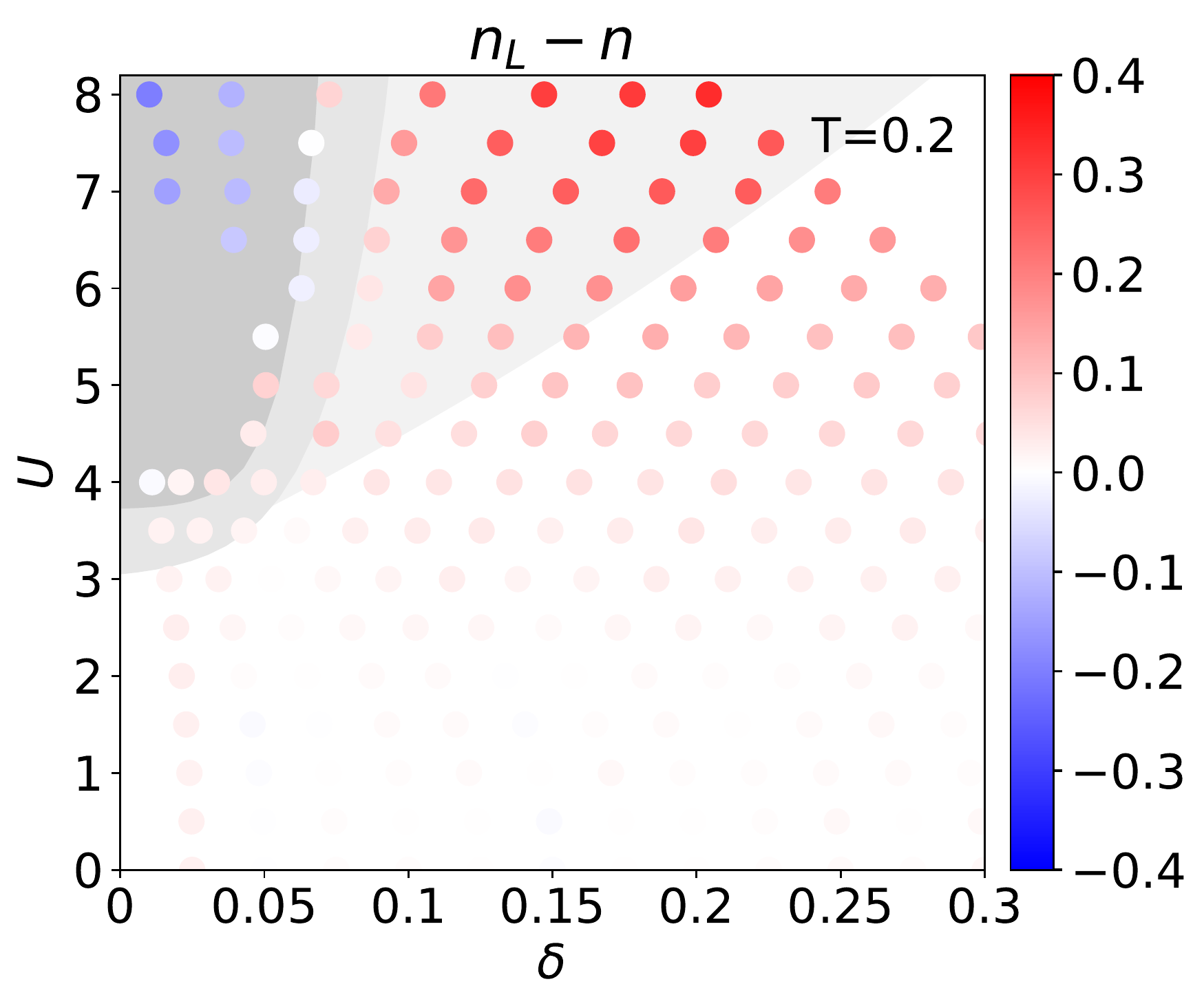}
\includegraphics[width=0.23\textwidth]{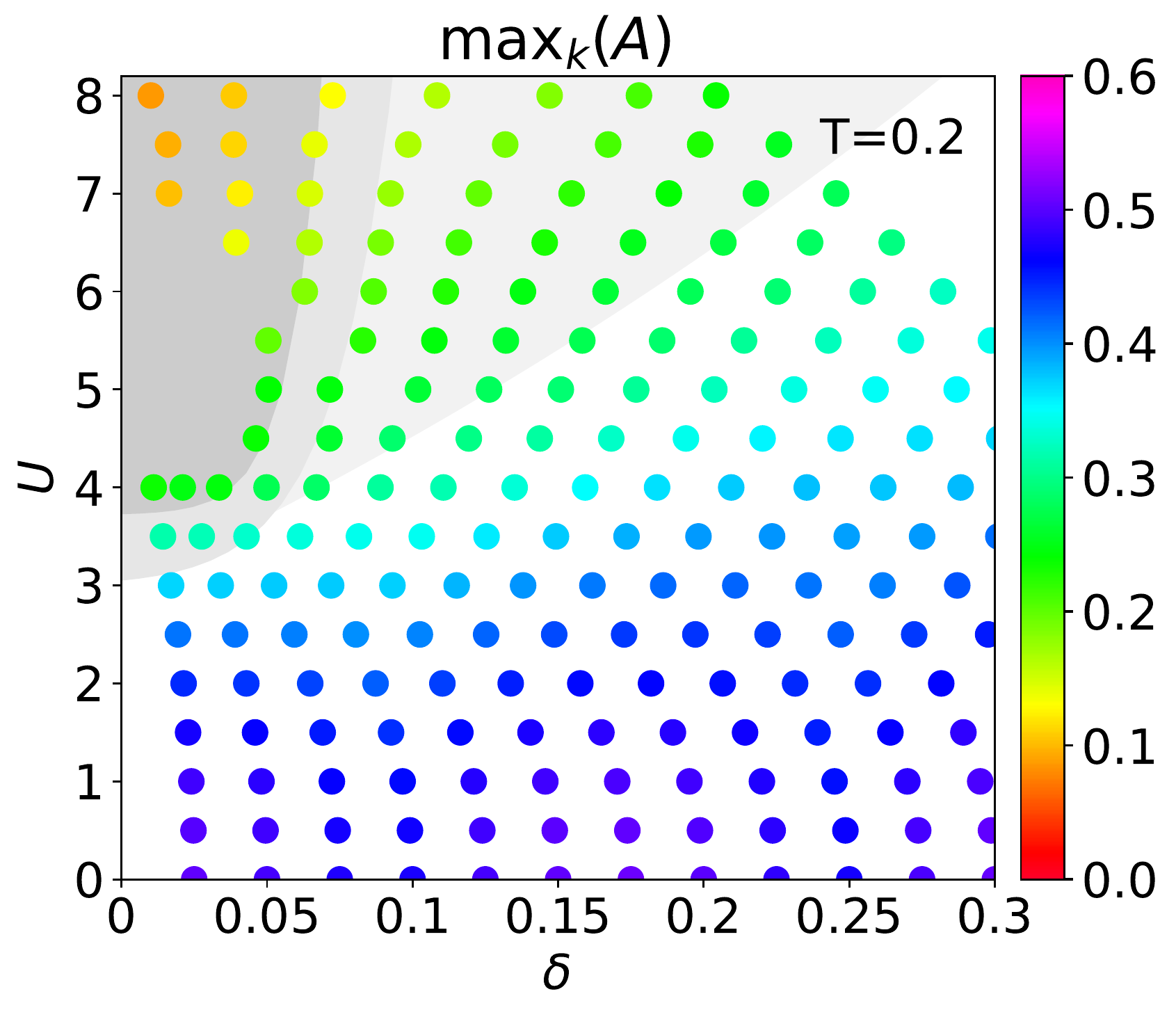}
\includegraphics[width=0.24\textwidth]{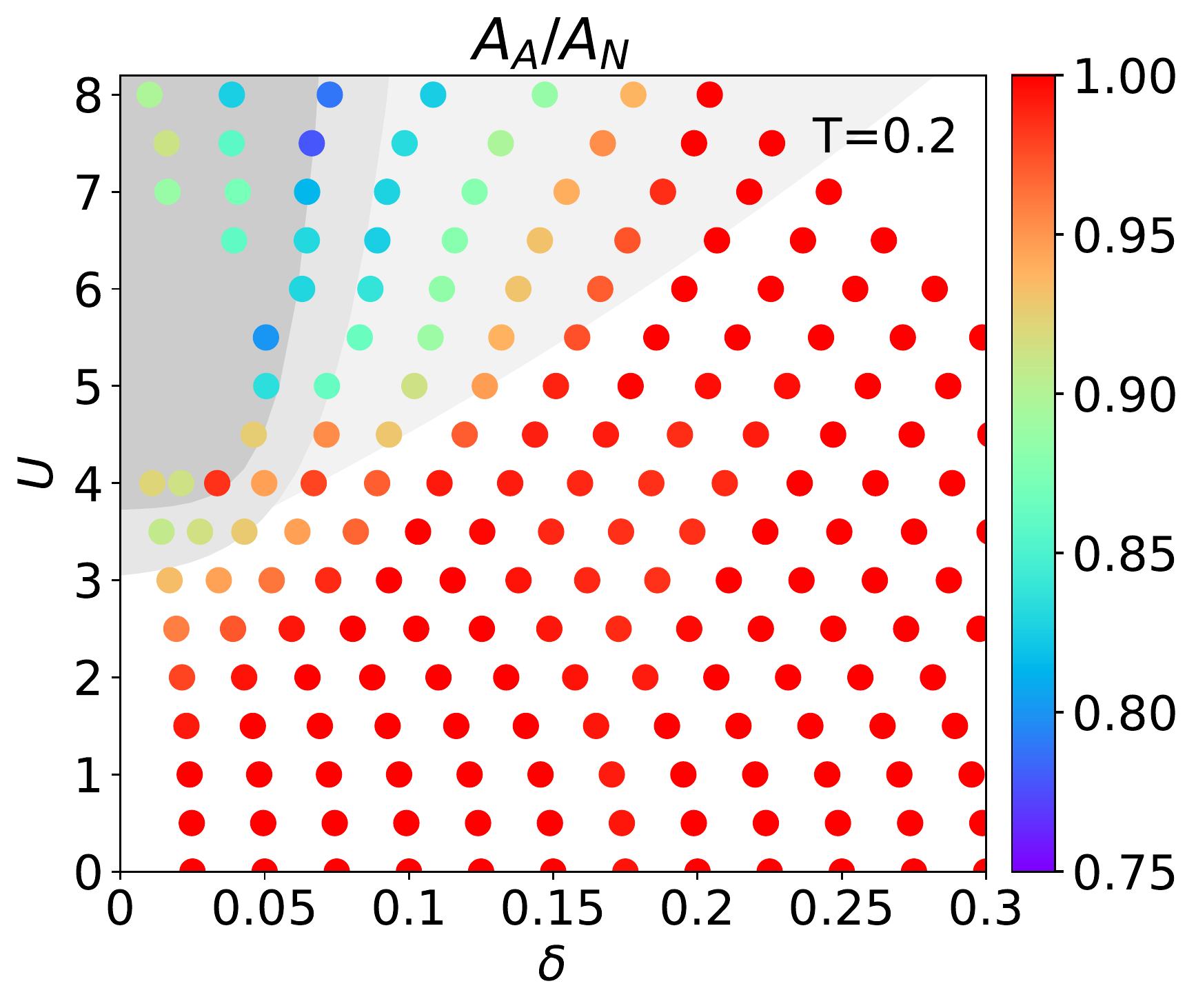}
\caption{Crossover phase diagrams at $T=0.2$. From left to right: The Luttinger volume $n_{\text{L}}$; The difference between $n_L$ and the density $n$; The maximum value of the spectral function in the Brillouin zone; The ratio between the values of the spectral function between the node (N) and the antinode (A). Each circle is a data point. Different shading corresponds to the distinct regions in Fig.\ref{fig_PD}}
\label{fig_A}
\end{figure}

In this section, we study the self-energy (in Fig.~\ref{fig_Sigma}) and spectral function (in Fig.~\ref{fig_A}) at constant temperature $T=0.2$ and across the $U$-$\delta$ crossover \emph{phase diagram}.

From the first panel (from the left) of Fig.~\ref{fig_Sigma} we deduce that the imaginary part of the self-energy is small ($<1$) throughout the weakly correlated metal region. The magnitude of the self-energy then grows with increasing interaction and decreasing doping. As it reaches values of about $\sim1$ we observe a crossover into either the strongly correlated metal as well as the weak-coupling pseudogap and continues to increase slowly thereafter. Only after we reach the strongly correlated pseudogap does the magnitude start to grow rapidly and reaches very large values of $\sim 4$. In the second panel we show the ratio between the self-energy at the center and edge momenta. In the weakly correlated metal the two are comparable and the magnitude of the edge is only slightly higher. In contrast, for the other three regions the difference is already significant and the ratio approaches zero deep inside the strong-coupling pseudogap. The positions of the maxima for the center (along the momentum line $(0,0)\to(\pi,\pi)$) and edge (along the line $(0,0)\to(0,\pi)\to(\pi,\pi)$) points also changes significantly by moving away from the Fermi surface of the half-filled model upon doping and upon an increase in interaction strength (two rightmost panels). Neither of the two, however, seems to follow the crossover lines. 

In Fig.~\ref{fig_A} we concentrate on the properties of the spectral function as a function of interaction and doping. In the first two panels we show the Luttinger volume $n_{\text{L}}$ (the area defined by the Fermi surface) and its difference to the density $n$. In the weakly correlated metal and weak-coupling pseudogap the Luttinger volume roughly follows the density. For both quantities we observe a maximum in the strongly correlated metal regime, which is also the only regime where $n_{\text{L}}>1$. In contrast, in the strong-coupling pseudogap the Luttinger volume sharply decreases and eventually becomes lower than the density. In the third panel we investigate the maximum of the spectral function over the Brillouin zone, which decreases steadily when either interaction is increased or the doping decreased. Finally, in the last panel we show the ratio between the spectral weight at the antinode and at the nodal momenta. In the weakly correlated metal and weak-coupling pseudogap they are essentially equal, with a ratio of $A_{\text{A}}/A_{\text{N}} \sim 1$. As one approaches the strong-coupling pseudogap, however, the ratio slowly decreases to about $\sim 0.8$. All of the above observations can be used as additional criteria in distinguishing between the weakly correlated metal and weak-coupling pseudogap and their respective strong-coupling counterparts.

\section{Comparison with dynamical mean-field theory}

\begin{figure}[h!]
\centering
\includegraphics[width=0.24\textwidth]{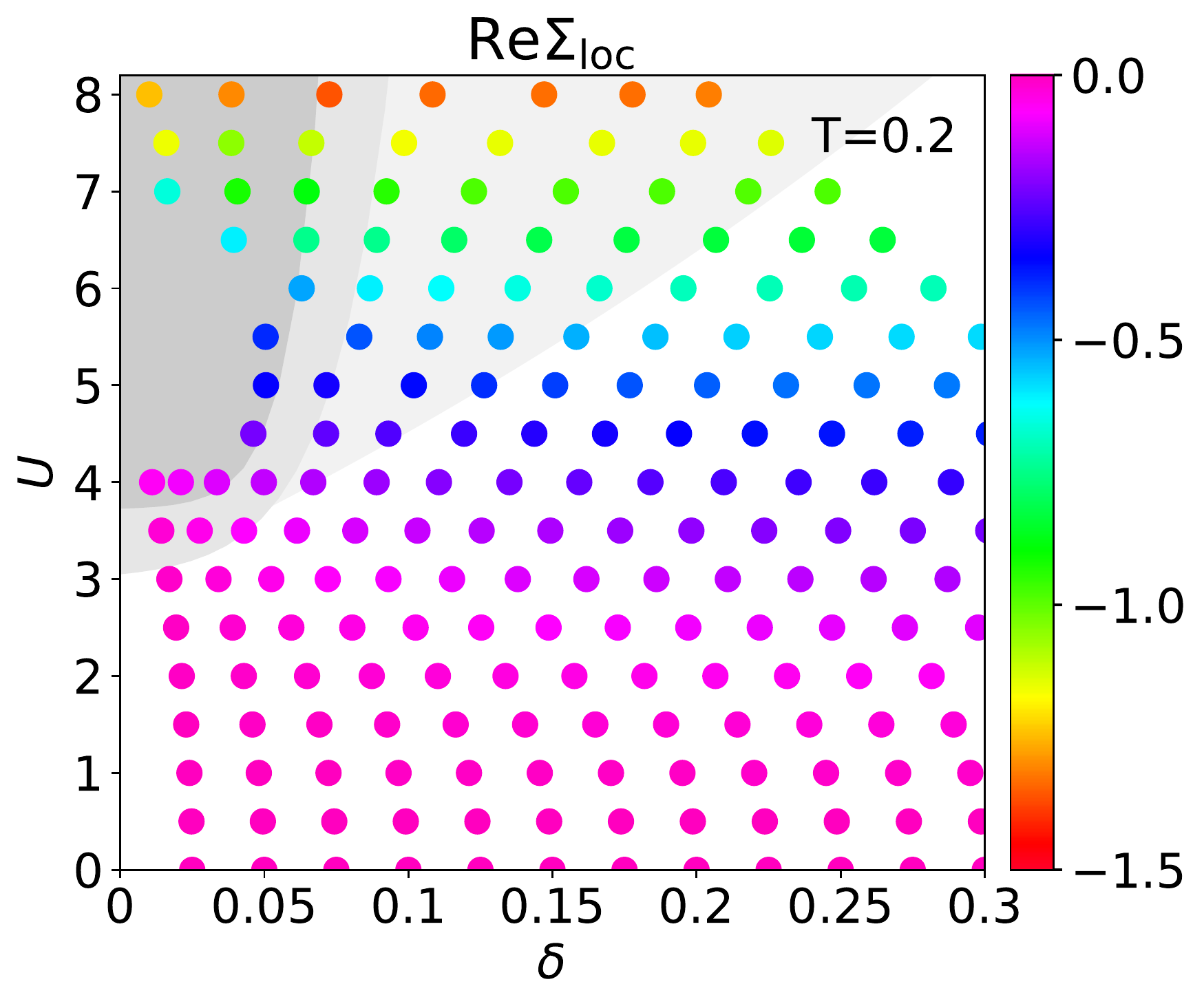}
\includegraphics[width=0.24\textwidth]{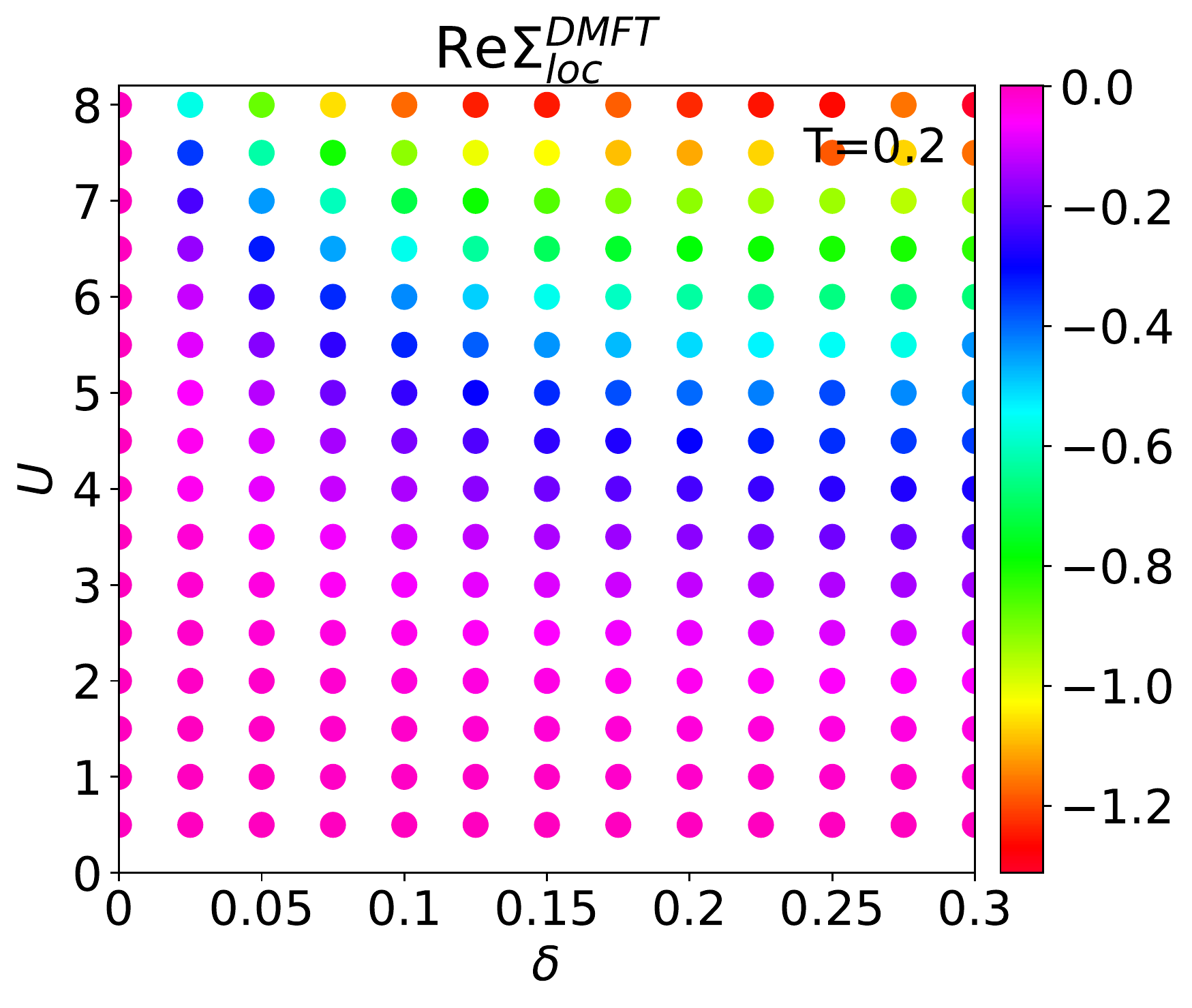}
\includegraphics[width=0.24\textwidth]{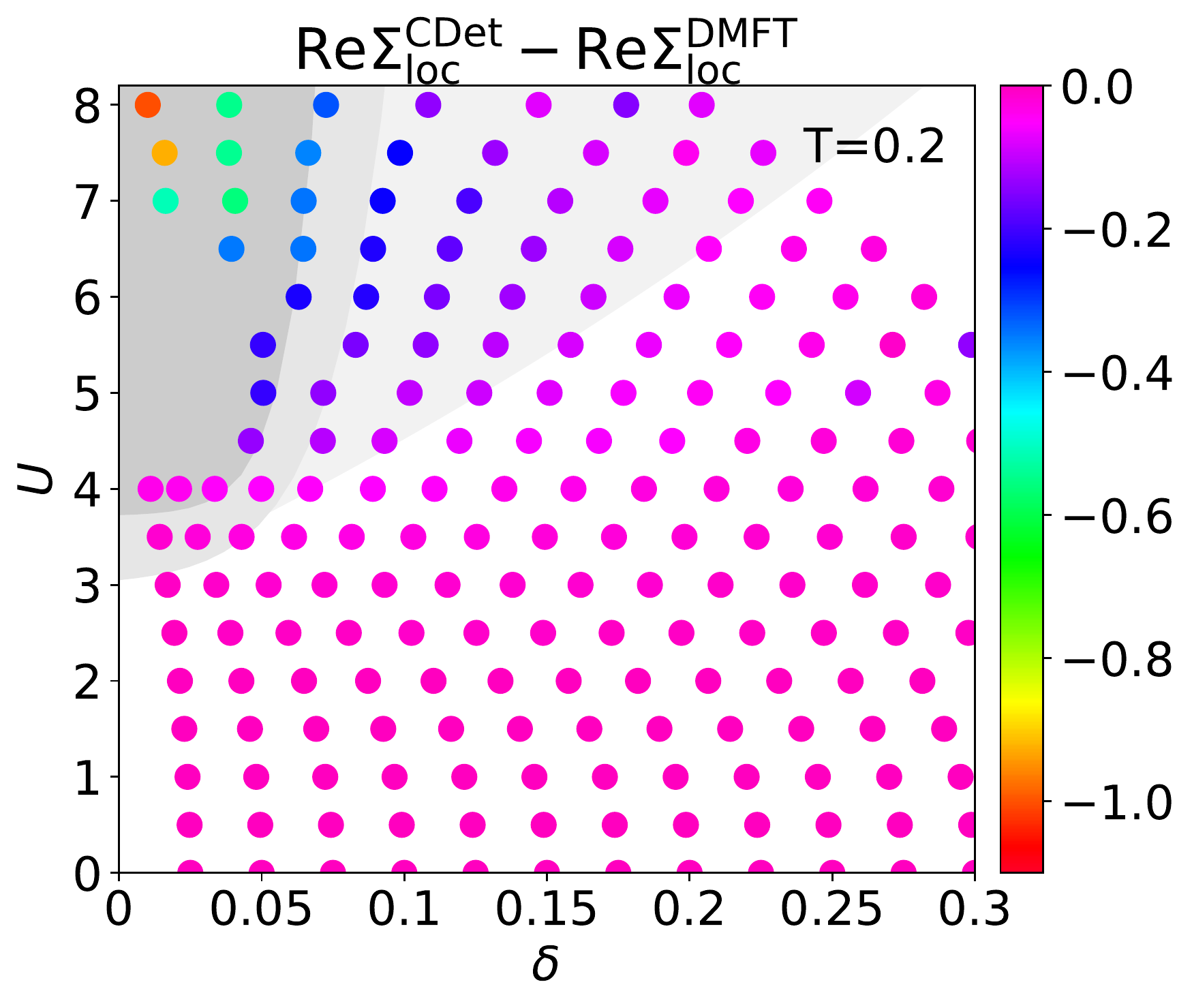}
\includegraphics[width=0.23\textwidth]{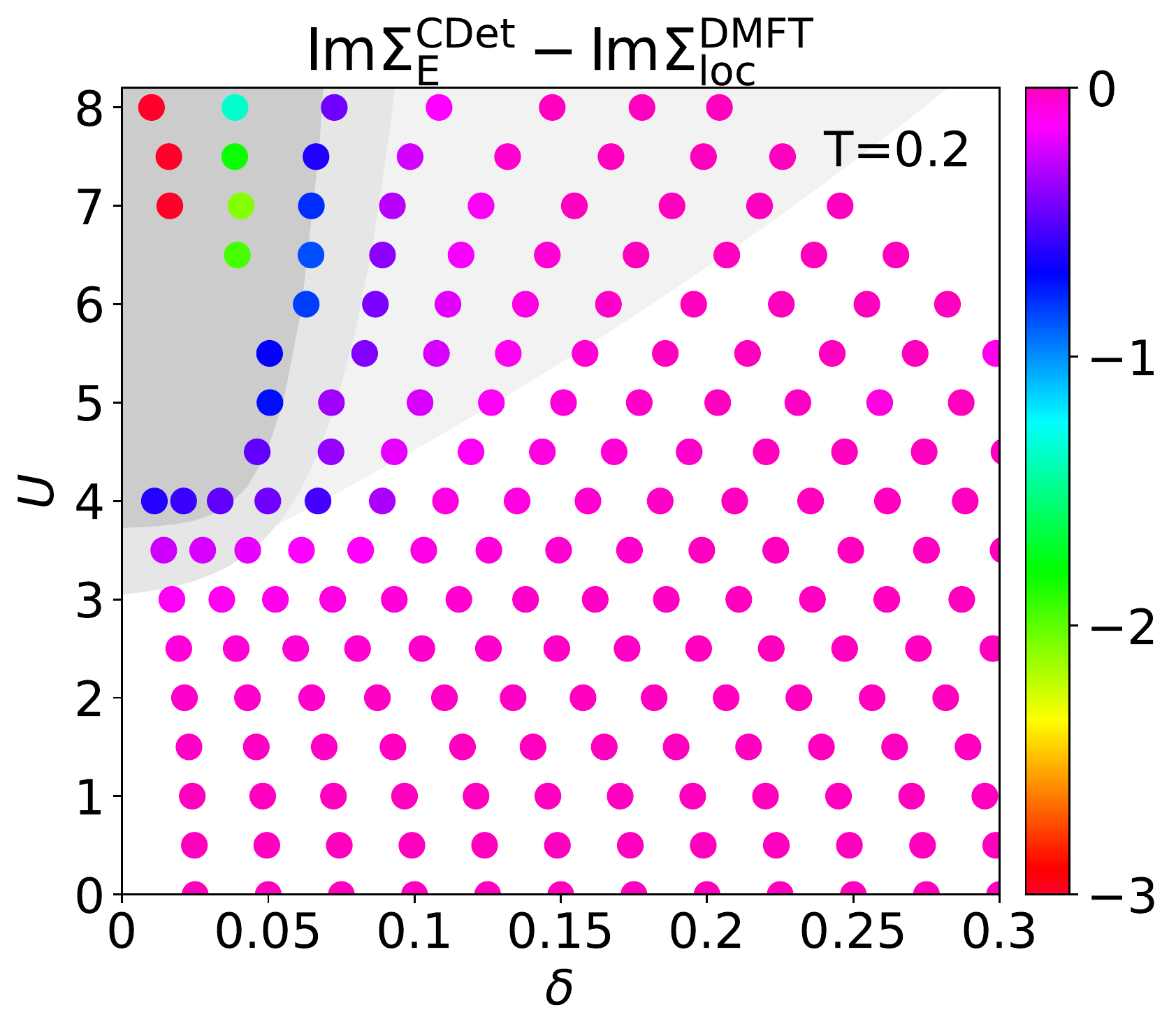}
\includegraphics[width=0.23\textwidth]{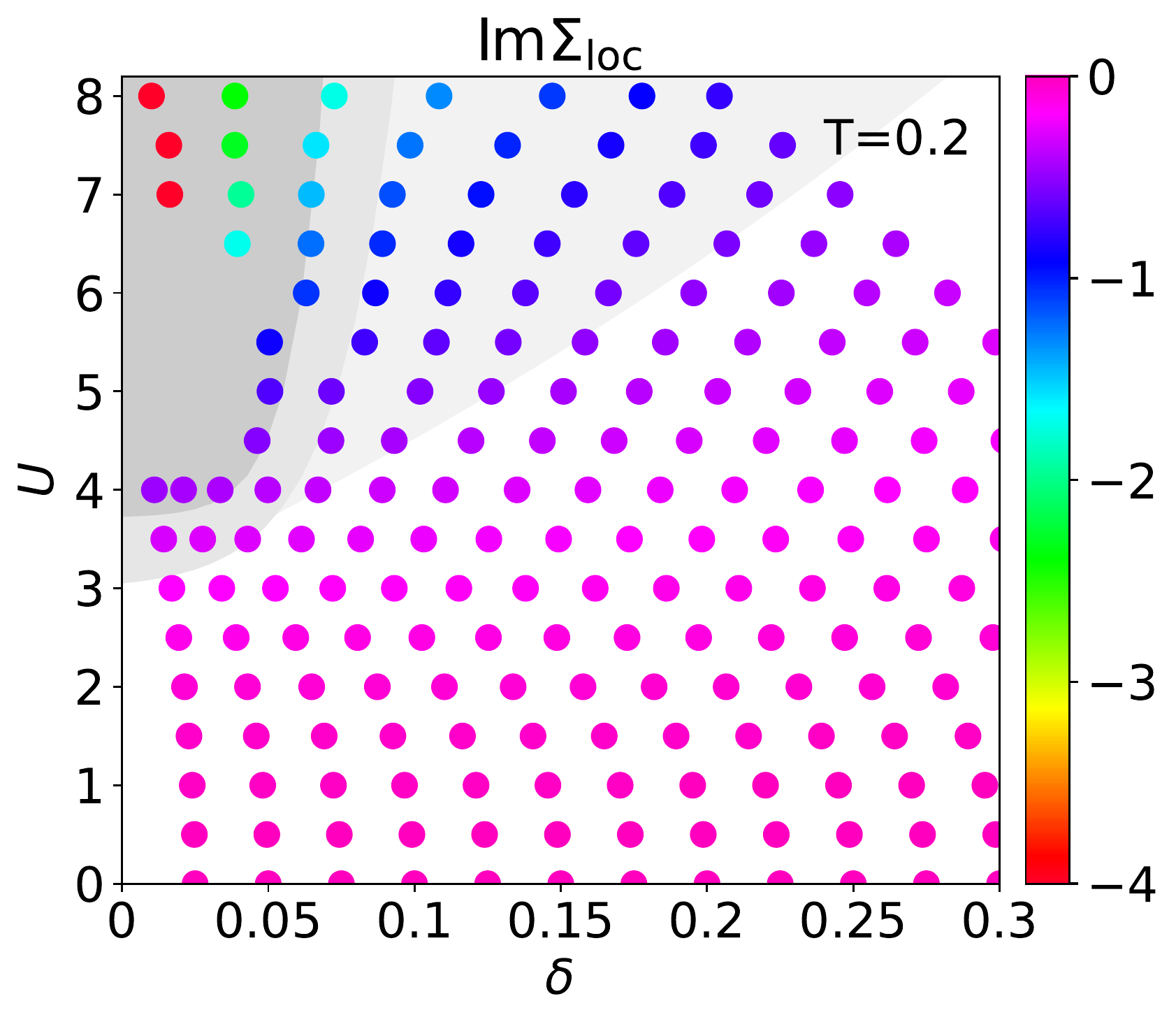}
\includegraphics[width=0.24\textwidth]{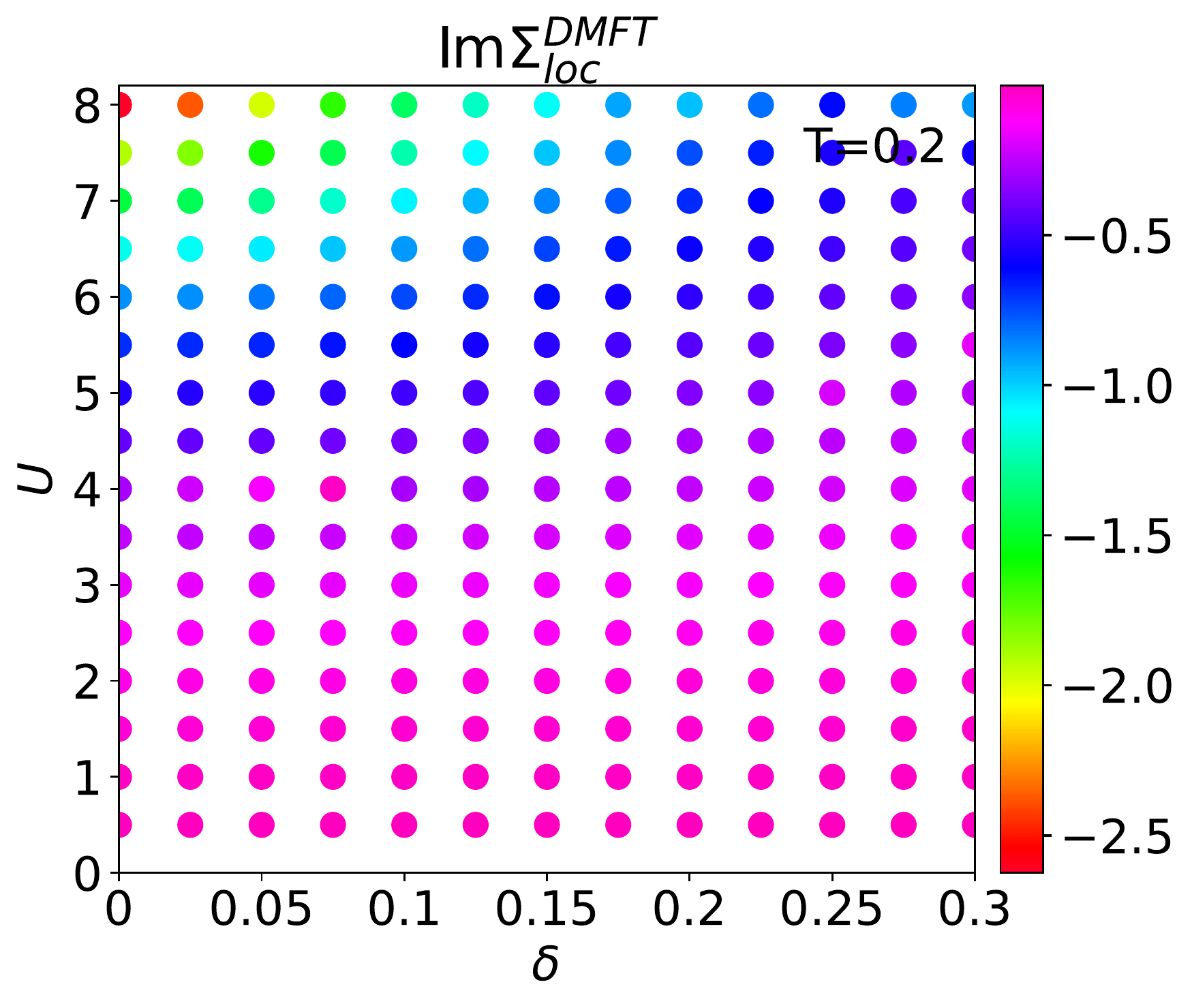}
\includegraphics[width=0.23\textwidth]{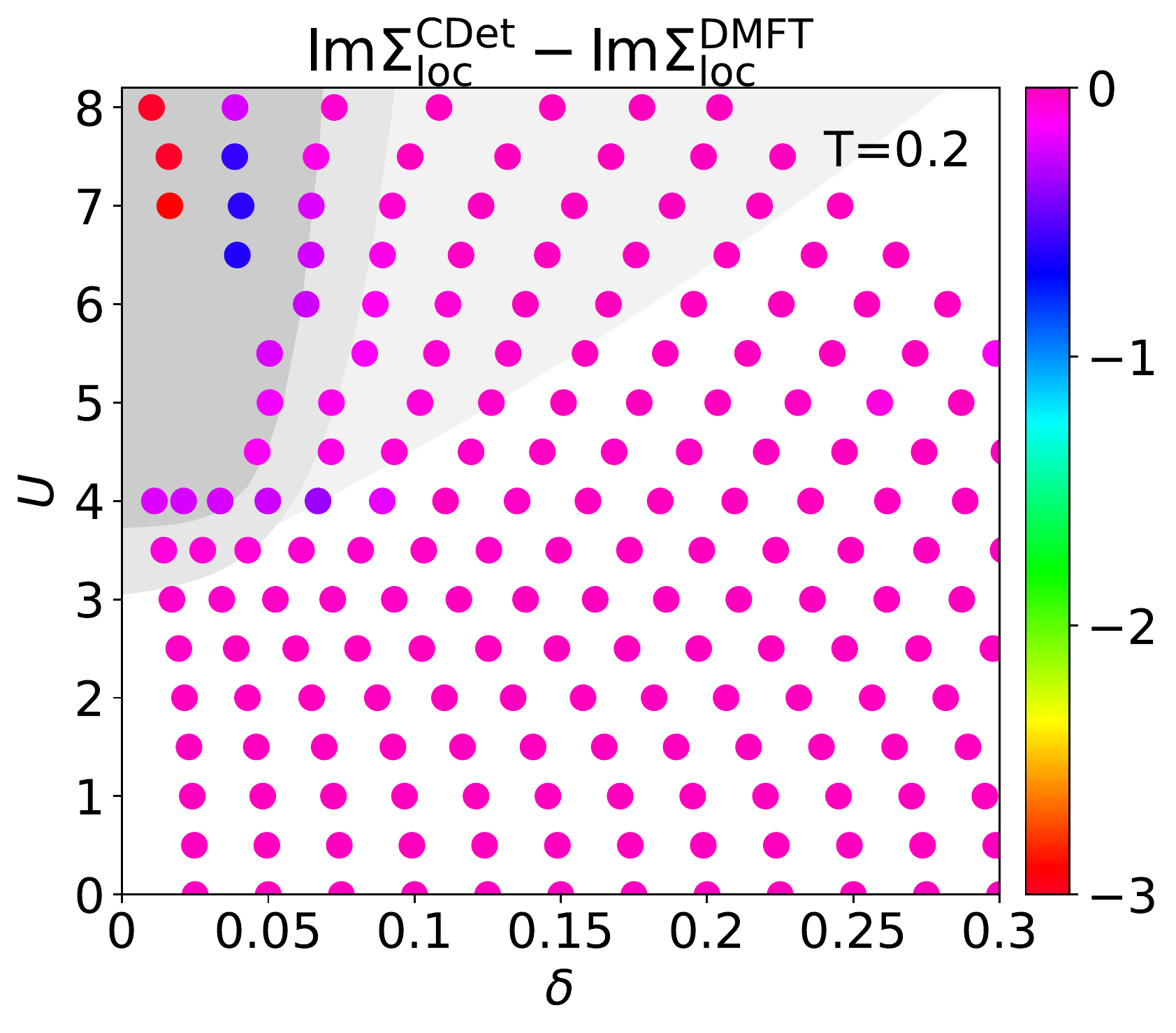}
\includegraphics[width=0.23\textwidth]{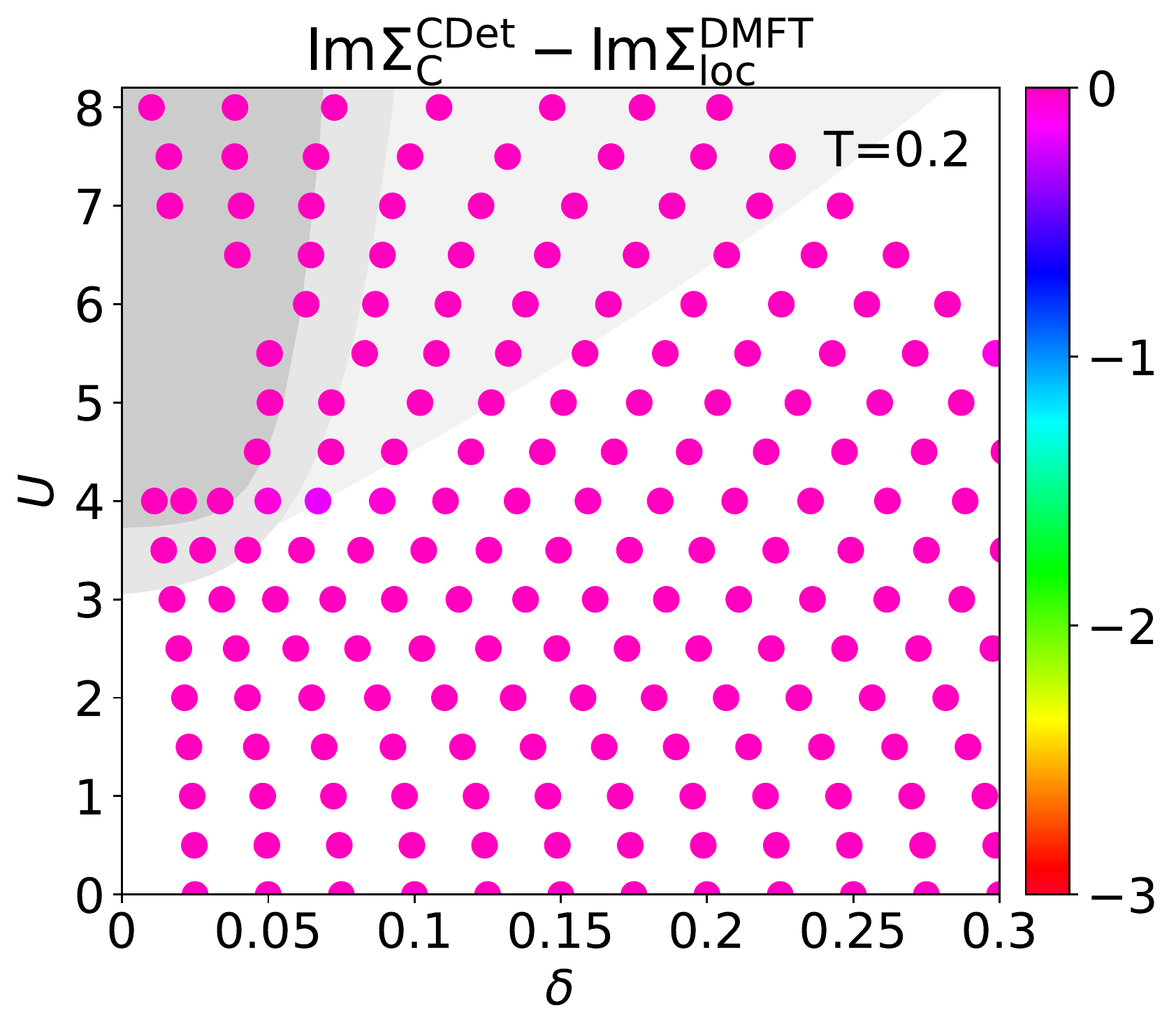}
\caption{Comparison between DMFT and CDet in terms of the real and imaginary parts of the local self-energy for $T=0.2$. Each circle is a data point. Different shading corresponds to the distinct regions in Fig.~\ref{fig_PD}.}
\label{fig_PD_DMFT}
\end{figure}

In this section, we compare the approximate result for the local self-energy obtained from the dynamical mean-field theory (DMFT), $\Sigma_{\text{loc}}^{\text{DMFT}}$, with the controlled result from diagrammatic Monte Carlo (CDet), $\Sigma_{\text{loc}}$. Our findings are summarized in the first three columns of Fig.~\ref{fig_PD_DMFT}, where results for the real part of the local self-energy in top row and the imaginary part in the bottom row. We find that in the weakly correlated metal and pseudogap regimes both local self-energies are essentially identical. In the strongly correlated metal this is still true for the imaginary part, however, the real parts start to differ. More specifically, DMFT underestimates its (negative) magnitude. This effect is further amplified in the strong-coupling pseudogap, where some differences also start to occur between the imaginary parts of the local self-energies. In the last column of Fig.~\ref{fig_PD_DMFT}, we compare the local DMFT self-energy with the momentum resolved CDet self-energy at the center (C) and edge (E). Surprisingly we find that, of the two momenta, the local DMFT self-energy coincides almost perfectly with the CDet self-energy at the center momentum over the full parameter range that has been analysed.

\section{Cross-benchmarking with the dynamical cluster approximation} \label{appendix_DCA}

\begin{figure}[h!]
\centering
\includegraphics[width=0.8\textwidth]{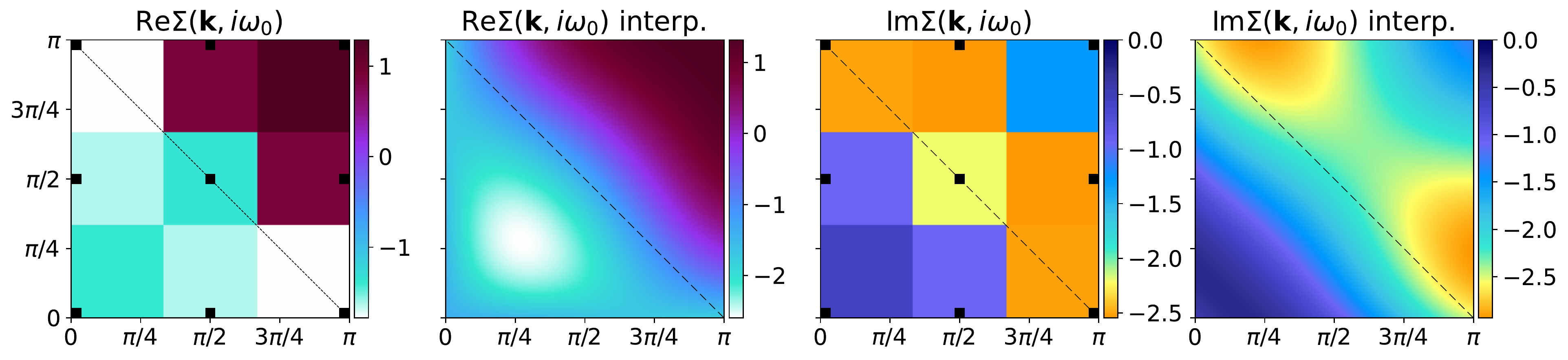}
\caption{The self-energy at $T=0.2$, $U=7$ and $n=0.975$ as obtained from 16-site DCA, with and without a momentum-space interpolation. Black squares signify actually computed momentum points.}
\label{fig_DCA_interp}
\end{figure}

Finally, we cross-benchmark our CDet results for the thermodynamic limit with 16-site dynamical cluster approximation (DCA) calculations, which is a cluster extension of DMFT. Ideally we would like to be able to compare the full momentum dependence of the two methods, however DCA only gives us access to six distinct momentum points. For this reason we use a cubic interpolation of the available points, as shown for the real and imaginary parts of the self-energy in Fig.~\ref{fig_DCA_interp}.

\begin{figure}[h!]
\centering
\includegraphics[width=0.49\textwidth]{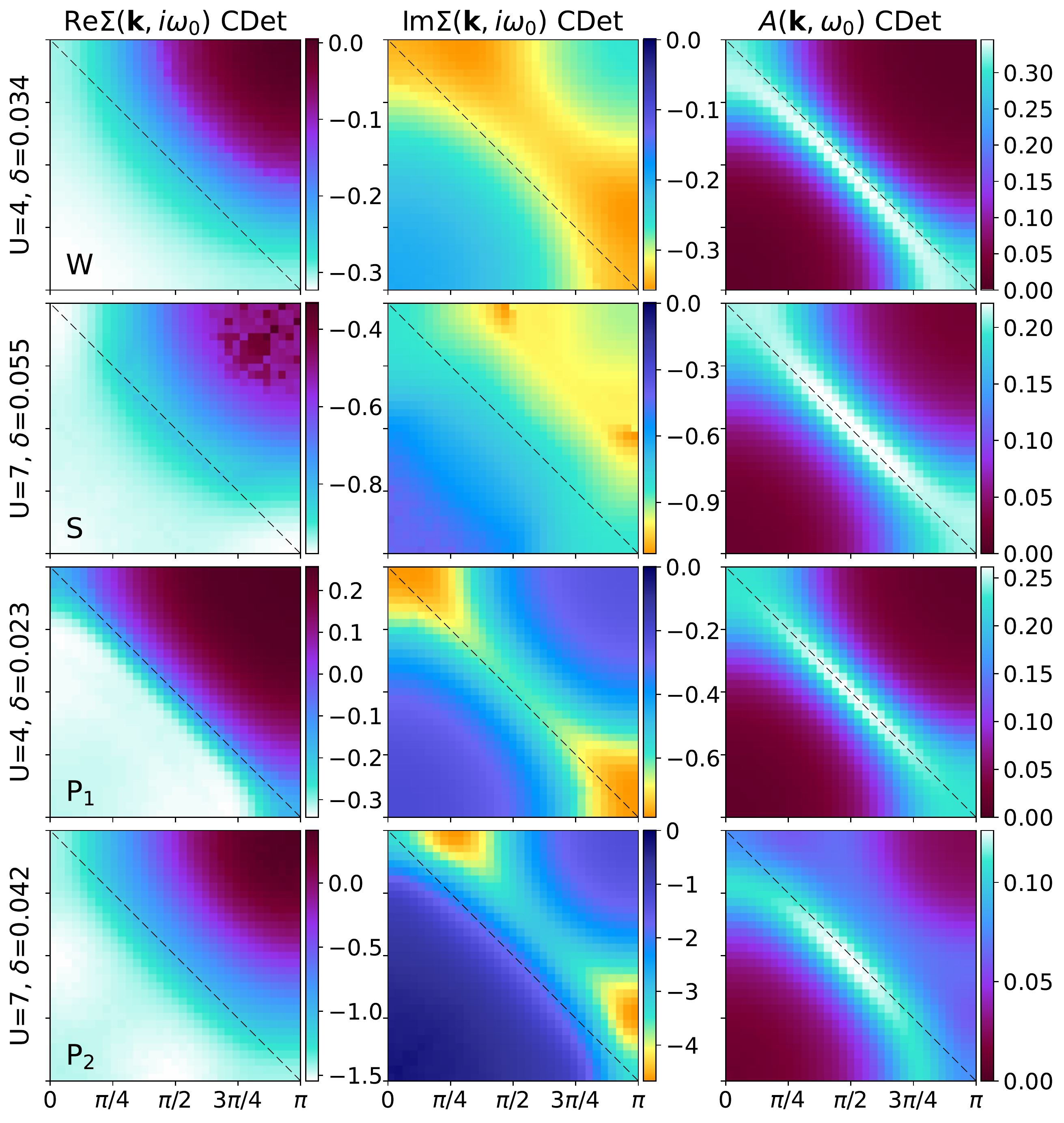}
\includegraphics[width=0.49\textwidth]{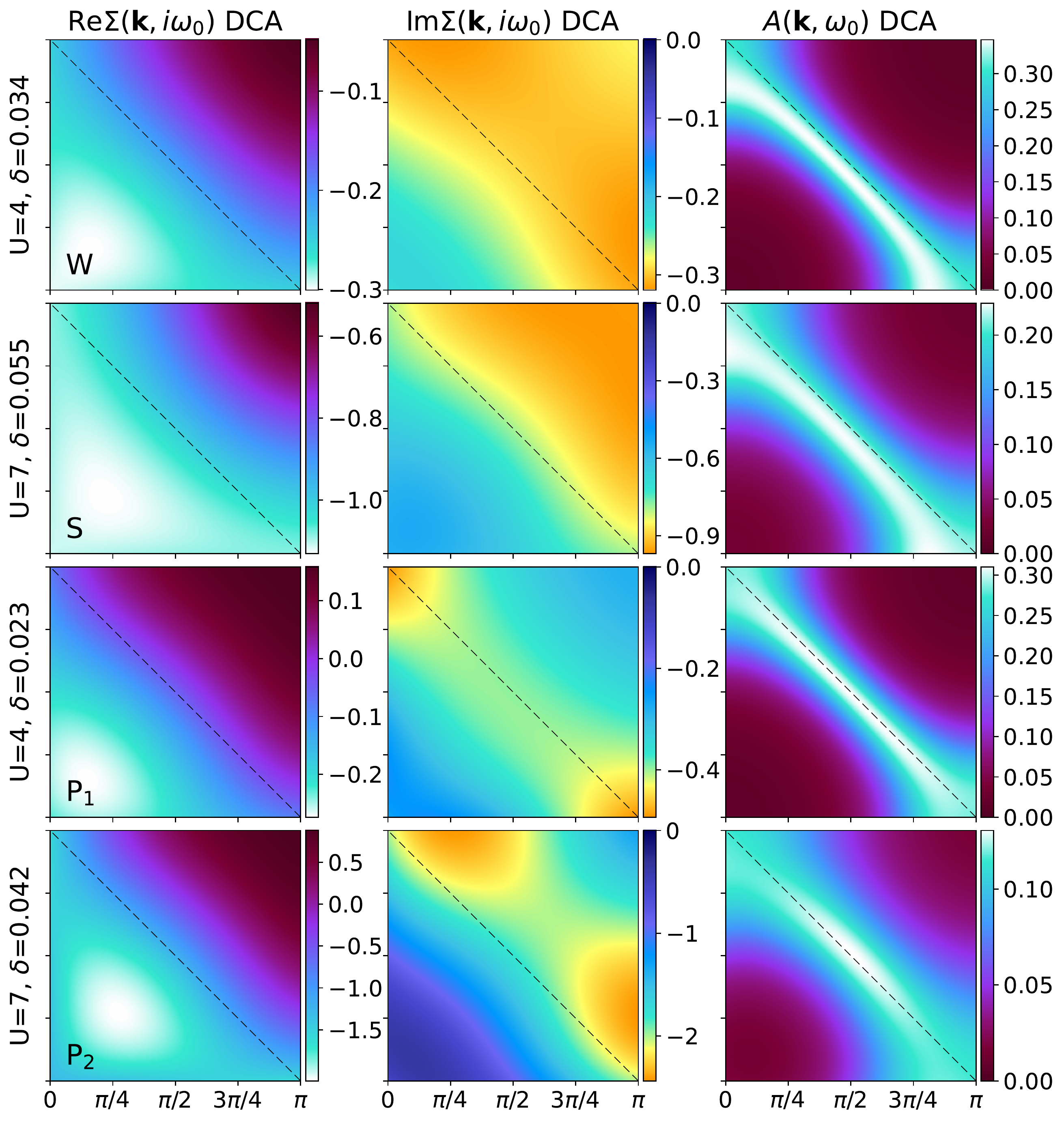}
\caption{A comparison of real and imaginary parts of the self-energy as well as the spectral function as computed by CDet (left) and interpolated 16-site DCA (right). Points were chosen to roughly correspond to Fig.~\ref{fig_regions} of the main text.}
\label{fig_DCA_vs_CDet}
\end{figure}

We proceed to compare our interpolated DCA to results for the self-energy and spectral function against their CDet counterparts from Fig.~\ref{fig_DCA_vs_CDet}, which includes all four finite-temperature regimes identified in the main text. The real part of the self-energies match rather very well for all regimes, both in terms of absolute magnitudes and momentum distribution. However, slight differences can be observed, in particular, the minima are found along the $(q,q)$ line for DCA and along the $(\pi,q)$ line for CDet. This is likely an artefact of the cubic interpolation which was used for DCA. For the imaginary self-energy we find that the location of the minima match between the two methods, however DCA underestimates their magnitude by up to a factor two in both pseudogap regimes. Given the lacking momentum resolution, DCA is also not able to distinguish fine features, which appear in the CDet data. We observe the most striking differences between the two methods in the spectral function. The only regime producing a good match is the weakly correlated metal. In the weak-coupling pseudogap regime, DCA overestimates the maximum spectral weight by about $20\%$ and fails to capture the suppression in the antinode region. This shortcoming gets even worse in the strongly correlated metal and pseudogap regimes and is mainly due to the smaller values of the imaginary self-energy produced by DCA. In both strong-coupling regimes this ultimately leads to stark differences in the Fermi surfaces identified by the two methods. 

We can further compare the positions of the pseudogap crossover lines obtained by the two methods from the spectral function criterion, see the left panel of Fig.~\ref{fig_PG_A_vs_Sigma}. It is evident that 16-site DCA underestimates the extent of the pseudogap region, although it finds qualitatively correctly the shape of the crossover region. We expect this discrepancy with respect to CDet to become smaller when the cluster size is further  increased, as has been shown in the case of the half-filled Hubbard model~\cite{vsimkovic2020extended}.   


\end{document}